\documentclass[12pt,letterpaper]{iopart}

\usepackage{graphicx}
\usepackage{subfigure}
\usepackage{siunitx}
\usepackage{placeins}
\usepackage{color}
\usepackage{standalone}
\usepackage{dcolumn}
\usepackage{bm}
\usepackage{microtype}
\usepackage{etoolbox}
\usepackage{amssymb}
\usepackage{accents}
\usepackage[normalem]{ulem}
\usepackage[dvipsnames]{xcolor}
\usepackage[colorlinks,urlcolor=NavyBlue,citecolor=NavyBlue,linkcolor=NavyBlue,pdfusetitle]{hyperref}
\usepackage[all]{hypcap}
\usepackage{enumerate}
\usepackage{enumitem}
\usepackage{gensymb}
\usepackage{comment}
\graphicspath{{./figs/}}

\newtoggle{commentsoff}
\newtoggle{highlight}

\ifdefined\nocomments
	\toggletrue{commentsoff}
\fi

\ifdefined\highlight
    \toggletrue{highlight}
\fi

\iftoggle{commentsoff}{
	\newcommand*{\tmp}[1]{}
	\newcommand*{\red}[1]{}
	\newcommand*{\ls}[1]{}
	\newcommand*{\owner}[1]{}
}{
	\newcommand*{\tmp}[1]{{\color{red} [{\bf Placeholder}: #1]}}
	\newcommand*{\red}[1]{{\color{red} #1}}
	\newcommand*{\ls}[1]{{\color{red} [{\bf LS}: #1]}}
	\newcommand*{\owner}[1]{{\color{blue} [{\bf Owner}: #1]}}
}

\iftoggle{highlight}{
	\newcommand*{\change}[1]{{\color{blue}{\bf #1}}}
}{
    \newcommand*{\change}[1]{#1}
}

\newcommand{\dcc}{LIGO--P1900245}

\begin{document}
\title[]{Characterization of systematic error in Advanced LIGO calibration}

\author{
Ling~Sun$^{1,2}$, 
Evan~Goetz$^{3}$,
Jeffrey~S.~Kissel$^{4}$,
Joseph~Betzwieser$^{5}$,
Sudarshan~Karki$^{6}$,
Aaron~Viets$^{7}$,
Madeline~Wade$^{8}$,
Dripta~Bhattacharjee$^{6}$,
Vladimir~Bossilkov$^{9}$,
Pep~B.~Covas$^{10}$,
Laurence~E.~H.~Datrier$^{11}$, 
Rachel~Gray$^{11}$,
Shivaraj~Kandhasamy$^{12}$,
Yannick~K.~Lecoeuche$^{4}$,  
Gregory~Mendell$^{4}$,
Timesh~Mistry$^{13}$,
Ethan~Payne$^{14}$,
Richard~L.~Savage$^{4}$,
Alan~J.~Weinstein$^{1}$,
Stuart~Aston$^{5}$,  
Aaron~Buikema$^{15}$, 
Craig~Cahillane$^{1}$,
Jenne~C.~Driggers$^{4}$,
Sheila~E.~Dwyer$^{4}$,
Rahul~Kumar$^{4}$,
and Alexander~Urban$^{16}$}

\address {$^{1}$LIGO, California Institute of Technology, Pasadena, CA 91125, USA}
\address{$^{2}$OzGrav-ANU, Centre for Gravitational Astrophysics, College of Science, The Australian National University, ACT 2601, Australia}
\address {$^{3}$University of British Columbia, Vancouver, BC V6T 1Z4, Canada}
\address {$^{4}$LIGO Hanford Observatory, Richland, WA 99352, USA }
\address {$^{5}$LIGO Livingston Observatory, Livingston, LA 70754, USA}
\address {$^{6}$Institute of Multi-messenger Astrophysics and Cosmology, Missouri Institute of Science and Technology, Rolla, MO 65409, USA}
\address {$^{7}$Concordia University Wisconsin, 12800 N Lake Shore Dr, Mequon, WI 53097, USA }
\address {$^{8}$Kenyon College, Gambier, OH 43022, USA }
\address {$^{9}$OzGrav, University of Western Australia, Crawley, Western Australia 6009, Australia}
\address {$^{10}$Universitat de les Illes Balears, IAC3---IEEC, E-07122 Palma de Mallorca, Spain}
\address {$^{11}$SUPA, University of Glasgow, Glasgow G12 8QQ, UK}
\address {$^{12}$Inter-University Centre for Astronomy and Astrophysics, Pune 411007, India }
\address {$^{13}$The University of Sheffield, Sheffield S10 2TN, UK }
\address {$^{14}$OzGrav, School of Physics \& Astronomy, Monash University, Clayton 3800, Victoria, Australia }
\address {$^{15}$LIGO, Massachusetts Institute of Technology, Cambridge, MA 02139, USA }
\address {$^{16}$Louisiana State University, Baton Rouge, LA 70803, USA }

\ead{ling.sun@ligo.org}

\vspace{10pt}
\begin{indented}
\item[]1 September 2020
\end{indented}

\begin{abstract}
	The raw outputs of the detectors within the Advanced Laser Interferometer Gravitational-Wave Observatory need to be calibrated in order to produce the estimate of the dimensionless strain used for astrophysical analyses.
The two detectors have been upgraded since the second observing run and finished the year-long third observing run.
Understanding, accounting, and/or compensating for the complex-valued response of each part of the upgraded detectors improves the overall accuracy of the estimated detector response to gravitational waves.
We describe improved understanding and methods used to quantify the response of each detector, with a dedicated effort to define all places where systematic error plays a role.
We use the detectors as they stand in the first half (six months) of the third observing run to demonstrate how each identified systematic error impacts the estimated strain and constrain the statistical uncertainty therein. 
For this time period, we estimate the upper limit on systematic error and associated uncertainty to be $< 7\%$ in magnitude and $< 4$~deg in phase (68\% confidence interval) in the most sensitive frequency band 20--2000~Hz.
The systematic error alone is estimated at levels of $< 2\%$ in magnitude and $< 2$~deg in phase.
\end{abstract}

%
%
%
%
%

\section{Introduction}
\label{sec:introduction}

The Advanced Laser Interferometer Gravitational-Wave Observatory (Advanced LIGO) detectors~\cite{LIGO2014} and the Virgo detector~\cite{Virgo2014} have directly observed transient gravitational waves from multiple binary black hole coalescences and one binary neutron star merger in the first and second observing runs~\cite{LVC-catalog}.
After a series of instrument upgrades to further improve the sensitivity, e.g., replacing test masses and optics, increasing laser power, and adding squeezed light~\cite{O3DetectorPaper}, the two LIGO detectors started the third observing run (O3), together with Virgo, on April 1st, 2019, and ended the first half of O3 (O3A) on Oct 1st, 2019 \cite{GW190425,GW190412,GW190814,O3A-Catalog,GraceDB}.

The time series of dimensionless strain, $h$, measured by each detector and used to determine the detection of a gravitational-wave (GW) signal and infer the properties of the astrophysical source, is reconstructed from \change{the raw, digitized electrical output} of each detector.
This reconstruction process, with an accurate and precise model of the detector's response to $h$, is referred to as ``calibration."
The accuracy and precision of $h$ are important for detecting gravitational wave signals and crucial for the reconstruction of their astrophysical parameters~\cite{Lindblom2009,LVC2017,Viets2018}.

We report the accuracy and precision of $h$ by estimating the upper and lower 68\% confidence interval bounds on the systematic error and uncertainty for each detector response. \change{Systematic error is defined to be the deviation of the estimated detector response from the true detector response at a given time, and is a combination of known and estimates of unknown errors.}
The error is quantified by propagating the \change{measured} error of each response component through the overall response \change{of a given} detector.
The \change{associated} uncertainty \change{of this collection of measured systematic errors} arises from either the random \change{statistical} noise in the measurements, repeated sampling of parameters from a random parent distribution, or \change{the uncertainty from quantifying a systematic error with unknown physical source.}
The resulting \change{systematic error} and \change{associated uncertainty intervals} of the error in the detector response, and thus in $h$, are  complex-valued, frequency-dependent functions.
Photon calibrators (Pcal), which independently use photon radiation pressure to produce strain within the detector~\cite{Karki2016,Pcalpaper-P2000113}, are the primary absolute reference used to validate the estimates of $h$ itself as well as the error and uncertainty of the detector response.
We expect the ratio of the estimated $h$ to the strain produced by the Pcal systems to lie within the error and uncertainty bounds 68\% of the time~\cite{LVC2017, Cahillane2017}.
\change{When and where possible, we correct for the errors in $h$, if the physical mechanism of the error is sufficiently understood and if the error can be quantified with minimal uncertainty. 
The final systematic error of the detector response presented in this paper is the estimate of what is left uncorrected in the reconstructed $h$ used for astrophysical data analyses in O3A.}

In the first and second observing runs of Advanced LIGO (O1 and O2), we achieved a combined error and uncertainty limit \change{(68\% confidence interval bounds)} of $\lesssim 5\%$ in magnitude and $\lesssim 3$~deg in phase in the most sensitive frequency band 20--2000~Hz.
The method to determine those estimates for O1 and O2 is presented in~\cite{Cahillane2017}. 
Details of how a complete model is used to compute and produce the $h$ data stream can be found in~\cite{Viets2018}.
In this paper, we update the discussion of the methods in~\cite{Cahillane2017}, use new studies of the upgraded O3A detectors to elucidate all sources of systematic error considered, and estimate the contribution of each source through measurements and Bayesian inference.

\begin{figure}
	\begin{center}
		\includegraphics[width=0.6\textwidth]{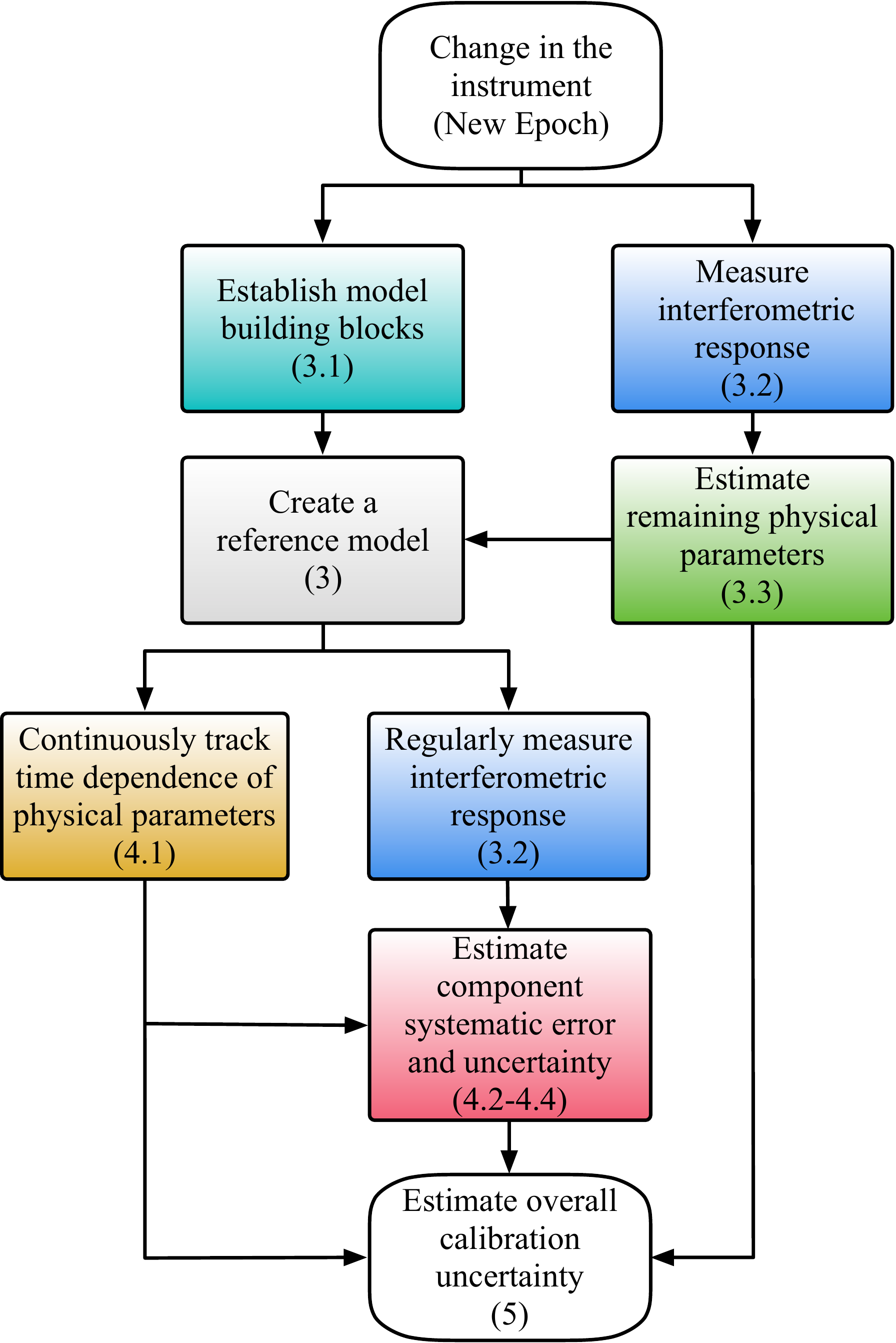}
	\end{center}
	\caption{Flowchart of how the detector systematic error and uncertainty estimate is produced. Parenthetical numbers in each box guide the reader to the corresponding section in the paper.} \label{fig:flowchart}
\end{figure}

The structure of the paper is as follows.
In \sref{sec:model}, we review the model components of the detector response to $h$, adding new qualifying details that are important to the O3 detectors, and discuss each component's contribution to the detector response.
In sections~\ref{sec:observables} and \ref{sec:syserrors}, we describe the procedure for creating a detector response model and estimating its error and uncertainty, following the workflow in \fref{fig:flowchart}:
With a verified absolute calibration reference, a model of the detector mechanical dynamics, and detailed measurement of the detector electronics (\sref{sec:precursory}), we estimate remaining detector response parameters through Markov chain Monte Carlo (MCMC) analysis of \change{a single set of} interferometric measurements \change{taken at the reference time to create a static, reference model} (sections~\ref{sec:meas} and \ref{sec:mcmc}).
We then discuss how continuous time dependence in model parameters within a given observation period are tracked and accounted for (\sref{sec:tdcf}), limitations of the O3A detector model components (\sref{sec:detuning}), estimation of residual frequency-dependent error and statistical uncertainty through Gaussian Process Regression (GPR) methods \change{using multiple sets of interferometric measurements} (\sref{sec:residual}), and the negligible and/or unaccounted for systematic errors in each component (\sref{sec:uncompensated}).
\change{Finally, after all the well-understood systematic errors are corrected for, the residual static (i.e., time-independent) and time-dependent errors and their associated uncertainties, as well as other statistical uncertainties in the measurements, are all collected from each modeled component and propagated to the detector response function. The final numerically estimated error and uncertainty for each detector response in O3A are presented in \sref{sec:results}.}
We summarize and conclude in \sref{sec:conclusion}.

\section{Model fundamentals}
\label{sec:model}

\begin{figure}
	\begin{center}
		\includegraphics[width=0.85\textwidth]{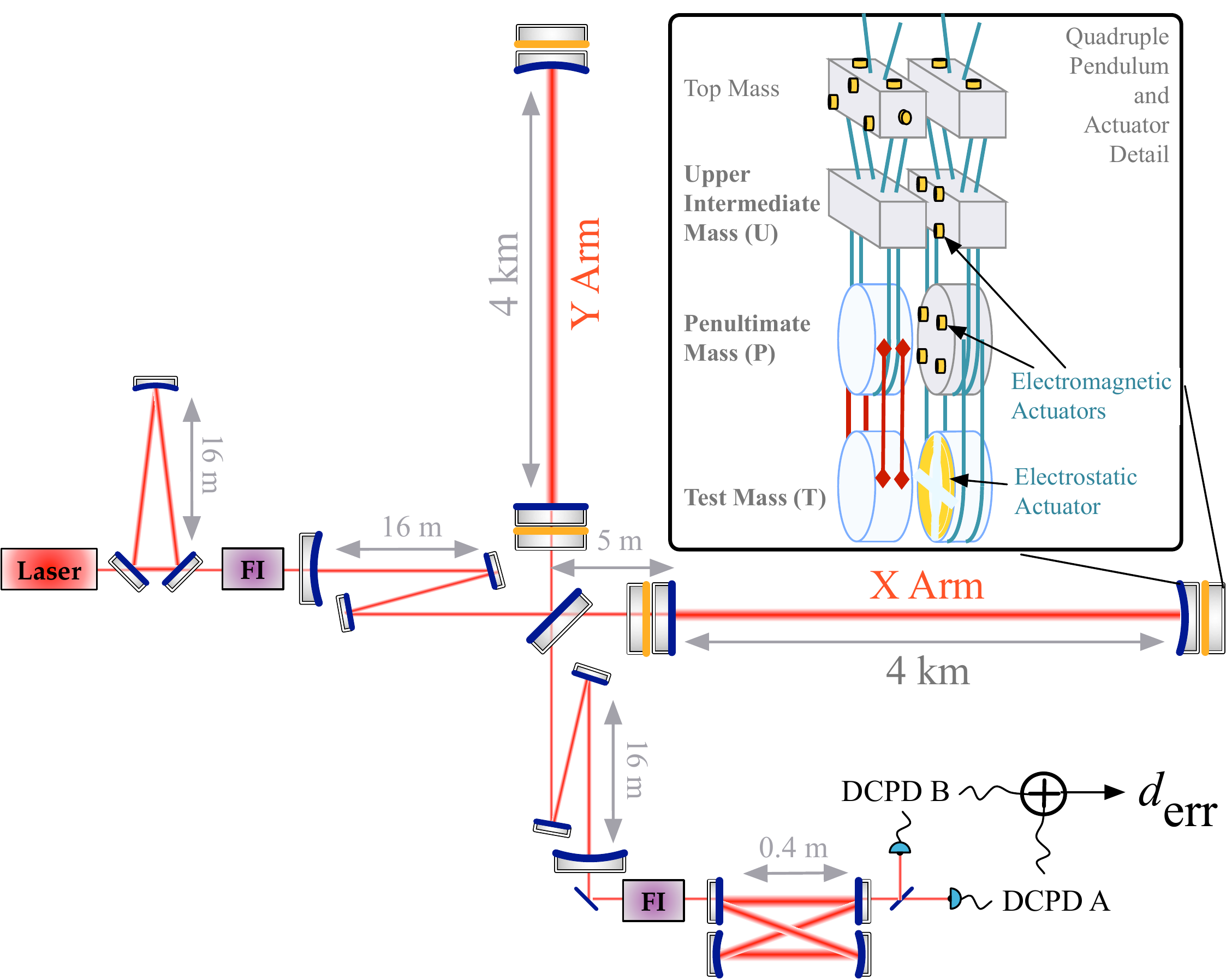}
	\end{center}
	\caption{Conceptual diagram of the optical configuration of the Advanced LIGO interferometers: dual-recycled, Fabry-P\'{e}rot Michelson. The X and Y arms are 4-km-long, Fabry-P\'{e}rot cavities formed by the highly reflective end test masses and partially transmissive input test masses. Pre-stabilized laser light enters the detector from the left, and is further stabilized using an input mode cleaner optical cavity. Cleaned light then enters the Power Recycling Cavity (formed by a partially transmissive input coupler and two high reflectors), is split by a 50/50 beamsplitter, and sent into the long arm cavities where the light interacts most with the potentially changing gravitational field. The light returning from the arm cavities interferes at the beamsplitter, and is then extracted from the beamsplitter's anti-symmetric port by the Signal Recycling Cavity (SRC), similarly formed by two high reflectors and a partially transmissive output coupler. Finally, light exiting the SRC is cleaned with an additional resonant cavity, referred to as the ``output mode cleaner''. Faraday Isolators (FI) are used for optical isolation of the main interferometer from the rest of the instrument. The transmitted light of the output mode cleaner is split onto two photodiodes, whose output current is turned to voltage, conditioned, digitized, de-conditioned digitally, and then linearly combined to form $d_{\rm err}$. Inset: one of the full quadruple pendulum suspension systems and its actuators. 
	\label{fig:IFOOpticalLayout}}
\end{figure}

While the instrument has been upgraded between O2 and O3~\cite{O3DetectorPaper}, the conceptional design of the Advanced LIGO detectors has not changed fundamentally since the first observation of gravitational waves \cite{GW150914}, as described in, e.g.,~\cite{abbott2016gw150914a}.
The optical configuration of the two LIGO detectors remain dual-recycled, Michelson interferometers with 4-km-long Fabry-P\'{e}rot resonant cavities (\fref{fig:IFOOpticalLayout}).
These detectors have been built to measure \change{a dimensionless strain incident upon them. This dimensionless strain, denoted by $h$,} is defined by the differential changes in arm length (DARM length) $\Delta L_{\rm free}$ divided by the average length of the arms $L$, 
\begin{equation}
\label{eqn:strain}
h = \frac{\Delta L_{\rm free} }{L}  = \frac{\Delta L_{\rm x}  -\Delta L_{\rm y} }{L}, 
\end{equation}
where $\Delta L_{\rm x}$ and $\Delta L_{\rm y}$ are the displacements in the two orthogonal arms, X and Y, respectively.
Due to the presence of noise and the desire to maintain the resonance condition of the optical cavities, the detectors do not directly measure $\Delta L_{\rm free}$.
Instead $\Delta L_{\rm free}$ is derived from the error and control signals of the DARM control loop, using methods described previously in~\cite{LVC2017,Viets2018,Cahillane2017,Tuyenbayev2016}.

\begin{figure}[!tbh]
	\centering
	\includegraphics[width=\textwidth]{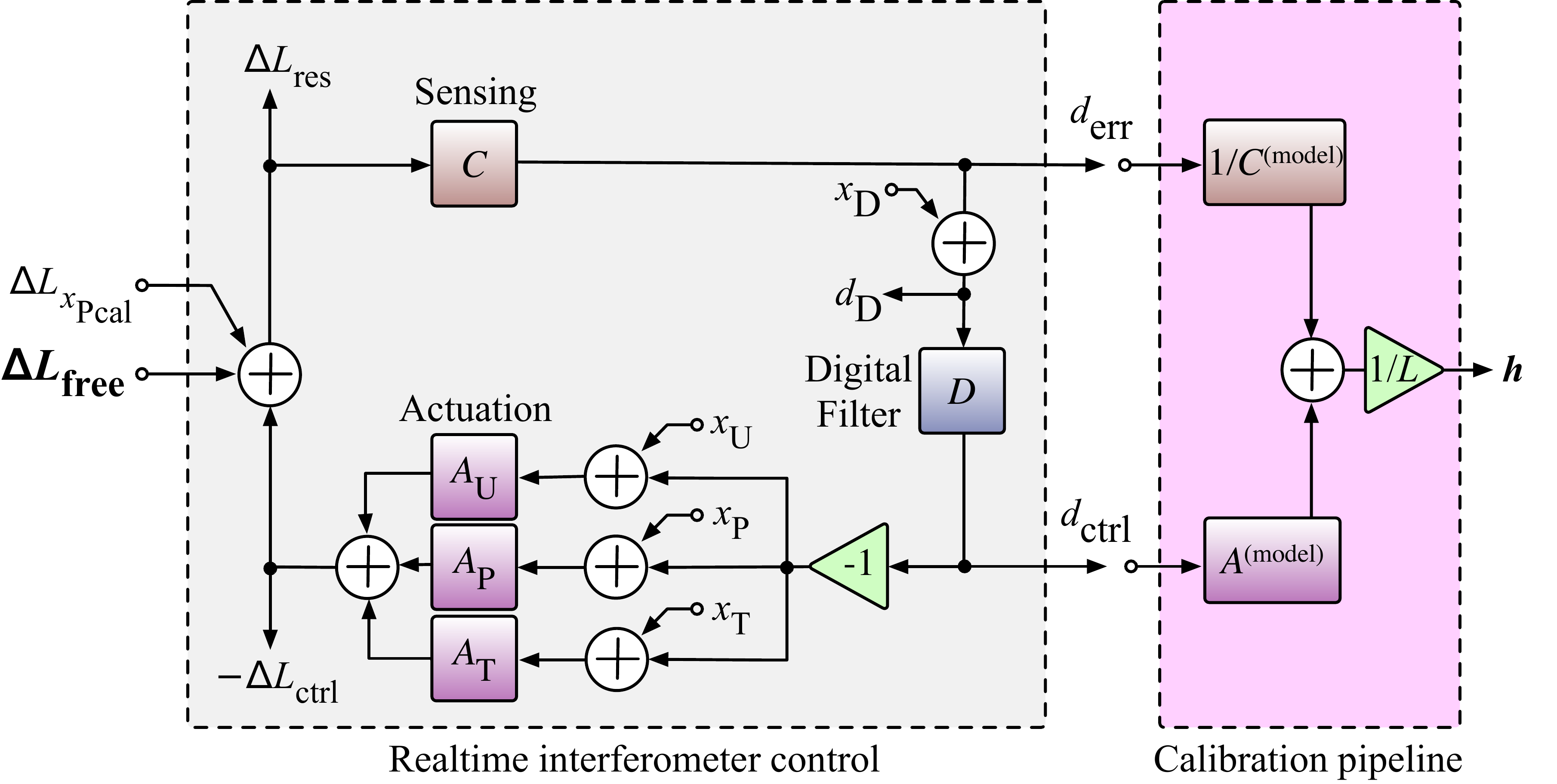}
	\caption[]{Advanced LIGO differential arm (DARM) length feedback control loop (gray box) and the generation of calibrated strain data (pink box). The sensing function $C$ converts the residual DARM displacement $\Delta L_{\rm res}$ to the digital error signal $d_{\rm err}$. The digital filter $D$ processes $d_{\rm err}$ and produces the digital control signal $d_{\rm ctrl}$. The actuation function \change{$A_i$ ($i=U,P,T$; refer to \fref{fig:IFOOpticalLayout} for the definitions of $U,P,T$)} converts $d_{\rm ctrl}$ \change{to the control force allocated to the test masses that form the arm cavities}, producing displacement $-\Delta L_{\rm ctrl}$ to suppress $\Delta L_{\rm free}$. During the time dedicated to loop characterization (see \sref{sec:meas}), DARM displacement excitations $\Delta L_{x_{\rm Pcal}}$ are added using the photon radiation pressure actuator system. Similarly, $x_{\rm D}$ and $x_i$ ($i=U,P,T$) are added using the quadruple pendulum actuator via the digital control system. In the presence of $x_{D}$,  the digital signal $d_{\rm D} = d_{\rm err} + x_{\rm D}$ may be used to characterize the DARM loop suppression. In the pink box, \change{the estimated DARM strain} $h$ is constructed using the sensing and actuation models, $C^{\rm (model)}$ and $A^{\rm (model)}$.}
	\label{fig:cal_model}
\end{figure}

\change{In this paper, we describe the procedure entirely in frequency domain. See \cite{Viets2018} for the discussion of reconstructing $h$ in time domain.}
\Fref{fig:cal_model} shows the interferometer DARM feedback control loop and the calibration process at a conceptual level. 
The loop contains the physical interferometer, analog electronics, analog-to-digital converters, a network of ``front-end'' computers, and digital-to-analog converters, as described in \cite{bork2011,bork2020advligorts}.
The residual DARM displacement $\Delta L_{\rm res}$ is converted by the sensing function $C$, to produce the digital output $d_{\rm err}$. 
The error signal is filtered through a set of digital filters $D$, creating the digital control signal, $d_{\rm ctrl}$ (i.e., $d_{\rm ctrl} = D d_{\rm err}$). 
The actuation function $A$ converts $d_{\rm ctrl}$ to the control displacement $-\Delta L_{\rm ctrl}$
that suppresses $\Delta L_{\rm free}$ caused by external stimuli, holding the optical cavities on resonance and leaving a small amount of $\Delta L_{\rm res}$ in the DARM loop. 
Conceptually, $\Delta L_{\rm free}$ is reconstructed with models of these functions as
\begin{equation}
\label{eqn:displacement}
\Delta L_{\rm free} = \Delta L_{\rm res} + \Delta L_{\rm ctrl} = \frac{1}{C^{\rm (model)}}  d_{\rm err} + A^{\rm (model)}  d_{\rm ctrl}.
\end{equation}
We can define a response function, ${R}^{\rm (model)}$, given by
\begin{eqnarray}
\label{eqn:response}
{R}^{\rm (model)} = \frac{1+{A}^{\rm (model)} {D} {C}^{\rm (model)}}{{C}^{\rm (model)}} = \frac{1+{G}^{\rm (model)}}{{C}^{\rm (model)}},
\end{eqnarray}
where ${G}^{\rm (model)} \equiv {A}^{\rm (model)} {D} {C}^{\rm (model)}$ is the DARM open loop gain, such that 
\begin{equation}
\label{eqn:dLfree}
h = \frac {R^{\rm (model)}d_{\rm err}}{L}.
\end{equation}
\change{Note that the estimated DARM strain $h$ output from the pink box in \fref{fig:cal_model} is not the GW strain.}

It is desirable to produce calibrated strain with low latency for quick electromagnetic follow-up. 
To fulfill this desire, a reasonably accurate, low-latency estimate of $h$ is created in near real-time. 
Later, a carefully-vetted, most-accurate estimate of $h$ is delivered within a few months after the raw data are stored.

The low-latency estimate of $h$ is produced from the model in two parts.
In the first part, the models $A^{\rm (model)}$ and $C^{\rm (model)}$ are \change{reproduced with moderate fidelity} by infinite impulse response (IIR) filters, which modify copies of $d_{\rm err}$ and $d_{\rm ctrl}$ in near real-time on a parallel computer within the network of the feedback control system \change{to create estimates of $\Delta L_{\rm res}$ and $\Delta L_{\rm ctrl}$.}
These estimates are summed to form a \change{crude} version of $\Delta L_{\rm free}$, and \change{all are} stored for later consumption. 
This ``front-end'' production of $\Delta L_{\rm free}$ is limited \change{in fidelity} by causality and the finite sample rate of the \change{computer network}, but good enough to assess the detector noise performance \change{in near real-time}. 
However, the systematic errors in the \change{moderate-fidelity,} front-end production of $\Delta L_{\rm free}$ \change{are too large to} be used in \change{detailed} astrophysical analyses.
As such, in the second part, $\Delta L_{\rm res}$ and $\Delta L_{\rm ctrl}$ are pulled from the front-end storage, modified further with finite impulse response (FIR) filters derived from the model, \change{and divided by $L$ to produce an high-fidelity estimate of $h$} with $\sim$10~seconds of latency \change{and manageable systematic error.}

This low-latency (online) estimate of $h$ uses the best models of the detector at the time of recording.  
Over the course of any observing run, data dropouts due to computer failures, mistakes in modeling $A^{\rm (model)}$ and $C^{\rm (model)}$, and unknown residual systematic errors are often identified. 
Further, methods may be developed at a later time to correct for systematic errors.
Finally, there are known model components excluded from the IIR and FIR reproductions of $A^{\rm (model)}$ and $C^{\rm (model)}$ for expediency, which create further, albeit small, systematic error in all online estimated $h$. 

Hence it motivates the creation of an additional high-latency (offline) estimate of $h$, allowing for improved accuracy, which uses the best models developed after the low-latency data are collected, stored, and understood.
The offline estimate of $h$ is created entirely with the FIR reproductions of $A^{\rm (model)}$ and $C^{\rm (model)}$, starting ``from scratch'' with $d_{\rm err}$ and $d_{\rm ctrl}$.
Further details of the computational software and methods for producing these versions of $h$ can be found in \cite{Viets2018}.

In this paper, we focus on the systematic error and uncertainty of the offline estimated $h$. The accuracy and precision of any estimate of $h$ for a given detector is quantified by comparing a large collection of independent measurements of the detector response using the actuation excitation paths at $x_{\rm Pcal}$, $x_{\rm D}$, and $x_i$ ($i=U,P,T$) against the model $R^{\rm (model)}$.
In sections~\ref{subsec:sensing} and \ref{subsec:actuation}, we detail the components and parameters of the sensing and actuation function models, $C^{\rm (model)}$ and $A^{\rm (model)}$, respectively.
\Sref{subsec:tdep} describes causes and impacts of slow time-variation of these frequency-dependent functions.
The frequency-dependent contribution of $C$ and each component in $A$ to the response $R^{\rm (model)}$ is discussed in \sref{subsec:contribution}. The systematic error in $R^{\rm (model)}$ and the impact from each component are defined in \sref{subsec:model_sys_err}.

\subsection{Sensing function}
\label{subsec:sensing}
The sensing function $C(f)$ is the response of the filtered, digitized combination of photo-detector output signals, i.e. $d_\textrm{err}$, to the residual DARM displacement, $\Delta L_\textrm{res}$.
This response is complex-valued (amplitude and phase), frequency-dependent, and slowly time-varying. 
It is comprised of a linear combination of several conceptually different parts: \change{(a)} the opto-mechanical, interferometric response to $\Delta L_{\rm res}$, producing power (in units of watts) at the output of the signal recycling mirror, \change{(b)} the opto-electronic processing of that power into photo-current, including any optical loss on the path to and through the output mode cleaner, the final beamsplitter ratio as the transmitted light is sent to the readout photodiodes, and the photodiodes' response, \change{(c)} the analog signal processing electronics for the photodiodes and analog-to-digital conversion process which turn photo-current into digital counts, and \change{(d)} the conditioning, re-combination, and linearization of those digitized counts into a suitable, single error signal for the DARM control loop. 

In O3, we retain the same frequency-domain model transfer function for this collection of conceptual parts (at a particular time) as in~\cite{Cahillane2017}, analytically given by,
\begin{equation}
{C}^{\rm (model)}(f) = \left(\frac{H_C}{1+if f_{cc}^{-1}}\right) \left(\frac{f^2}{f^2+f_s^2 - if f_s Q^{-1}}\right)C_R(f)\exp(-2\pi i f \tau_C).  
\label{eqn:static_sensing}
\end{equation}
The overall gain of the sensing function, $H_C$, is the product of the scalar gains from each component in all four parts (in units of digital counts of $d_{\rm err}$ per meter of DARM length).
Aside from $H_{C}$, the frequency dependence of part \change{(a)}, represented by the first two parenthetical terms, defines the response of the coupled Fabry-P\'{e}rot arm cavities and signal recycling cavity (SRC) to $\Delta L_{\rm free}$.
In the first parenthetical term, $f_{cc}$ is the differential coupled-cavity pole frequency.
In the second term, the numerator represents two zeros at $0$~Hz, and $f_s$ and $Q$ in the denominator are, respectively, the pole frequency and quality factor. 
Collectively the zeros and poles of the second parenthetical term represent the optical spring response created by any detuning present between the SRC and the arm cavities. 
The approximations and deficiencies within the first two terms are described in sections~\ref{sec:detuning} and \ref{sec:uncompensated}.
The collective frequency response of the analog electronics described in parts \change{(b)--(d)} are addressed in two ways.
Some portions of the response are paired with corresponding inverse digital filters, applied after the photodiode signal is digitized. 
Thus, they compensate the analog response within the DARM loop itself (``in-loop") and are not explicitly included in \eref{eqn:static_sensing}. 
The portions in parts \change{(b)--(d)} not compensated in-loop are collected within $C_R$ (see further discussion in sections~\ref{sec:precursory} and \ref{sec:uncompensated}). 
The collection of analog and digital time delays from all four parts is denoted by $\tau_C$ in the final term.

\subsection{Actuation function}
\label{subsec:actuation}

The actuation function, $A$, is the response of the control DARM displacement, $\Delta L_{\rm ctrl}$, to the requested digital control signal, $d_{\rm ctrl}$. Like the sensing function described above, it is composed of several components.
We first qualify the O3 actuator model by extending the discussion in previous work~\cite{Viets2018,Cahillane2017}.

First, we consider the DARM control system only in the frequency band above 5~Hz.
Below 5~Hz, actuation from absolute references (such as the Pcal) cannot be sufficiently resolved in the detector noise in $\Delta L_{\rm free}$.
Hence, any further allocation of $\Delta L_{\rm ctrl}$ to other actuators below 5~Hz, e.g., to the first, top-most stage of the quadruple suspension, is ignored.

Second, while $\Delta L_{\rm ctrl}$ may be induced by actuating on any stage of any of the four arm cavity optics quadruple pendulum systems~\cite{robertson2002quadruple, aston2012update}, we reduce complexity by only modeling the DARM control actuator as the bottom three stages of a single quadruple pendulum.
In other words, if the upper two suspension stages of the X arm end test mass and the final test mass stage in the Y arm are used in combination to produce $\Delta L_{\rm ctrl}$, each stage is measured and modeled independently, and the actuation from each stage is summed as though created by single quadruple pendulum.

Third, each detector has many other cavities length and angle degrees of freedom that must be controlled. 
Some of those control systems also use the quadruple suspension systems as actuators. 
These auxiliary control loops will only impact the DARM loop response if there is cross-coupling from $\Delta L_{\rm ctrl}$ to the auxiliary degree of freedom and from the auxiliary degree of freedom back to $\Delta L_{\rm res}$. 
Reducing potential auxiliary cross-coupling to DARM is an essential element of the Advanced LIGO detectors collective control system design~\cite{LIGO2014, hall2017long}. Further, the O3 detectors only use three actuation stages among the six available lower stages of the end test mass suspensions to create $\Delta L_{\rm ctrl}$.
The actuation model does not include any cross-coupling with auxiliary degrees of freedom.

Finally, each quadruple pendulum system is actually a pair of closely adjacent quadruple suspensions, with the ``main chain'' holding the suspended test mass, and the ``reaction chain'' suspending equally isolated masses upon which the actuators are mounted (see the inset of \fref{fig:IFOOpticalLayout} and~\cite{robertson2002quadruple, aston2012update}).
Among the lowest three stages of each quadruple suspension, the upper intermediate (UIM), and penultimate (PUM), are driven by magnetic coil actuators.
The lowest stage of the suspension, named the test mass (TST) stage, is driven by an electrostatic actuator system.
The force from the actuators on the reaction chain is considered to be applied directly to the center of the mass at each stage of the main chain.
For the purpose of estimating the displacement of the test mass, 
only the dynamic response of the main chain is modeled;
it is not necessary to take into consideration the added complexity of the reaction chain.

With these qualifying remarks, the response of each actuator stage is modeled as \change{(a)} the digital distribution system which allocates $d_{\rm ctrl}$ (i.e., the filtered $d_{\rm err}$) to the computer that controls the three end test mass suspension stages, where subsequent digital filtering (i.e., the assignment of frequency-dependent control authority) and signal conditioning occurs, \change{(b)} the digital-to-analog converters and associated signal processing electronics that convert the conditioned digital signal into electrical signal suitable for that stage's actuator, \change{(c)} the mechanical pendulum dynamics of the stage's actuator itself, and \change{(d)} the mechanical, force-to-displacement response of quadruple pendulum suspension system in the DARM direction from the given stage to the optic.

Thus, the total actuation model (at a particular time) is similar to that in~\cite{LVC2017,Cahillane2017,Viets2018}, with only slight modifications,
\begin{eqnarray}
\nonumber
	{A}^{\rm (model)}(f) &=& F_U(f)H_U{A}_U(f)\exp(-2\pi i f \tau_{U}) \nonumber\\
	& & + F_P(f)H_P{A}_P(f)\exp(-2\pi i f \tau_{P}) \nonumber\\
	& & + F_T(f)H_T{A}_T(f)\exp(-2\pi i f \tau_{T}),
	\label{eqn:static_actuation} 
\end{eqnarray}
where $U$, $P$, and $T$ represent the UIM, PUM, and TST stages, respectively (see the inset of \fref{fig:IFOOpticalLayout}).
For each stage ($i=U,P,T$),
$F_i(f)$ is the digital, frequency-dependent filter which allocates $d_{\rm ctrl}$ to the appropriate stage, $H_i$ is the overall gain, the product of the scalar gains of each component in all four parts (in units of meters of DARM length per digital count of $d_{\rm ctrl}$), and $\tau_{i}$ is the total time delay in the digital-to-analog conversion.
Similar to the sensing function, some portions of the analog electronics frequency response in part \change{(b)} are paired with inverse digital filters applied before converting the digitized signal to analog voltage, and thus not explicitly included in the model. 
Thus, ${A}_i(f)$ includes the dynamical force-to-displacement frequency response of the quadruple pendulum and the residual response from any uncompensated analog electronics (for further discussion, again, see sections~\ref{sec:precursory} and \ref{sec:uncompensated}).
Note that \eref{eqn:static_actuation} differs from the equivalent expressions in~\cite{LVC2017,Viets2018,Cahillane2017} only in the generalization of $\tau_{i}$ to be an arbitrary delay at each actuator stage, instead of a common delay for all stages.
Limitations of this model are discussed in \sref{sec:uncompensated}.

\subsection{Time dependence}
\label{subsec:tdep}

The static, reference models described by \eref{eqn:static_sensing} and \eref{eqn:static_actuation} are constructed with parameters $H_C$, $f_{cc}$, $f_s$, and $Q$ for ${C}^{\rm (model)}$ and $H_i$ ($i=U,P,T$) for ${A}^{\rm (model)}$ that are measured at a given time.
Some parameters, however, are slowly varying over time due to various physical mechanisms~\cite{TDCF-T1700106}.
Sensing function parameters $H_C$, $f_{cc}$, $f_s$, and $Q$ fluctuate on a time-scale of minutes due to the variations of optical alignment in the arm cavities, the relative alignment between the arm cavities and the SRC, and the laser power.
The overall strength of the TST electrostatic actuator changes slowly on the time-scale of days to weeks due to the slow accumulation of static charges around the test mass and reaction mass.
The overall strengths of the UIM and PUM magnetic coil actuators are expected to be static, but occasional changes in actuator electronics in the path often require compensation.
The time-dependent sensing and actuation functions are virtually identical to those in~\cite{LVC2017,Cahillane2017,Viets2018,Tuyenbayev2016}, and are summarized here:
\begin{eqnarray}
\nonumber
{C}(f;t) =& \kappa_C(t) \left(\frac{H_C}{1+if f_{cc}^{-1}(t)}\right) \left(\frac{f^2}{f^2+f_s^2(t) - if f_s(t) Q^{-1}(t)}\right) \\
&\times C_R(f) \exp(-2\pi i f \tau_C), \label{eqn:tdcf_sensing}
\end{eqnarray}
where $\kappa_C(t)$ is a dimensionless, real-valued, scalar gain factor characterizing the frequency-independent variations of $H_C$, and
\begin{eqnarray}
\nonumber
{A}(f;t) &=& \kappa_U(t)F_U(f)H_U{A}_U(f) \exp(-2\pi i f \tau_{U}) \nonumber\\
  &  & + \kappa_P(t) F_P(f)H_P {A}_P(f)\exp(-2\pi i f \tau_{P}) \nonumber\\
  &  & + \kappa_T(t) F_T(f)H_T{A}_T(f)\exp(-2\pi i f \tau_{T}),  \label{eqn:tdcf_actuation}
\end{eqnarray}
where $\kappa_U(t)$, $\kappa_P(t)$, and $\kappa_T(t)$ are similar dimensionless scalar gain factors (though in this case complex) for the UIM, PUM, and TST stages, respectively, with the real parts varying about unity.
In O1 and O2, the fluctuations in UIM and PUM stages were tracked with a combined factor $\kappa_{PU}(t)$. 
In O3, $\kappa_{PU}(t)$ is replaced by separate scalar factors $\kappa_U(t)$ and $\kappa_P(t)$ to provide more accurate tracking of temporal variation of the actuation functions. We refer to these time-dependent parameters in \eref{eqn:tdcf_sensing} and \eref{eqn:tdcf_actuation}, $\kappa_C(t)$, $f_{cc}(t)$, $f_s(t)$, $Q(t)$, and $\kappa_i(t)$ ($i = U,P,T$), as time-dependent correction factors (TDCFs).
Additional details are provided in \sref{sec:tdcf}.

\subsection{Contribution to the response function}
\label{subsec:contribution}

\begin{figure}[!tbh]
	\centering
	\subfigure[]
	{
		\label{fig:h1_contri}
		\scalebox{0.3}{\includegraphics{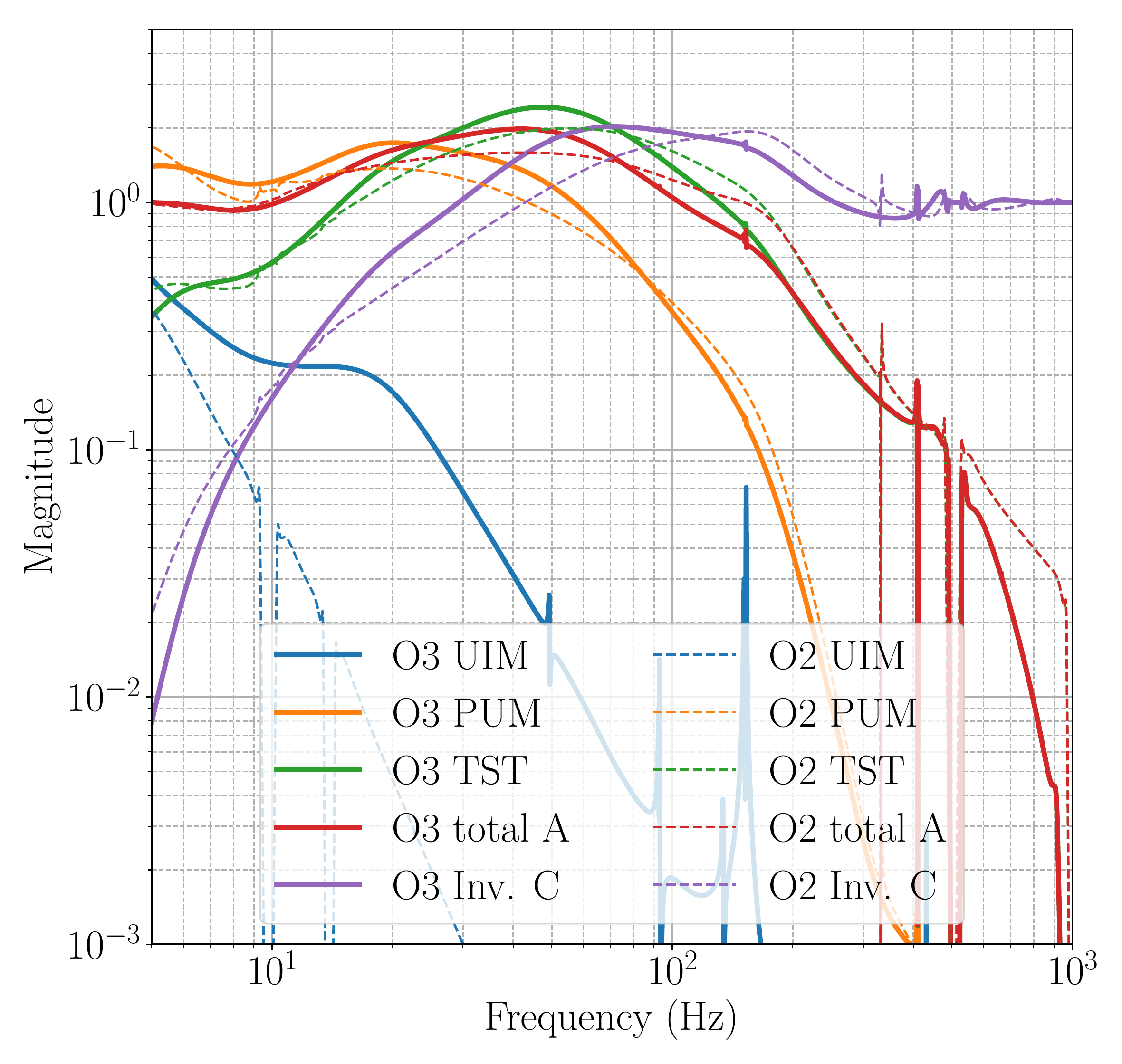}}
	}
	\subfigure[]
	{
		\label{fig:l1_contri}
		\scalebox{0.3}{\includegraphics{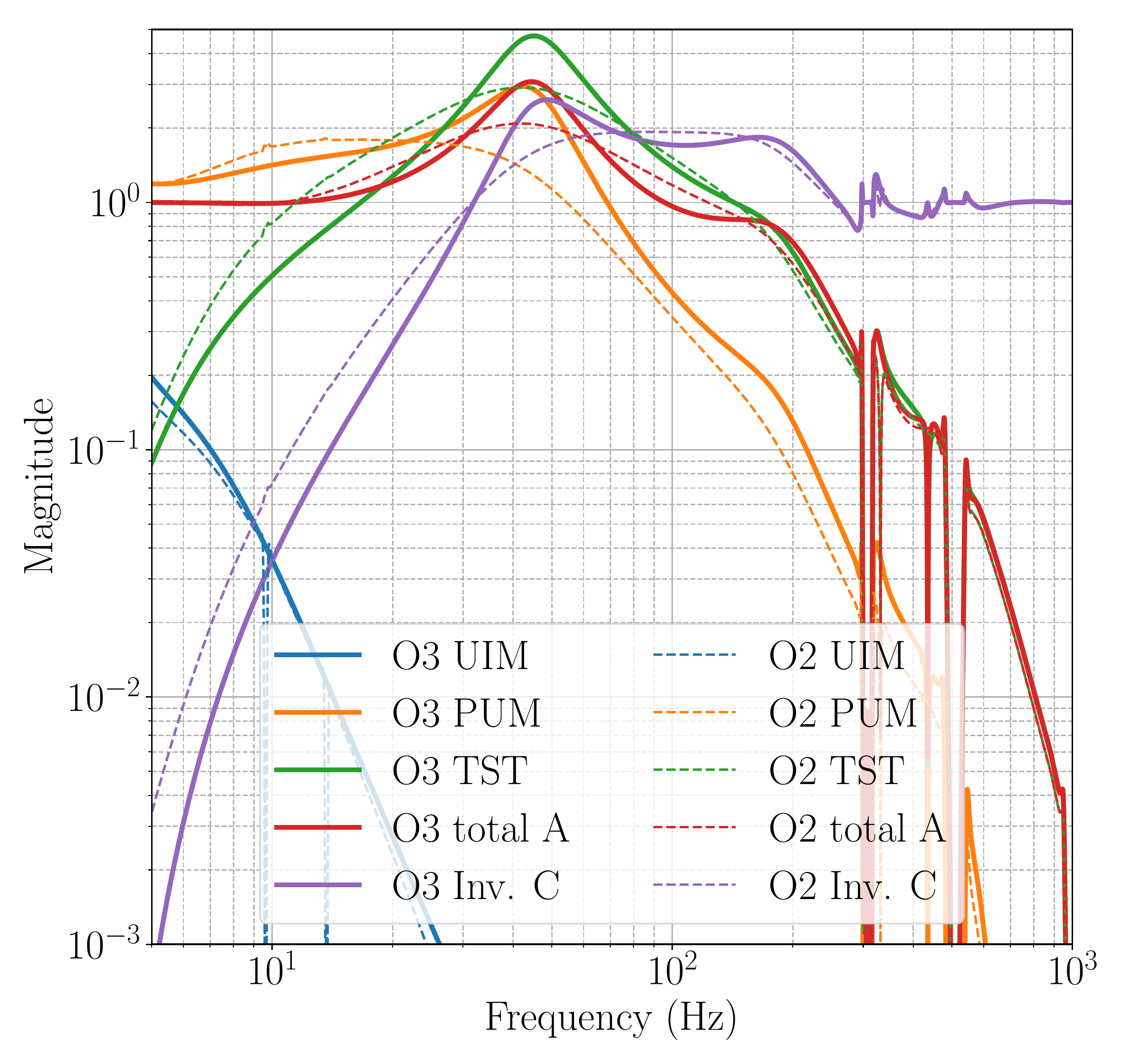}}
	}
	\caption[]{Contributions of each stage of the actuators, the overall actuation, and the inverse sensing to the response function at (a) Hanford and (b) Livingston. The solid and dashed curves indicate the static, reference models used towards the end of O3A and O2, respectively.}
	\label{fig:contri}
\end{figure}

The DARM loop response function is dominated by the actuation and sensing components at low and high frequencies, respectively.
The exact frequency dependence is determined by choices made in the digital filtering, i.e., in the shape of $D$ and $F_i$~($i=U,P,T$), as well as the physical setup and state of the detector.
It is important to quantify the frequency-dependent contributions to the response function from each component to determine how each component contributes to the uncertainty and systematic error.
\Fref{fig:contri} shows the magnitudes of these contributions, i.e., $F_i H_i{A}_i^{\rm (model)} {D}/{R}^{\rm (model)}$ ($i=U,P,T$ for each individual suspension stage), ${A}^{\rm (model)} {D}/{R}^{\rm (model)}$, and $1/({C}^{\rm (model)} {R}^{\rm (model)})$. A similar figure for phase contributions is not shown here for brevity.

The solid curves in \fref{fig:contri} are computed from the reference model used in September 2019, towards the end of O3A.
As described in \sref{subsec:tdep}, the strength of actuators and alignment of optical cavities can change, resulting in different contributions. In addition, different choices may be made and evolve with time to accommodate new detector parameters and/or to improve detector noise performance.
For comparison, the O2 model values are shown as dashed curves, reflecting the impacts due to these changes.
In particular, the contribution from the TST actuator at Livingston has increased in O3A compared to O2.
The same amount of error and uncertainty in modeling ${A}_T^{\rm (model)}$ in O3A as in O2 results in a larger overall calibration uncertainty around 50~Hz (see \sref{sec:results}).

\subsection{\label{subsec:model_sys_err}Systematic error definition}
The frequency-dependent systematic error of the response function, equivalent to the systematic error in estimated $h$, is defined by
\begin{equation}
{\eta}_R \equiv \frac{{R}}{{R}^{\rm (model)}} = \frac{\delta {R}}{{R}^{\rm (model)}} + 1\,,
\end{equation}
where $\delta {R}/{R}^{\rm (model)} = {\eta}_R-1$ is the relative error in the response function as defined in~\cite{Cahillane2017}.
By applying ${\eta}_R$ to the model response function, we obtain the true response ${\eta}_R {R}^{\rm (model)}$.
A systematic error in ${C}$, defined by ${\eta}_C \equiv {C}/{C}^{\rm (model)}$, will impact the response function systematic error as
\begin{equation}
{\eta}_{R;C} = \frac{1}{{R}^{\rm (model)}} \left[\frac{1}{ {\eta}_C {C}^{\rm (model)}}  +{A}^{\rm (model)}{D} \right]\,.\label{eqn:etaC}
\end{equation}
Similarly, a systematic error in ${A}_i$ ($i=U,P,T$), defined by ${\eta}_{A_i} \equiv {A}_i/{A}_i^{\rm (model)}$, will impact the response function systemic error as
\begin{equation}
{\eta}_{R;A_i} = \frac{1}{{R}^{\rm (model)}}  \left[  \frac{1}{{C}^{\rm (model)}} +  \left(  {\eta}_{A_i}  {A}_i^{\rm (model)} + \sum\limits_{j \neq i} {A}_j^{\rm (model)}\right) {D}\right]\,.\label{eqn:etaA}
\end{equation}
These definitions are employed in sections~\ref{sec:electronics}, \ref{sec:residual}, and \ref{sec:results}.

\section{Construct a reference model}
\label{sec:observables}
In this section, we describe the method and procedure of constructing a static reference DARM loop model. 
\Sref{sec:precursory} discusses the tools prepared and measurements made before constructing the model: (a) the photon calibrator absolute reference, (b) a verified model of the quadruple pendulum mechanical dynamics, and (c) a characterization of all actuator and photodiode signal processing electronics present in $A$ and $C$.
\Sref{sec:meas} describes the measurements of the remaining model parameters, which are only measurable when the detectors are in their nominal low-noise configuration, and \sref{sec:mcmc} explains how these parameters and their associated uncertainties are computed.

\subsection{Essential building blocks}
\label{sec:precursory}
The systematic error and uncertainty associated with the absolute reference of the DARM loop model and other essential ``building blocks'' are discussed in this subsection. 
Prior to O3A, the dynamics of the quadruple pendulum and response of signal processing electronics were included in the DARM loop model without considering their contributions to the uncertainty or systematic error in the detector response. 
However, as our understanding of the detectors improves, we now consider them to be a potential source of systematic error, and thus their fundamentals are described in more detail here.

\subsubsection{Photon calibrator absolute reference}

The displacement fiducials upon which all estimates of $h$ depend are generated by the Pcal systems~\cite{Karki2016, karki2019accurate}. 
These systems employ power-modulated auxiliary lasers with beams reflecting from end test masses to displace the mirrors via photon radiation pressure.  
Pcal systems are deployed on both end test masses of each interferometer. 
Their functionality is summarized here along with system updates relevant to the O3A observing run.

Each Pcal system has a 2-watt laser operating at 1047-nm wavelength, housed in a transmitter module located outside the vacuum envelope.  
A feedback control loop that uses an acoustic-optic modulator to vary the laser power in response to a digital excitation signal, $x_{\rm Pcal}$, generates a power-modulated output waveform that reproduces the excitation waveform.
The modulated laser light is directed into the vacuum envelope and reflects from the surface of the end test mass, producing true DARM displacement, $\Delta L_{x_{\rm Pcal}}$ (see \fref{fig:cal_model}).  
The reflected light is directed to a laser power sensor that uses an integrating sphere and photodetector to generate a digital signal, $d_{\rm Pcal}$, proportional to the received laser power.  
The bandwidth of this laser power control servo is approximately 100~kHz.   
The Pcal systems can thus produce arbitrary time-dependent forces resulting in $\Delta L_{x_{\rm Pcal}}$ similar to those that can be produced by the actuators of the quadruple suspension system. 
We estimate the induced DARM displacement from digitized photodiode signal $d_{\rm Pcal}$ with a bank of digital filters $H_{\rm Pcal}$, 
\begin{equation}
	\Delta L_{\rm Pcal} = H_{\rm Pcal} d_{\rm Pcal},
\end{equation}
where $\Delta L_{\rm Pcal}$ is the estimate of the true DARM displacement $\Delta L_{x_{\rm Pcal}}$.
See (1) in \cite{Karki2016} for details of $H_{\rm Pcal}$.   

Though the test mass displacement decreases as the square of the modulation frequency, with a maximum modulated power of approximately 1~W, the Pcal systems can generate $\Delta L_{x_{\rm Pcal}}$ that is orders of magnitude larger than the $\Delta L_{\rm free}$ noise floor across the most sensitive band of the detector. The digital control system allows for arbitrary excitation waveforms, but two specific waveforms for $x_{\rm Pcal}$ are typically used. 
The first is a sequence of monochromatic sinusoidal length modulations, referred to as a {\em swept-sine} excitation. 
The second is a colored random noise modulation used to probe more sensitive frequency regions with high frequency resolution.

The $1\sigma$ uncertainties of $H_{\rm Pcal}$, and thus $\Delta L_{\rm Pcal}$ as an estimate of $\Delta L_{x_{\rm Pcal}}$, for all four LIGO Pcal systems during O3A are 0.54\%.
They are dominated by unknown systematic errors rather than by statistical variations in measured values (\change{see details in} \cite{Pcalpaper-P2000113}).
At the end of O3A, the $H_{\rm Pcal}$ value was refined using  system characterization measurements carried out during the six months of the run, as well as the correction of errors in the masses of the end mirrors. 
These updates are accounted for by a multiplicative correction factor, $\eta_{\rm Pcal}$, applied to $H_{\rm Pcal}$ for each Pcal system.
For the Hanford reference Pcal system on the Y arm end test mass, we have $\eta_{\rm Pcal} = 1.0043$. For the Livingston reference system, also on the Y arm end test mass, we have $\eta_{\rm Pcal} = 1.0031$~\cite{Pcal-G1902259}.
Accounting for these systematic errors and uncertainties is discussed in \sref{sec:results}.

\subsubsection{Dynamics of the quadruple suspension}

Preliminary models of the quadruple suspension system rigid-body, force-to-displacement transfer functions were developed from first principles well before the Advanced LIGO interferometers were installed~\cite{barton2002quadmodel,barton2008quadmodel}.
These preliminary models aided analysis and diagnostics of the early prototype quadruple suspension systems~\cite{shapiro2010quadmodel}.
Refinements were added to the model in order to match them to the first production suspension system and improve their accuracy~\cite{shapiro2014quadmodel}.
The refined model was later used to verify the function of all production quadruple suspensions installed in each LIGO detector.

The model parameters are kept up-to-date as the installed suspensions are modified (e.g., between O2 and O3, small, few-gram damping mechanisms were added to the test masses~\cite{biscans2019suppressing}).  
The change in model parameters can be typically quantified to high accuracy (e.g. each $\sim 40$~kg test mass can be measured to an accuracy of $\sim 10$~g), 
and the subsequent updated model parameters are revalidated to high precision through many local and interferometric measurements of the dynamical response.
These models are used as the basis for the frequency dependence of force-to-displacement transfer functions in ${A}_{i}$.
Beyond these rigid-body dynamics, we have found the need for additional, non-rigid body modifications to these transfer functions in order to improve the model accuracy. The impact of the additional modifications is discussed in sections~\ref{sec:residual} and \ref{sec:uncompensated}.

\subsubsection{\label{sec:electronics}Signal processing analog electronics}

The responses of all signal processing electronics are measured independently in advance and modeled as transfer functions with poles and zeros at well-determined frequencies. 
Within the sensing function, these electronics conditioning the current produced by the readout photodiodes are: the transimpedance amplifiers of the photodiodes, ``whitening'' filters (i.e., frequency-dependent, signal pre-amplification or noise reduction filters), and anti-aliasing filters.
Within the actuation function, requested voltage at each stage is conditioned through anti-imaging filter electronics and sent to either the magnetic coil current drivers or electrostatic voltage drivers, depending on the actuator type of the given stage of the quadruple suspension.
The actuator drivers also have frequency-dependent response for noise reduction much like the readout photodiode transimpedance amplifier and whitening filters.

The pole and zero frequencies for the responses of these electronic components range from as low as 0.5~Hz to as high as 50~kHz, all of which need to be included in $A^{\rm (model)}$ and $C^{\rm (model)}$ to produce accurate estimates of $h$.
For example, to minimize the contribution to systematic error in the response function near the DARM loop unity gain frequency ($\sim$100~Hz), the phase of $A^{\rm (model)}$ and $C^{\rm (model)}$ must be accurate to a level of $\lesssim$0.1~deg. Such accuracy cannot be achieved if any of the poles or zeros, even those at $\sim 50$~kHz, are excluded in the model.
Measurements of the response of each electronic component from 0.1~Hz to 100~kHz are made using an analog spectrum analyzer.
Pole and zero frequencies are determined by fitting the measured response of each electronic component to a model consisting of poles and zeros \cite{IIRational}.
Only the poles and zeros frequency response is needed at this point, and later the gain of each path is measured in $H_{C}$ and $H_{i}$ ($i=U,P,T$) using techniques described in \sref{sec:meas}.

The pole and zero frequencies of the electronic components are used in different ways throughout the production of $h$.
As described in sections~\ref{subsec:sensing} and \ref{subsec:actuation}, poles and zeros below $\sim$500~Hz are used to design digital IIR filters within the DARM loop that replicate the inverse response of the electronics.
The DARM loop would be negatively impacted without including these poles and zeros.
Each pair of analog response and compensating digital inverse response occurs before constructing $d_{\rm err}$ within $C$ and after $d_{\rm ctrl}$ is distributed through $A$.
Thus, the low-latency or offline estimate of $h$ need only further include the higher-frequency poles and zeros.
Poles and zeros above $\sim$500~Hz but below the Nyquist frequency of the real-time system ($\sim$7000~Hz) are included in the front-end IIR reproductions of $A^{\rm (model)}$ and $C^{\rm (model)}$ (outside the DARM loop) to produce the roughly calibrated $\Delta L_{\rm free}$.
Limitations of the front-end IIR filter construction result in growing systematic error approaching the Nyquist frequency and prevent the inclusion of any response above the Nyquist frequency.
The second part of the low-latency pipeline repairs any distorted high-frequency response of IIR models of $A^{\rm (model)}$ and $C^{\rm (model)}$ and includes the response of super-Nyquist poles and zeros to form the low-latency estimate of $h$.
The most accurate, offline estimate of $h$ includes all poles and zeros above $\sim$500~Hz (i.e., those not compensated within the DARM loop) in the FIR reproductions of $A^{\rm (model)}$ and $C^{\rm (model)}$.

\begin{figure}[!tbh]
	\centering
	\subfigure[]
	{
		\label{fig:omc_mag}
		\scalebox{0.3}{\includegraphics{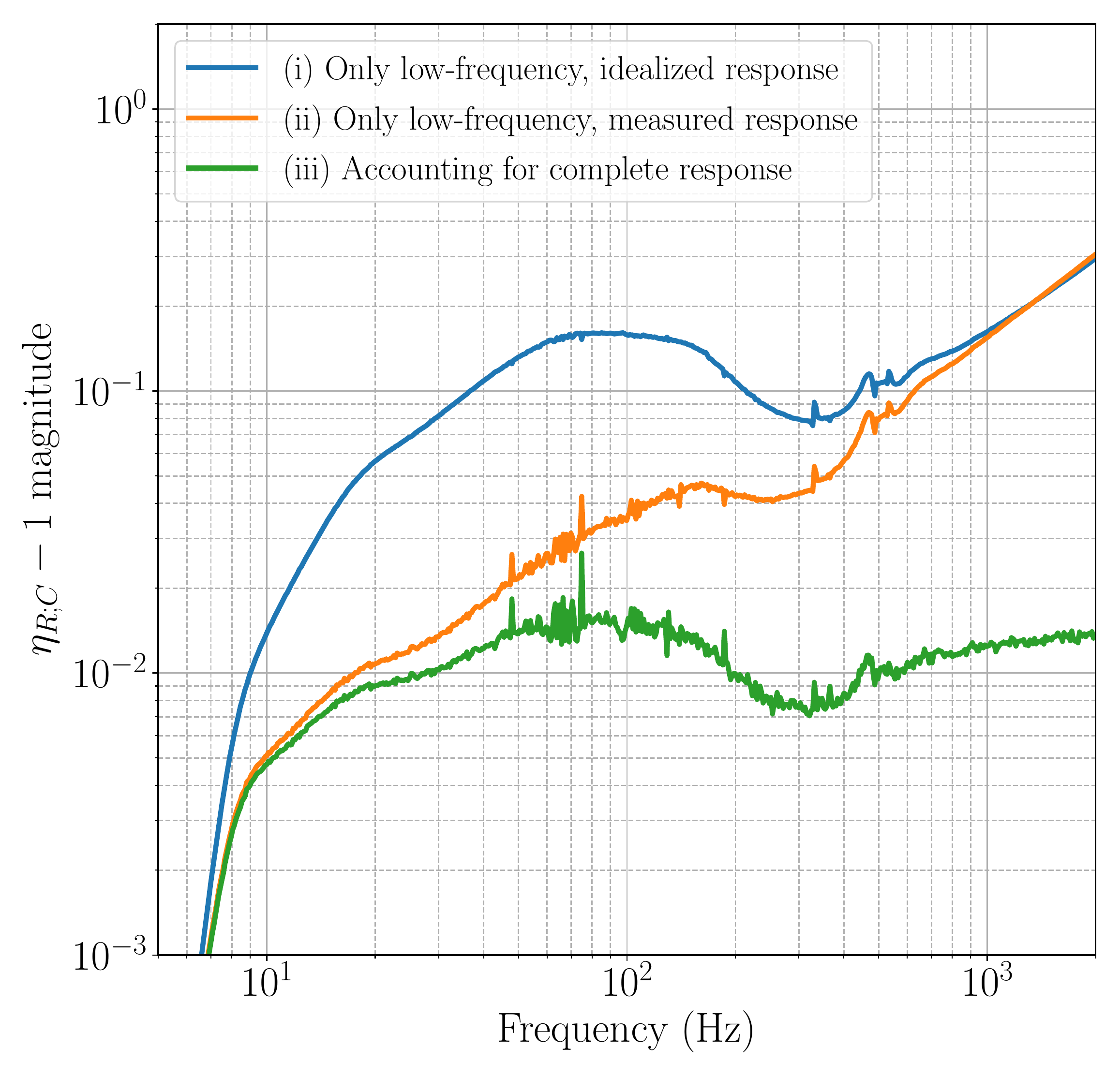}}
	}
	\subfigure[]
	{
		\label{fig:omc_phase}
		\scalebox{0.3}{\includegraphics{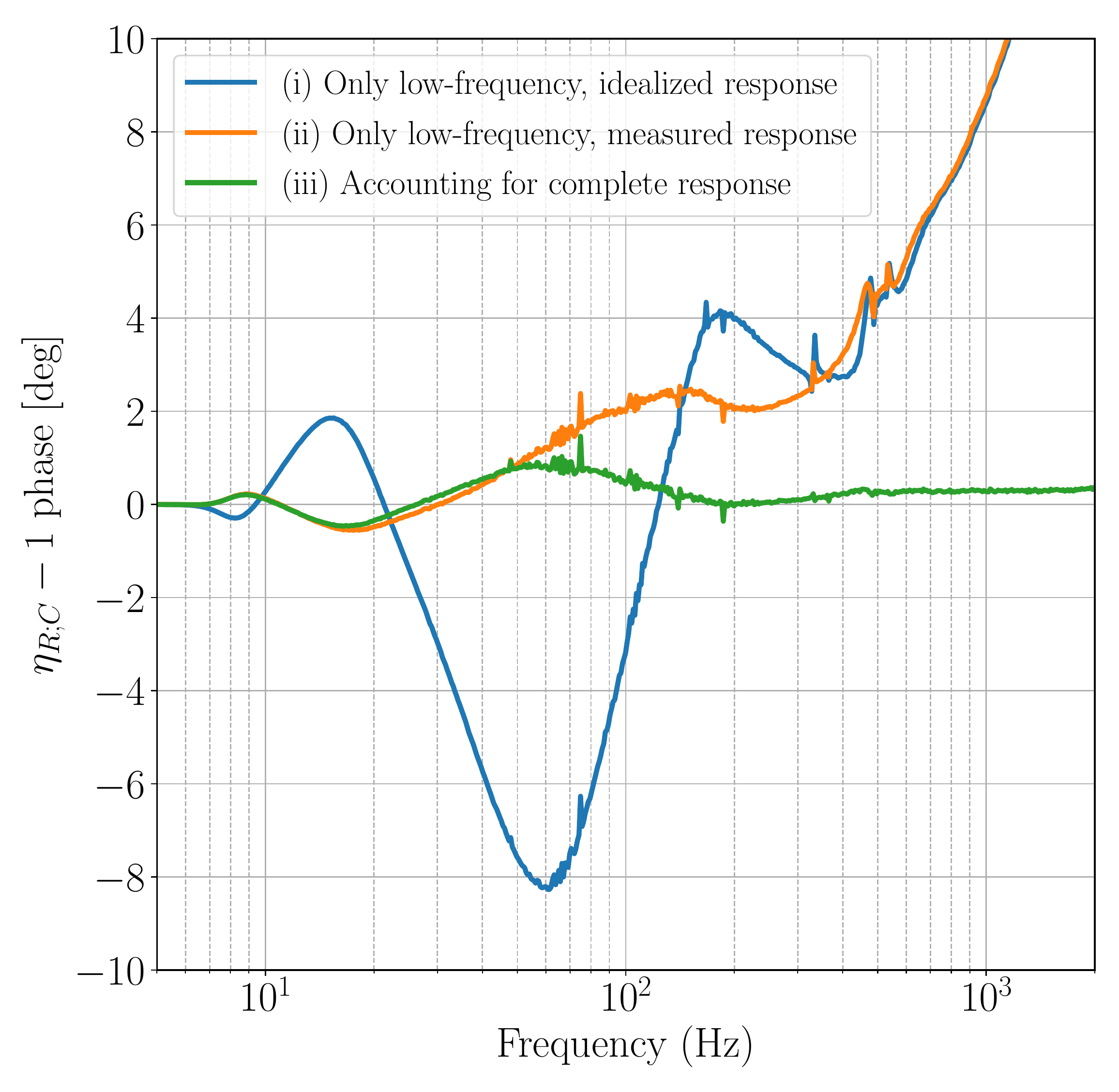}}
	}
	\caption[]{\change{Relative error in the response function (${\eta}_{R;C}-1$)} in (a) magnitude and (b) phase from the systematic error in sensing function whitening filter electronics alone. The colored curves indicate the resulting \change{${\eta}_{R;C}-1$}, if (i) the in-loop compensation filters are informed only by design specifications, and high-frequency poles and zeros are excluded from $A^{\rm (model)}$ and $C^{\rm (model)}$ (blue), (ii) the in-loop compensation filters are informed by measurements, but high-frequency poles and zeros are still excluded (orange), and (iii) all measured pole and zero frequencies are included (green) as in the final estimate of $h$.}
	\label{fig:omc}
\end{figure}

The contribution to the systematic error in $h$ from each electronic component in $C$ and $A$ is evaluated by ${\eta}_{R;C}$ or ${\eta}_{R;A}$ in \eref{eqn:etaC} and \eref{eqn:etaA}, respectively.
An example is given in \fref{fig:omc} for the sensing function whitening filter electronics alone, assuming all other electronics are modeled perfectly.
To emphasize the need for careful measurements, the blue curves indicate the resulting \change{${\eta}_{R;C}-1$} if the in-loop compensation filters are designed using only pole and zero frequencies below $\sim$500~Hz as reported in the design specifications; the poles or zeros above $\sim$500~Hz are not included. 
The orange curves show the resulting \change{${\eta}_{R;C}-1$} if in-loop compensation filters are designed using pole and zero frequencies obtained via fitting but do not account for poles or zeros above $\sim$500~Hz.
Finally, the green curves correspond to \change{${\eta}_{R;C}-1$} remaining in the low-latency and offline estimates of $h$ if the in-loop compensation filters are designed with the measured poles and zeros below $\sim$500~Hz and include all other poles and zeroes above 500~Hz in the FIR reproductions of $A^{\rm (model)}$ and $C^{\rm (model)}$.



\subsection{Interferometric  measurements}
\label{sec:meas}

At frequencies much lower than the acoustic resonance frequencies of the test masses ($\lesssim 1$~kHz), the excitations from the Pcal systems $x_{\rm Pcal}$ cause DARM displacement $\Delta L_{x_{\rm Pcal}}$, equivalent to $\Delta L_{\rm free}$.
Thus, measuring the DARM loop error signal $d_{\rm err}$ in the presence of these Pcal excitations, while the detector is otherwise fully functional, is a direct measure of the (inverse) response function, 
\begin{equation}
\frac{{d}_{\rm err}}{{\Delta L}_{\rm Pcal}} =  \frac{1}{{R}^{\rm (meas)}} = \frac{{C}^{\rm (meas)}}{1 + {A}^{\rm (meas)} D {C}^{\rm(meas)}},
\label{eq:responsemeas}
\end{equation}
where the superscript ``(meas)" stands for the measurement.
To measure the sensing function ${C}^{\rm (meas)}$, an additional separate measurement of the loop suppression is required (see \fref{fig:cal_model}), 
\begin{equation}
\frac{{d}_{\rm D}}{{x}_{\rm D}} = \frac{1}{1 + {A}^{\rm (meas)} {D} {C}^{\rm(meas)}},
\label{eq:loop_suppression}
\end{equation}
where $x_{\rm D}$ indicates the displacement excitations added using the quadruple pendulum actuator system, and $d_{\rm D}$ is the sum of $d_{\rm err}$ and $x_{\rm D}$.
Measurements in \eref{eq:responsemeas} and \eref{eq:loop_suppression} need to be taken sufficiently close in time, such that the time dependence of $A$ and $C$ can be ignored.
Combining \eref{eq:responsemeas} and \eref{eq:loop_suppression}, we obtain the sensing function directly as
\begin{eqnarray}
{C}^{\rm (meas)}(f) = \left(\frac{{d}_{\rm err}(f)}{{\Delta L}_{\rm Pcal}(f)}\right) \left(\frac{{x}_{\rm D}(f)}{{d}_{\rm D}(f)}\right).
\label{eq:sensmeas}
\end{eqnarray}

For frequencies above 1~kHz, where much longer averaging is required to obtain appreciable signal-to-noise (SNR) with respect to the detector's noise floor, the sensing function is measured by introducing discrete sinusoidal excitations from 1~kHz to 4~kHz, at 500~Hz intervals. 
Excitations at a single frequency are left on for 24 hours of ``observation-ready" (i.e., nominal operating configuration) time before moving to the next frequency in the sequence. 
At frequencies above 1~kHz, where the detector's response is determined by the sensing function, we have $|{A}^{\rm (meas)} {D} {C}^{\rm(meas)}|\lesssim 10^{-4}$ and thus the independent measurement of the loop suppression in \eref{eq:sensmeas} is not necessary, i.e., $|{x}_{\rm D}/{d}_{\rm D} -1| \lesssim 10^{-4}$. 
Therefore, for these frequencies, the sensing function is well-approximated by
\begin{eqnarray}
{C}^{\rm (meas)}(f) \approx \frac{{d}_{\rm err}(f)}{{\Delta L}_{\rm Pcal}(f)},
\label{eq:sensmeas_hf}
\end{eqnarray}
\change{calculated from the average of 30-minute fast Fourier transforms (FFTs) over 24 hours at each frequency.}
Due to the long duration of the measurements, the resulting sensing function must be corrected for time dependence using $\kappa_{C}(t)$ and $f_{cc}(t)$ (see \sref{sec:tdcf}).

The Pcal systems are also used to determine the actuator strength, $H_{i}$, for each stage of the quadruple suspension. 
Measuring $d_{\rm err}$ caused by the excitations from the suspension actuators, ${x}_i(f)$, gives
\begin{equation}
\frac{{d}_{\rm err}}{{x}_{i}} = \frac{{A}_{i}^{\rm (meas)}{C}^{\rm (meas)}}{1 + {A}^{\rm (meas)} D {C}^{\rm(meas)}}.
\label{eq:meas_A}
\end{equation}
Combining \eref{eq:responsemeas} and \eref{eq:meas_A}, we can extract ${A}_i^{\rm (meas)}$~($i=U,P,T$), as in \cite{LVC2017,Cahillane2017},
\begin{eqnarray}
\nonumber
{A}_i^{\rm (meas)}(f) &=& \left[{F}_{i}(f)H_i{A}_i(f) \exp(-2\pi if\tau_{i})\right]^{\rm (meas)} \\
&=& \left(\frac{{\Delta L}_{\rm Pcal}(f)}{{d}_{\rm err}(f)}\right) \left( \frac{{d}_{\rm err}(f)}{{x}_i(f)} \right). \label{eq:actmeas}
\end{eqnarray}

The relative magnitude uncertainty and absolute phase uncertainty of the measured transfer functions at each frequency point is given by~\cite{Cahillane2017,bendat2011random}
\begin{equation}
	\sigma^{\rm (meas)} (f) = \sqrt{\frac{1-\gamma^2(f)}{2 N_{\rm avg} \gamma^2(f)} },
	\label{eq:unc_meas}
\end{equation}
where $N_{\rm avg}$ is the number of values averaged. The coherence, $\gamma^2(f)$, between the excitation $x$ and readout $d$ at frequency $f$ is calculated by~\cite{Viets2018}
\begin{equation}
\gamma^2(f) = \frac{|\langle {x}^*(f) {d}(f)\rangle|^2}{\langle |{x}(f) |^2\rangle  \langle |{d}(f) |^2\rangle},
\label{eq:gamma2}
\end{equation}
where the angled brackets denote averaging, and the asterisk denotes complex conjugation.

Measurements described by (\ref{eq:sensmeas}) and (\ref{eq:actmeas}) are repeated with a weekly cadence throughout the run. A single set of them is used to construct a reference model (see \sref{sec:mcmc}). The collection of weekly measurements is used to assess static, frequency-dependent systematic errors (see sections \ref{sec:detuning} and \ref{sec:residual}).

\subsection{Model parameter estimation}
\label{sec:mcmc}

Remaining parameters in the DARM reference model are determined from one set of measurements (\sref{sec:meas}) taken at the reference time, after dividing out all known frequency dependence from \sref{sec:electronics}, using MCMC fitting algorithms~\cite{emcee-citation}.
The remaining parameters are 
\begin{equation}
\bm{\lambda}^C = \left[H_C,  f_{cc},  f_s,  Q  , \delta \tau_C\right],
\end{equation}
for the sensing function, and
\begin{equation}
\bm{\lambda}^A_{i} = \left[H_i,  \delta \tau_i\right],
\end{equation}
for each $i$th stage of the actuation function, where $\delta \tau_C$ and $\delta \tau_i$ are the residual time delays of $\tau_C$ and $\tau_i$, respectively.
We note that only measurement data at frequencies below 1~kHz are used for parameter estimation. 
The high-frequency measurements are used for studying the static, residual systematic error and statistical uncertainty above 1~kHz (see \sref{sec:residual}).
The MCMC method produces the posterior distributions of the multivariate parameters assuming normally distributed priors for $H_C$, $f_{cc}$, $H_i$, and flat (uniform) priors for $f_s$, $Q^{-1}$, $\delta \tau_C$, and $\delta \tau_i$ ($i=U,P,T$). 
The maximum a posteriori (MAP) values, $\bm{\lambda}_{\rm MAP}^C$ and $\bm{\lambda}_{\rm MAP}^{A}$, are adopted to create the DARM response model, 
\begin{equation}
{R}^{\rm (model)}(f) = \frac{1}{{C}^{\rm (model)}(\bm{\lambda}_{\rm MAP}^C;f)} + {A}^{\rm (model)}(\bm{\lambda}_{\rm MAP}^A;f) {D}(f)\,.
\end{equation}
When any physical change of the interferometer is too large to be corrected by the TDCFs, or a precursory component has changed, we create a new calibration ``epoch." 
It is likely that in any new epoch one or more parameters in the existing $\bm{\lambda}_{\rm MAP}^C$ and $\bm{\lambda}_{\rm MAP}^{A}$ (and hence ${R}^{\rm (model)}$) are no longer valid. 
The MCMC parameter estimation process is repeated using new measurements, $A^{\rm (meas)}$ and $C^{\rm (meas)}$, in order to create an updated reference model and account for the precursory changes. 
\Tref{tab:epoch} lists the O3A epochs in both detectors and the main changes associated with each. We quantify the calibration systematic error and statistical uncertainty for each epoch in \sref{sec:results}.

\begin{table}[!tbh]
	\caption{\label{tab:epoch}O3A calibration epochs and the main changes in each epoch.}
	\begin{indented}
		\item[]\begin{tabular}{@{}ll}
			\br
		    Hanford epoch & Changes  \\
			\mr
			(a) Mar 28--Jun 11 & Start of the run   \\
			(b) Jun 11--Aug 28 & Input power increased; angular control system modified \\
			(c) Aug 28--Oct 1 & Added a microscopic length offset to SRC to relieve detuning \\
			\br
			Livingston epoch & Changes \\
			\mr
			(a) Mar 28--Jun 11 & Start of the run  \\
			(b) Jun 11--Oct 1 & Adjusted the gain in the TST actuator due to a 4\% drift  \\
            \br
		\end{tabular}
	\end{indented}
\end{table}

\begin{figure}
	\begin{center}
		\includegraphics[width=0.8\textwidth]{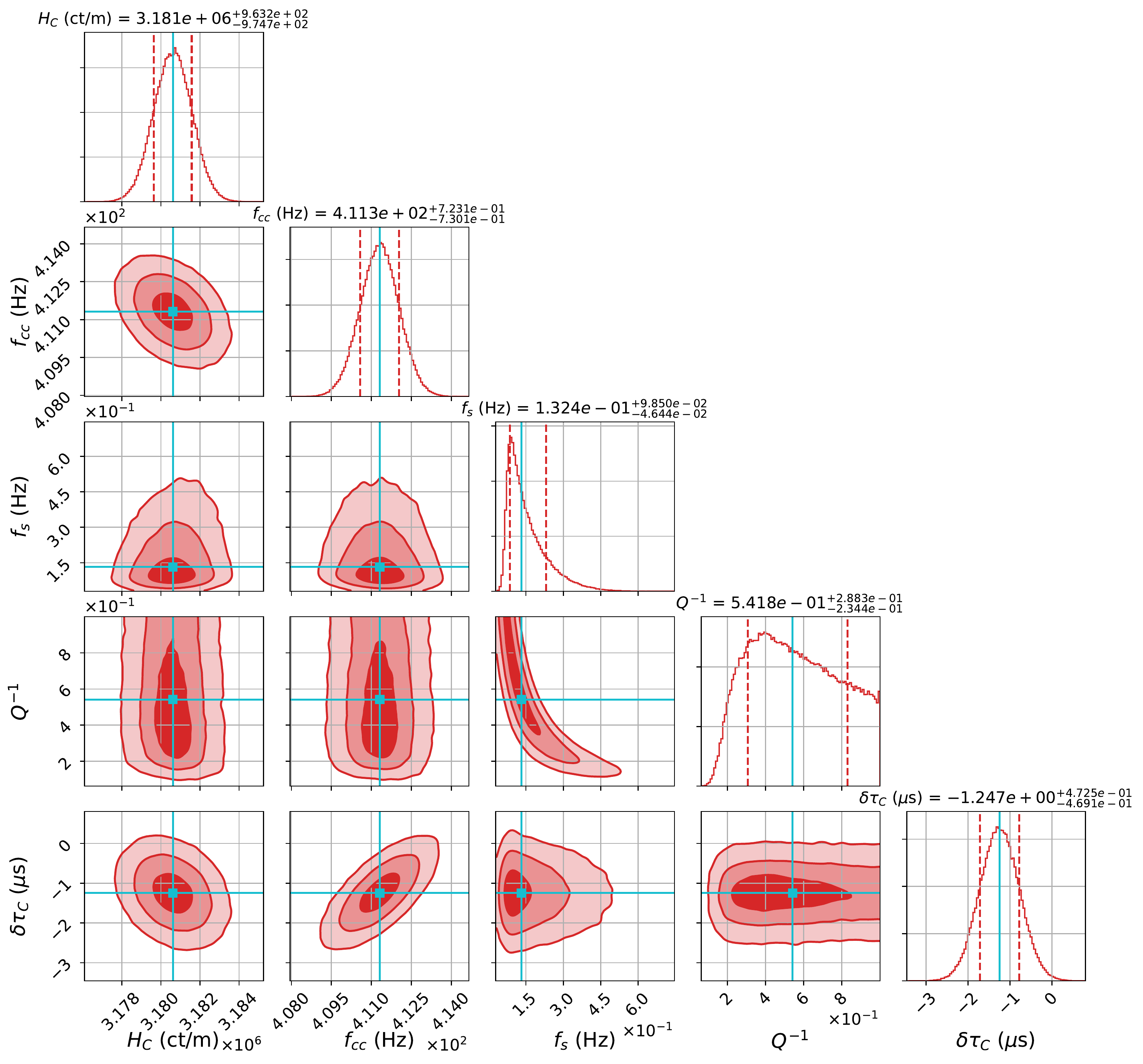}
	\end{center}
	\caption{Corner plot of posterior distributions of the sensing parameters, $\bm{\lambda}^C$, at Hanford. The one-dimensional histograms along the diagonal are the posterior distributions for the optical gain $H_C$, Fabry-P\'{e}rot coupled cavity pole frequency $f_{cc}$, SRC optical spring frequency $f_s$, inverse optical spring quality factor $Q^{-1}$, and residual time delay $\delta \tau_C$, from top left to bottom right. The off-diagonal two-dimensional histograms show the covariance of two parameters; $1\sigma$, $2\sigma$, and $3\sigma$ levels are delineated by contours (from dark to light). The cyan lines indicate the MAP values for each parameter. The dashed red lines in the 1-D histograms indicate the $1\sigma$ values in the distribution.} \label{fig:mcmc_corner}
\end{figure}

\begin{figure}
	\begin{center}
		\includegraphics[width=0.8\textwidth]{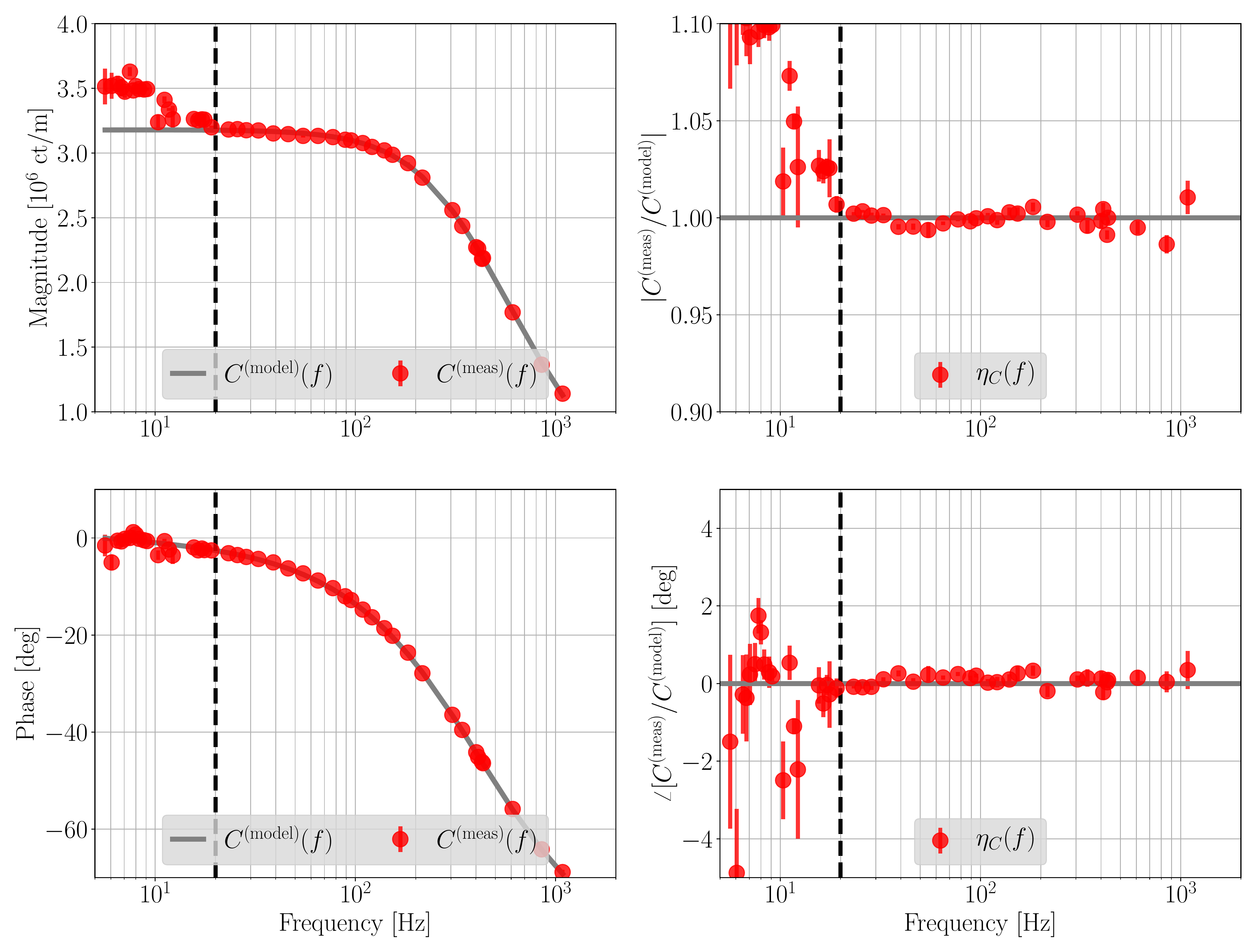}
	\end{center}
	\caption{Sensing measurement ${C}^{\rm (meas)}(f)$ and the reference model ${C}^{\rm (model)}(f)$ at Hanford created with the MAP parameters in \fref{fig:mcmc_corner} (left column), and the fractional residual between the two (right column). The gray curves and red markers indicate the reference model and the measurements, respectively. Vertical error bars indicate uncertainties of the measurements. The top and bottom rows display the magnitude and phase, respectively. The MAP values in \fref{fig:mcmc_corner} are inferred from the measurements above 20\,Hz only (on the right side of the vertical dashed lines; see explanations in text). 
	} \label{fig:mcmc_res}
\end{figure}

An example of the MCMC fitting for the sensing function at Hanford is given in figures~\ref{fig:mcmc_corner} and \ref{fig:mcmc_res}. A set of measurements is taken in the frequency band 5--1084\,Hz, and passed to the MCMC algorithm. The five-dimensional fitting results are shown in \fref{fig:mcmc_corner}. Posterior distributions of five parameters in $\bm{\lambda}^C$ are shown in the diagonal panels. 

The reference sensing model, created using the MAP parameters shown in \fref{fig:mcmc_corner}, is then compared to the original measured data points, plotted in \fref{fig:mcmc_res}. The left column shows the magnitude and phase of both the reference model (grey curve) and measurement (red points). The right column displays the residual between the two. 
The units of the sensing function are shown in digital counts of $d_{\rm err}$ per meter change in the DARM length. 
The deviation between the measurement and the model below 20~Hz is due to a poorly-modeled effect (discussed in \sref{sec:detuning}), and hence the measurements below 20~Hz are not used to inform the MCMC fit (as denoted by the dashed vertical lines).

A similar procedure is repeated to generate the model of the three end test mass actuator stages.
The Hanford detector produces $\Delta L_{\rm ctrl}$ with all three end test mass suspension stages on the X arm during O3A. 
The Livingston detector uses the TST stage on the Y arm, and the PUM and UIM stages on the X arm.\footnote{In O1 and O2, $\Delta L_{\rm ctrl}$ was produced entirely by the three end test mass actuator stages on the Y arm at Hanford, and the three stages on the X arm at Livingston.}
In this ``split actuator'' configuration for the Livingston detector, one must include information in the model reflecting that different computers and digital-to-analog converters are used to create $\Delta L_{\rm ctrl}$.
This is done by allowing for a different computational time delay in the model for each stage.
Thus, $\tau_i$ is fit independently and included in the model.
After accounting for these different delays, we find that all remaining residuals of $\tau_i$ are consistent with zero. 
The uncertainty of $\tau_i$ is discussed in \sref{sec:uncompensated}.

\section{Understanding of systematic errors}
\label{sec:syserrors}
\change{In this section, we first discuss the time-dependent systematic errors that can be corrected using TDCFs and related special issues in O3A in \sref{sec:tdcf}. \Sref{sec:detuning} describes a unique low-frequency feature in the Hanford detector, which cannot be simply addressed by TDCFs. In \sref{sec:residual}, we present how to account for unknown residual systematic errors, including the special feature described in \sref{sec:detuning}. Finally, we list and quantify uncompensated systematic errors from multiple sources in \sref{sec:uncompensated}.}

\subsection{Time dependent systematic error}
\label{sec:tdcf}
The TDCFs, $\kappa_{C}$, $f_{cc}$, $f_s$, $Q$, $\kappa_U$, $\kappa_P$, and $\kappa_T$ from \eref{eqn:tdcf_sensing} and \eref{eqn:tdcf_actuation}, are monitored by a collection of monochromatic, high-SNR sinusoidal excitations (``calibration lines'') injected into the DARM control loop by both Pcal and suspension actuators.
After demodulating the magnitude and phase of these calibration lines in $d_{\rm err}$, the TDCFs are calculated from \eref{eq:sensmeas}--\eref{eq:actmeas} (see complete derivation in~\cite{Tuyenbayev2016,Viets2018,TDCF-T1700106}) and applied to the appropriate components in $\Delta L_{\rm res}$ and $\Delta L_{\rm ctrl}$.
The uncertainties for all TDCFs are computed using \eref{eq:unc_meas}.

\Tref{tab:tdcf_corr} shows how time-dependent corrections are applied in O1, O2, and O3, for each of the three calibration pipelines. In O1 and O2, no TDCF was applied to correct for systematic errors in the front-end calibration, and only the \change{scalar gain factors} were applied in the low-latency, online calibration. The factor $f_{cc}$ was only applied in the high-latency, offline calibration, producing the final, corrected strain data a couple of months after the data were acquired. In O3, both the scalar gain factors and $f_{cc}$ are applied in all of the three calibration pipelines.
The front-end pipeline computes and applies the TDCFs separately and independently from the online and offline pipelines for real-time detector performance assessment and consistency checks.
The remaining two TDCFs related to SRC detuning, $f_s$ and $Q$, are monitored but not applied to the data in any of the pipelines. 
Static reference values for these \change{two parameters} are taken from $\bm{\lambda}_{\rm MAP}^C$ and used by the pipelines.
\change{The impact from these two uncorrected TDCFs remains below 20~Hz. In O3, there has been additional challenge in modeling the sensing function at low frequencies at Hanford. See detailed discussion below and in \sref{sec:detuning}.}

\begin{table}[!tbh]
	\caption{\label{tab:tdcf_corr}Time-dependent correction factors applied in each of the calibration pipelines in the three observing runs. Recall that the gains of the UIM and PUM suspension stages were tracked by the combined factor $\kappa_{PU}$ in O1 and O2.}
	\begin{indented}
	\item[]\begin{tabular}{@{}lll}
	\br
	Calibration pipeline & O1 and O2 & O3 \\
	\mr
	Front-end & None  & $\kappa_{C}, \kappa_{U}, \kappa_{P}, \kappa_T, f_{cc}$  \\
	Low-latency (online)  & $\kappa_{C}, \kappa_{PU}, \kappa_T$  & $\kappa_{C}, \kappa_{U}, \kappa_{P}, \kappa_T, f_{cc}$   \\
	High-latency (offline) & $\kappa_{C}, \kappa_{PU}, \kappa_T, f_{cc}$  & $\kappa_{C}, \kappa_{U}, \kappa_{P}, \kappa_T, f_{cc}$  \\
	\br
	\end{tabular}
	\end{indented}
\end{table}

The impacted frequency bands and level of systematic errors from all TDCFs are slightly different at Hanford than at Livingston due to differences in the design of digital filters, as shown in the contribution curves in \fref{fig:contri}.

\begin{figure}[!tbh]
	\centering
	\scalebox{0.3}{\includegraphics{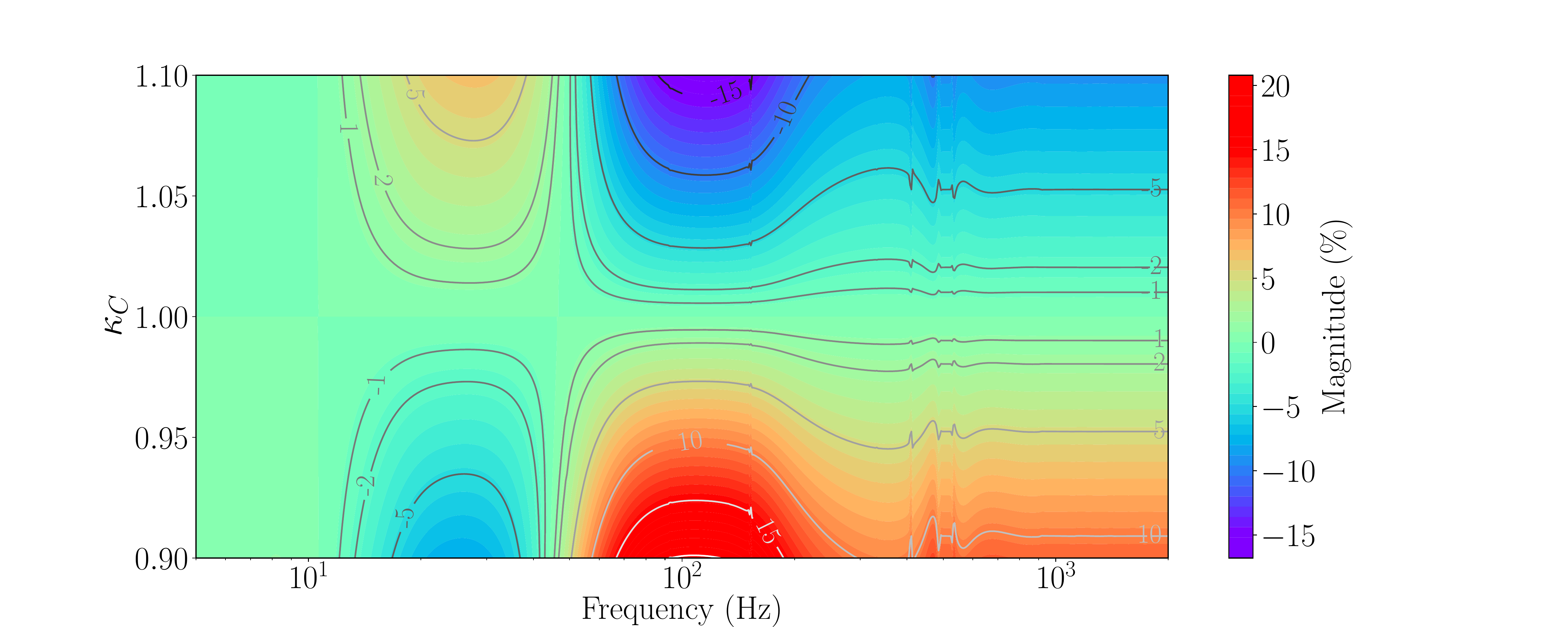}}
	\scalebox{0.3}{\includegraphics{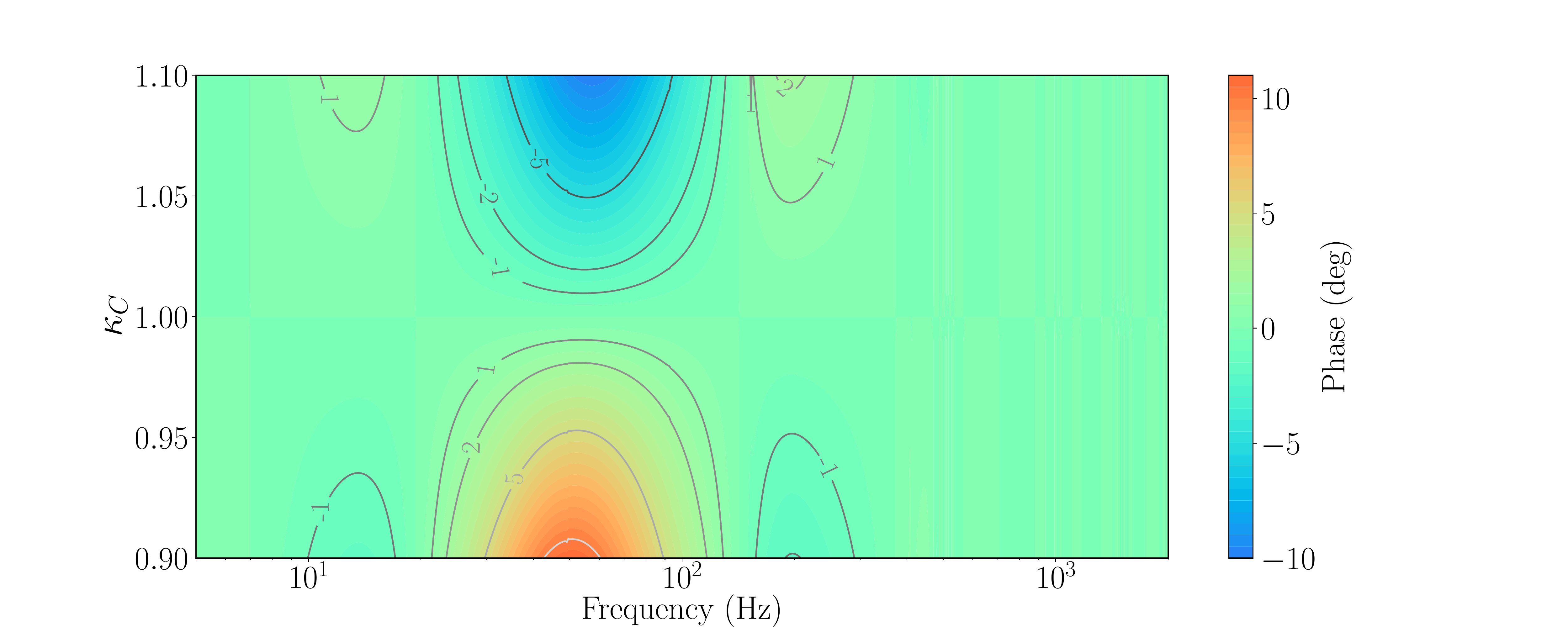}}
	\caption[]{Magnitude (top) and phase (bottom) of the fractional error ${\eta}_{R;C}-1$ in the Hanford detector response [O3A Epoch (c)] as a function of frequency due to uncorrected gain variations in the sensing function, tracked by the scalar time-dependent factor, $\kappa_{C}$. }
	\label{fig:carpet_kc}
\end{figure}

As an example to show the necessity of time-dependent corrections, we quantify the systematic errors that would be present in the O3A Hanford model, if the TDCFs were not applied.  
Similar to studies in~\cite{Tuyenbayev2016}, \fref{fig:carpet_kc} shows the estimated systematic errors (colored contours) introduced in ${R}^{\rm (model)}$ (i.e., ${\eta}_{R;C}-1$) if the time-variation of $H_{C}$, tracked by the factor $\kappa_{C}$, is not corrected. 
The top and bottom panels correspond to magnitude and phase of ${\eta}_{R;C}-1$, respectively.
The measured fractional variation of $\kappa_{C}$ is typically at the level of 1\%--2\%, and can be as large as $\sim$10\% in either detector.
As shown in the figure, an uncorrected $\sim$5\% change in $\kappa_{C}$ will result in $\sim$10\% systematic error in the magnitude of ${R}^{\rm (model)}$, near 100~Hz.
Similar plots for impacts of uncorrected $f_{cc}$, $\kappa_{T}$, $\kappa_{P}$, and $\kappa_{U}$ are shown in~\ref{appendix:carpet}.

\begin{figure}[!tbh]
	\centering
	\scalebox{0.29}{\includegraphics{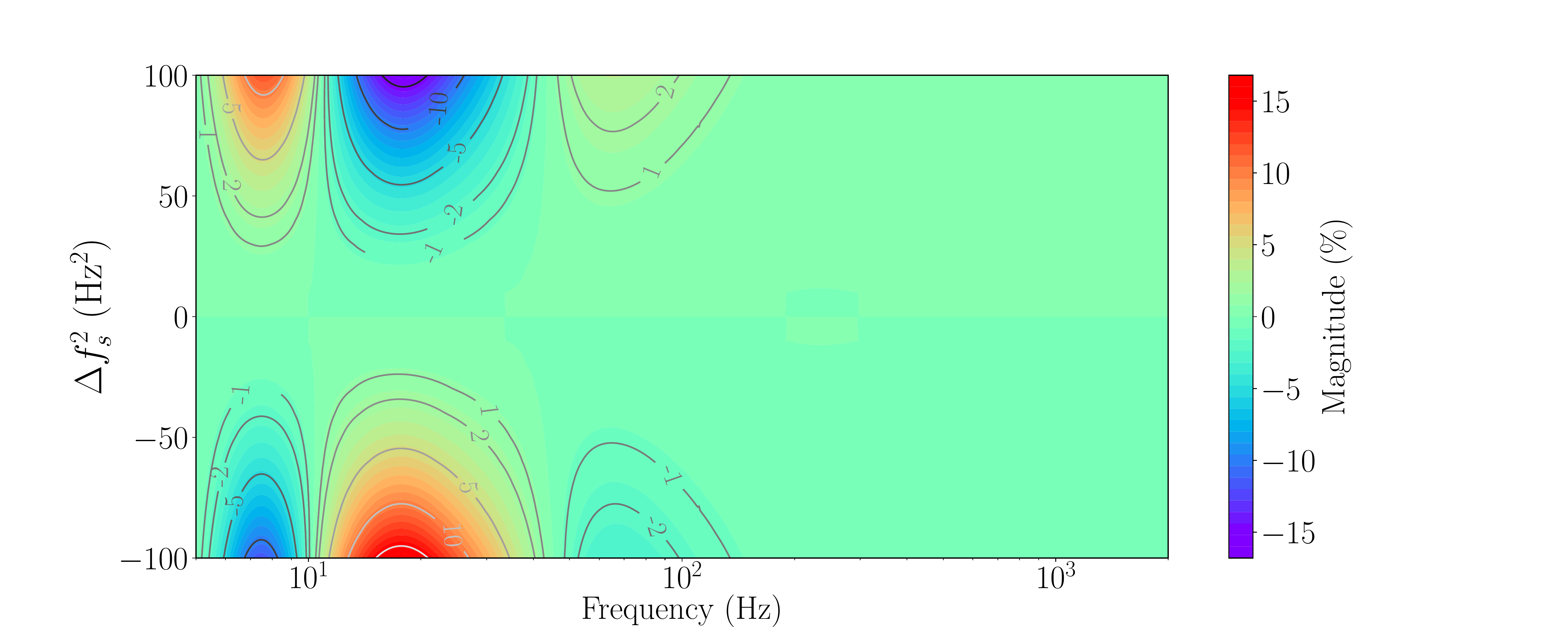}}
	\scalebox{0.29}{\includegraphics{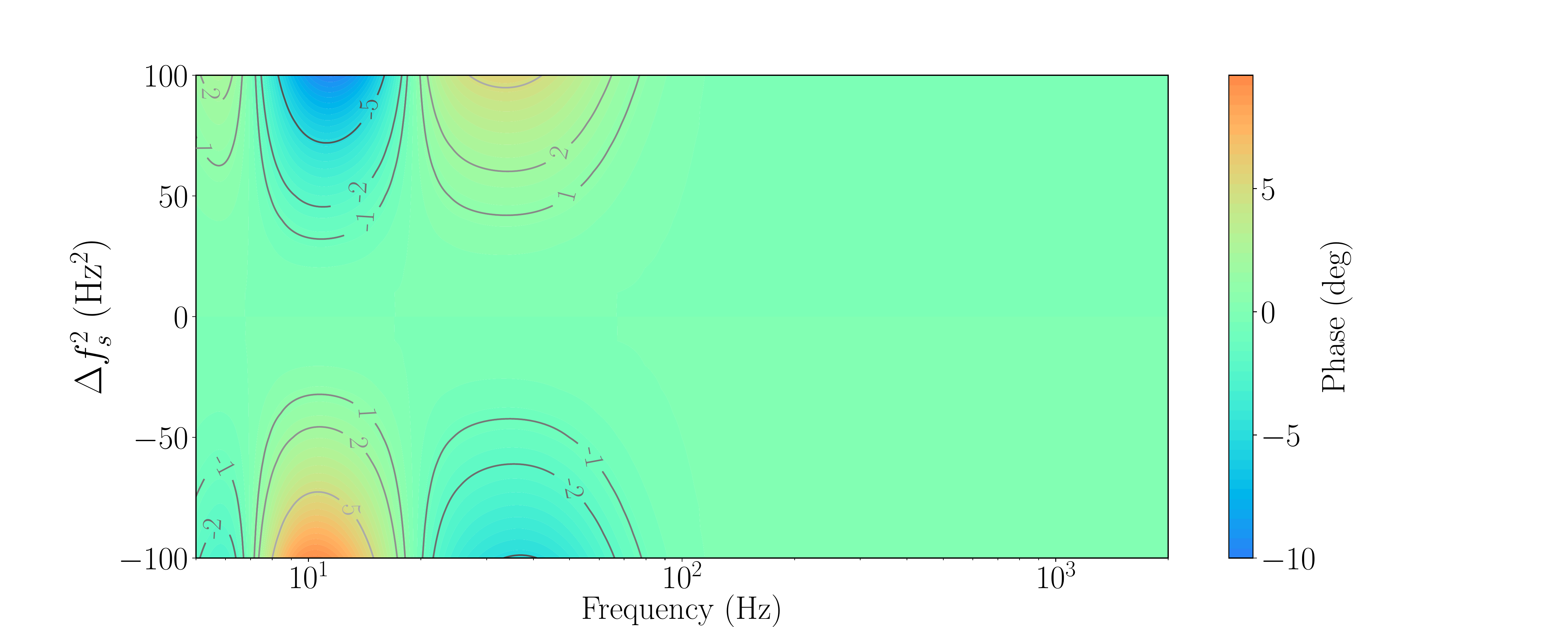}}
	\caption[]{Magnitude (top) and phase (bottom) of the fractional error ${\eta}_{R;C}-1$ in the Hanford detector response [O3A Epoch (c)] as a function of frequency due to uncorrected time-dependent SRC optical spring frequency, $f_s$ (with fixed Q=52.14). }
	\label{fig:carpet_fs}
\end{figure}

The impact of TDCFs related to SRC detuning, $f_{s}$ and $Q$, was not discussed in~\cite{Tuyenbayev2016}.
To study the impact from time-varying SRC detuning effect, we create an example with a perfectly tuned SRC reference model (i.e., $f_s=0$~Hz) and vary only $f_{s}$ ($Q$ is fixed at 52.14). 
In \eref{eqn:tdcf_sensing}, the time-varying $f_s$ value is always positive but can be real or imaginary, corresponding to an anti-spring-like or a spring-like detuned optical response, respectively. 
We quantify the spring-like or anti-spring-like effect with $f_{s}^{2}$ for simplicity; i.e., $f_{s}^{2} < 0$ is a spring response, and $f_{s}^{2} > 0$ is an anti-spring response.
\Fref{fig:carpet_fs} shows ${\eta}_{R;C}-1$ in colored contours if the time-variation of $f_s^{2}$ is not corrected. 
The variation of $f_s^{2}$, denoted by $\Delta f_s^2$ on the vertical axis, covers both the anti-spring-like and spring-like detuned optical responses.
For $|\Delta f_s^2| \lesssim 25$~Hz$^2$, the impact is generally negligible.
It has been found that occasionally we have $|\Delta f_s^2| \gtrsim 50$~Hz$^2$, resulting in an error of $\gtrsim5\%$ in the magnitude of ${R}^{\rm (model)}$.
See further discussion and treatment of the resulting systematic error in \sref{sec:detuning}.

\change{In the rest of this section, we provide additional details about the time-dependent corrections during an exceptional period in early O3A.}
There are two issues: (1) Prior to April 16, 2019, at Hanford, $\kappa_P$ and $\kappa_U$ were not applied because the estimates were computed using calibration lines separated by $\sim$20~Hz.
Such large frequency separations invalidated approximations used to compute $\kappa_P$ and $\kappa_U$.
The systematic error introduced by this issue is accounted for when reporting the overall accuracy of the estimated $h$ in \sref{sec:results}.
(2) From April 1 to June 11, 2019 at Livingston and from April 16 to June 11, 2019 at Hanford, the complete complex values for all actuator TDCFs, $\kappa_T$, $\kappa_P$, and $\kappa_U$, rather than only the real values were applied to $h$. 
Applying the complex actuator TDCFs was found to cause an overall increase in the systematic error in $h$.
For all other time periods in O3A, only the real part of $\kappa_T$, $\kappa_P$, and $\kappa_U$, were applied to $h$. The impact of (2) is discussed as follows.

By design, all of the actuator TDCFs at the reference time equal $1+0\textrm{i}$. 
At other times, the actuator TDCFs can take different values, typically with the real term $1\pm0.05$ and imaginary term $(0\pm0.01)\textrm{i}$.
Non-zero imaginary terms may arise due to small physical effects that change during the observing period or because approximations used to estimate the TDCFs break down~\cite{Viets2019}.
If the former is the case, then we expect that applying the full complex-valued TDCFs should reduce the measured systematic error in $h$.
If the latter, then applying the full complex-valued TDCFs does not correctly compensate since there was no actual physical change, so we expect the systematic error in $h$ to increase.
The application in early O3A described above was found to have typically increased measured systematic error, indicating that the latter is the problematic element during this period.

We characterize the response function systematic error in two cases: the full complex-valued actuator TDCFs are applied and only the real-valued actuator TDCFs are applied.
The method for computing the response function systematic error and associated uncertainty is described in \sref{sec:results}.
For those results presented in \sref{sec:results}, we have only considered the actuator TDCFs as real-valued.
Figures showing side-by-side results from applying only real-valued and full complex-valued actuator TDCFs are provided in~\ref{appendix:imag_kappa}.
The impact on the systematic error alone is $\lesssim 1\%$ in magnitude and $\lesssim 0.2$~deg in phase, and remains within the overall associated uncertainty.
Work is underway to implement an improved method of computing the actuation TDCFs that does not suffer this breakdown of approximations \change{so that} the full complex-value may be trusted to reflect true physical effects~\cite{Viets2019}.

\subsection{Deficiencies in the sensing function model at low frequencies}
\label{sec:detuning}

Understanding the results from weekly measurements of the Hanford sensing function below 20~Hz has posed a challenge unique in the advanced detector era. 
This section describes the results observed.

In the previous O1 and O2 observing runs, measurements at Hanford showed evidence for slight detuning of the SRC with respect to the arm cavities~\cite{Cahillane2017}.
Detuning of the SRC can be caused by either misalignment or mode mismatch with the arm cavities; both misalignment and mode mismatch can change as the thermal lenses in the input test masses change~\cite{Miyakawa2006}.
To account for the impact on the sensing function, an invertible, phenomenological, analytic representation of an optical spring was included in the model, parameterized by $f_{s}$ and $Q$, as in \eref{eqn:static_sensing} (see derivation in~\cite{hall2017long}; \change{also see discussions about the time-dependent $f_{s}$ and $Q$ in \sref{sec:tdcf}}). 
A fixed, positive value of $f_{s}^{2}$ was sufficient to describe the ensemble of sensing function measurements at Hanford throughout O1 and O2. 
The measurement ensemble of the Livingston detector in O1 and O2 showed some evidence of detuning but at sufficiently low frequency, and hence $f_{s}$ was set to 0~Hz. 
In O3A, a fixed, negative value of $f_{s}^{2}$ was sufficient to describe detuning in the Livingston detector.
The Hanford detector measurements, however, now show clear evidence for detuning responses with both $f_{s}^{2} > 0$ and $f_{s}^{2} < 0$.

There is also evidence for two-way cross-coupling between the DARM and angular control systems at Hanford in O3A, further modifying the response below 20~Hz.
To avoid point defects~\cite{Brooks2020}, the Hanford detector alignment scheme has been modified to position the laser light impinging on arm cavity optics away from the center of the optics. 
Angular motion of the optics will therefore be sensed as DARM length change, and actuators used for angular control create DARM length change.
When there is a second cross coupling from DARM length to the angular sensors, the angular control loop response impacts the measured sensing function. 
In that case, $C^{\rm (meas)}$ shows a complex, low-frequency response inconsistent with detuning and \eref{eqn:static_sensing}.

While the Hanford sensing function is more complicated than \eref{eqn:static_sensing} in O3A, we nevertheless use continuous measurements of $f_{s}$ to monitor changes in the sensing function.
These measurements show a consistent evolution of $f_{s}^{2}$ from positive to negative over the first $\sim$2~hours after the detector achieves ``observation-ready" performance but before reaching thermal equilibrium.
Once thermal equilibrium is achieved, the continuously monitored value stabilizes and shows only small variations at the level of $|\Delta f_s^2| \lesssim 1$~Hz$^2$.

\begin{figure}[tbh!]
	\begin{center}
		\includegraphics[width=0.8\textwidth]{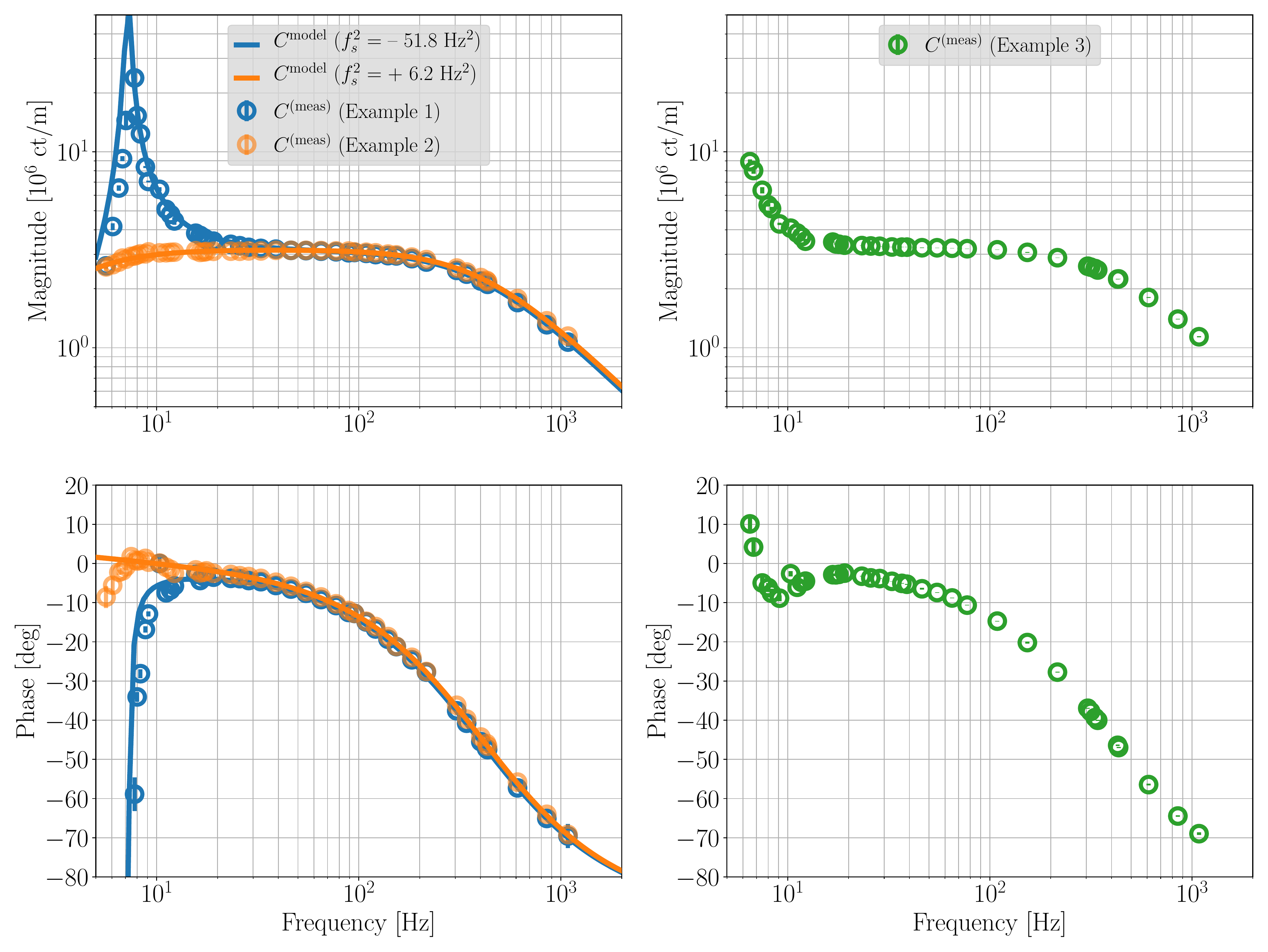}
	\end{center}
	\caption{Examples of O3A weekly sensing function measurements at Hanford in magnitude (top) and phase (bottom). The left panels show two measurements that behave like \eref{eqn:static_sensing}, with positive and negative $f_{s}^{2}$ values indicated in the legend. The right panels show a measurement that cannot be explained by \eref{eqn:static_sensing}. Vertical error bars crossing the markers indicate uncertainties of the measurements, most of which are too small to be seen by eye.}
	\label{fig:SRCLOffset} 
\end{figure}

\Fref{fig:SRCLOffset} shows several examples of sensing function measurements at Hanford in O3A. The left panels show the comparison between \eref{eqn:static_sensing} and two response measurements dominated by detuning. The right panels show an example measurement in which the low frequency response is dominated by angular cross-coupling.

We conclude that the Hanford sensing function low-frequency response depends on the complicated interaction between detuning, cross-coupling, and the thermal state of the detector as exemplified by the measurements presented in this section.
Since these effects are not modeled and poorly monitored, we use the discrepancy between model and the collection of weekly measurements to represent this deficiency as an unknown systematic error below 20~Hz using GPR described in \sref{sec:residual}.
Further limitations of \eref{eqn:static_sensing} at frequencies above 20~Hz are discussed in \sref{sec:uncompensated}.

\subsection{Accounting for unknown static frequency dependence}
\label{sec:residual}

Unknown systematic errors are accounted for by computing the
complex-valued residuals between the model and measurements of the sensing and actuation functions.
Weekly measurements taken throughout each epoch, including data at frequencies above 1~kHz and over relevant frequency bands (discussed in further detail below), are taken into consideration.
Each measurement of these interferometer components has all known loop sub-components and all known TDCFs applied such that only unidentified systematic and measurement statistical uncertainty remain, resulting in measures of ${\eta}_C$ and ${\eta}_{A_i}$.
The complex-valued frequency dependence and uncertainty of these residuals are characterized using the GPR method~\cite{scikit-learn,GPR}.
The posterior results from the GPR are then used as part of the overall response function uncertainty calculation~\cite{Cahillane2017}.

The GPR trains on the residual data using a physically motivated covariance kernel, defined as
\begin{equation}
k\left(\log(f),\log(f^\prime)\right) = \gamma_1^2 + 
\gamma_2^2\exp\left(-\frac{\left(\log(f)-\log(f^\prime)\right)^2}{2\ell^2}\right)\,,
\end{equation}
where $\{\gamma_1,\gamma_2,\ell\}$ are the hyperparameters of the
covariance kernel with the following bounding values \cite{GPR};
$\gamma_1\in[0.9,1.1]$, $\gamma_2\in[0.1,2.0]$, and
$\ell\in[0.5,1.5]$, \change{which respectively represent the magnitude scale of the residual (ideally unity), the amount of frequency-dependent correlation (ideally none), and if present, the ``length'' (in log scale) over which adjacent frequency points are correlated.}
Previous analysis in O1 and O2 determined the covariance kernel hyperparameters via GPR of the magnitude and phase residuals separately for each of the sensing and actuation functions~\cite{Cahillane2017}.
For O3, we determined the covariance kernel for the complex-valued residuals such that correlations between magnitude and phase are preserved in the residuals for a given model.
In addition, previous analysis provided more hyperparameters of the covariance kernel and did not restrict the parameter values away from unphysical regions of parameter space (e.g., covariance between complex-valued residuals of nearby frequency points should be preserved whereas residuals from widely spaced frequencies should have very small covariance).
This updated kernel, hyperparameter ranges, and use of the complex-valued residuals addresses all of these issues.

\begin{figure}[tbh!]
	\begin{center}
		\includegraphics[width=0.8\textwidth]{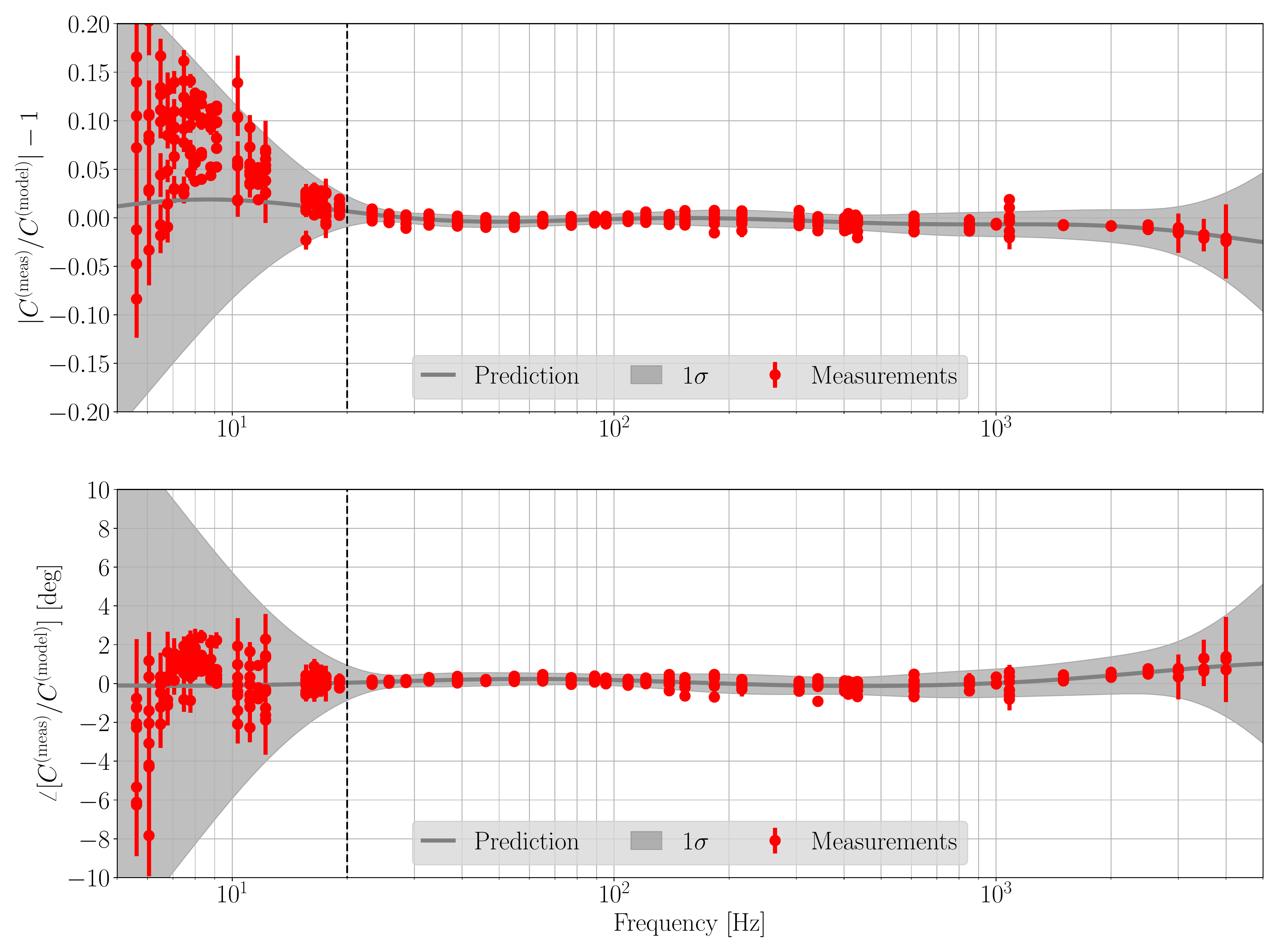}
	\end{center}
	\caption{Gaussian Process Regression results of the systematic error in Hanford sensing model [O3A Epoch (c)]. The red markers are the residuals between all the sensing measurements and the model in magnitude (top) and phase (bottom). The dark grey curve is the best prediction of the systematic error. The light grey shaded region indicates the $1\sigma$ uncertainty on the systematic error. \change{Only the residual data points to the right of the dashed vertical line are used in the GPR process.}} \label{fig:gpr_sensing}
\end{figure}

Measurements of the Hanford sensing function during O3A have shown significant deviations from the reference model at frequencies $\lesssim$20~Hz (see \sref{sec:detuning}).
It has proven difficult to model and track these changes a priori, so the variations are included as part of the residuals.
\Fref{fig:gpr_sensing} shows the measured residuals together with the GPR posterior confidence intervals.
\change{Regular measurements shown in \fref{fig:gpr_sensing} are taken after the Hanford detector reaches thermal equilibrium so that they are not under special conditions with the presence of the extreme low-frequency response discussed in \sref{sec:detuning}.
In other words, these regular measurements cannot fully represent the Hanford low-frequency response.
Data points below 20~Hz (to the left of the black dashed vertical line) are therefore excluded from the GPR process so that the sensing response in that frequency range is treated as entirely unknown.
In addition, for the Hanford sensing function residuals only, the allowed range for the frequency correlation length (in log scale) is modified to be shorter ($\ell\in[0.01,0.5]$) so that the low-frequency uncertainty appropriately represents the expected features as seen in \sref{sec:detuning}.}

Although the Hanford sensing function residuals and, in turn, the posteriors from the GPR, are larger at frequencies below 20~Hz, the actual impact on the response function uncertainty is small.
This is because the contribution to the response function by the sensing function is smaller at low frequencies than the contribution from actuation stages [i.e., below 20~Hz, the values of the purple curve in \fref{fig:h1_contri} are at least a factor of 2 times smaller than the total $A$ curve].
The Hanford actuation measurements do not show such residual variations as those seen in the sensing measurements.
No such variation is seen in any interferometric measurements of the Livingston detector.

\begin{figure}[tbh!]
	\begin{center}
		\includegraphics[width=0.8\textwidth]{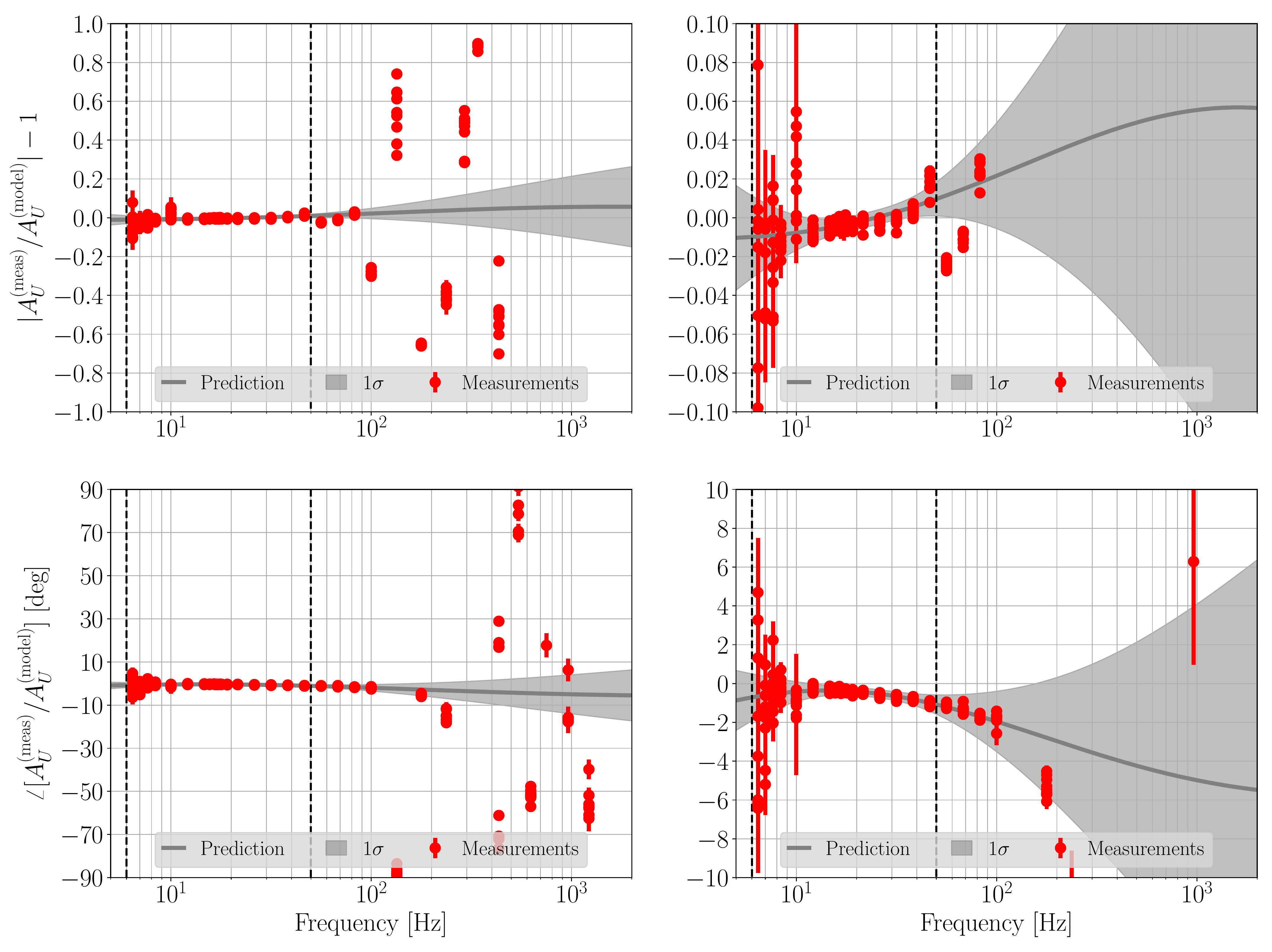}
	\end{center}
	\caption{Gaussian Process Regression results of the systematic error in Hanford UIM actuation model [O3A Epoch (c)]. The panels on the right are zoomed in on the vertical axis to display the error and uncertainty in the band of interest. The red markers are the residuals between all the UIM actuator measurements and the model in magnitude (top) and phase (bottom). The dark grey curve is the best prediction of the systematic error. The light grey shaded region indicates the $1\sigma$ uncertainty on the systematic error. \change{Only the residual data points in between the two dashed vertical lines are used in the GPR process.}} \label{fig:gpr_uim}
\end{figure}

In addition to the sensing function, it is instructive to consider the GPR for the UIM stage.
Measurements for UIM actuation stages at both detectors are consistent with model expectations between 6 and 50~Hz, so only data within this frequency range are used by the regression (see, for example, Hanford results in \fref{fig:gpr_uim}).
Similar to the sensing function below 20~Hz, the UIM contribution to the response function above 50~Hz is negligible (see \fref{fig:contri}).
Outside this band, especially above 50~Hz, measurements do not agree with the model. Restricting the regression from 6 to 50~Hz may not accurately reflect the potentially large systematic error between the model and measurements above 50~Hz.
The impact of neglecting this systematic effect above 50~Hz, however, is negligible because of the small UIM contribution. This is discussed further in \sref{sec:uncompensated}.

\subsection{Quantifying uncompensated systematic errors}
\label{sec:uncompensated}
Some \change{features of the detector response are known but excluded,} even in the most-accurate, offline production of $h$.
The resulting errors from excluding these features are small enough that they do not significantly contribute to the systematic error in the response function at frequencies between 20 and 2000~Hz, and only contribute appreciably at frequencies above 2000~Hz, or in narrow frequency bands within the 20--2000~Hz region.
We discuss and quantify these features in this section for completeness \change{and potential future importance, but do not include their impact in the final numerical estimate of the systematic error and uncertainty in $h$ for simplicity.}
 
\change{We name and enumerate the negligible sources of errors as follows: (a) FIR filters used to reproduce the offline $h$ data stream do not perfectly recreate the model at all frequencies; (b) intentionally applied low-pass and high-pass filters for improved data handling distort the data below 10~Hz; (c) the cross-coupling with auxiliary degrees of freedom in the actuation or sensing models is excluded; (d) the model of the UIM force-to-displacement transfer function is imperfect; (e) there are not well quantified residual time delays between actuator stages; (f) the amount of SRC detuning may impact the approximated coupled-cavity-pole-like response; (g) the periodic response of the Fabry-P\'{e}rot cavities to length changes is excluded; (h) measurements of the sensing function can be confused by the non-rigid-body displacement of the test masses in the presence of Pcal forces; (i) the uncertainty in the timing synchronization between multiple elements of the DARM control system is excluded.}

\begin{figure}[!tbh]
	\centering
	\scalebox{0.5}{\includegraphics{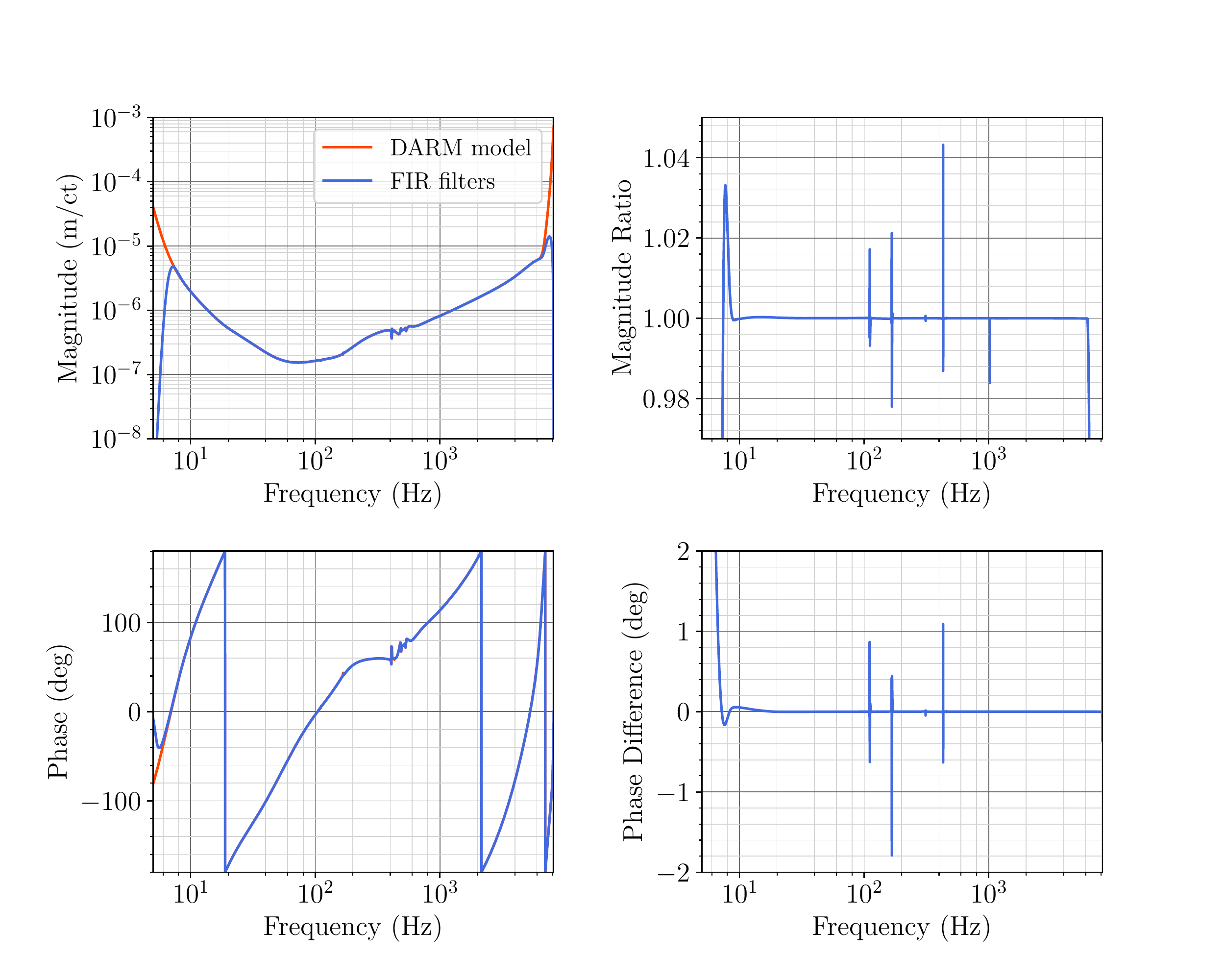}}
	\caption[]{Comparison between the effective response function implemented by the FIR calibration filters (blue) and the frequency-domain model of the response function (red) at Hanford. The left panels show the response functions, and the right panels show the residuals. The top and bottom panels are for magnitude and phase, respectively. Note that the sharp, narrow-band feature from the UIM actuator dynamics (e.g., at $\sim$150~Hz) is difficult to model with short FIR filters (see detailed discussion in text).}
	\label{fig:FIR_response}
\end{figure}

\begin{enumerate}[label=(\alph*)]
\item{FIR filters cannot perfectly reproduce all details of the DARM loop model across all frequencies.
In the frequency band from 10~Hz to 6~kHz, however, these errors are generally less than 0.1\% in magnitude and 0.1~deg in phase.
\Fref{fig:FIR_response} shows a comparison between the FIR filter implementation of ${R}^{\rm (model)}(f)$ in the offline calibration pipeline and the frequency-domain DARM model.
Some narrow-band systematic errors can be seen in the residual (right panels), caused by sharp spectral features that are difficult to resolve using FIR filters only a few seconds in duration.
A Kaiser window is applied to the FIR filter in the time domain, resulting in a frequency resolution of $\sim$3~Hz.
Such errors mostly originate from the filters that model the actuation system, especially the UIM stage at Hanford.
\Fref{fig:UIM_filter} displays the comparison between the frequency response of the UIM FIR filter and the frequency-domain model at Hanford. 
Narrow-band systematic errors caused by the limitation of short-duration FIR filters are left uncompensated.
This is a compromise between the data-loss due to FIR impulse response settling and the accuracy of reproducing the sharp spectral features.}

\begin{figure}[!tbh]
        \centering
        \scalebox{0.5}{\includegraphics{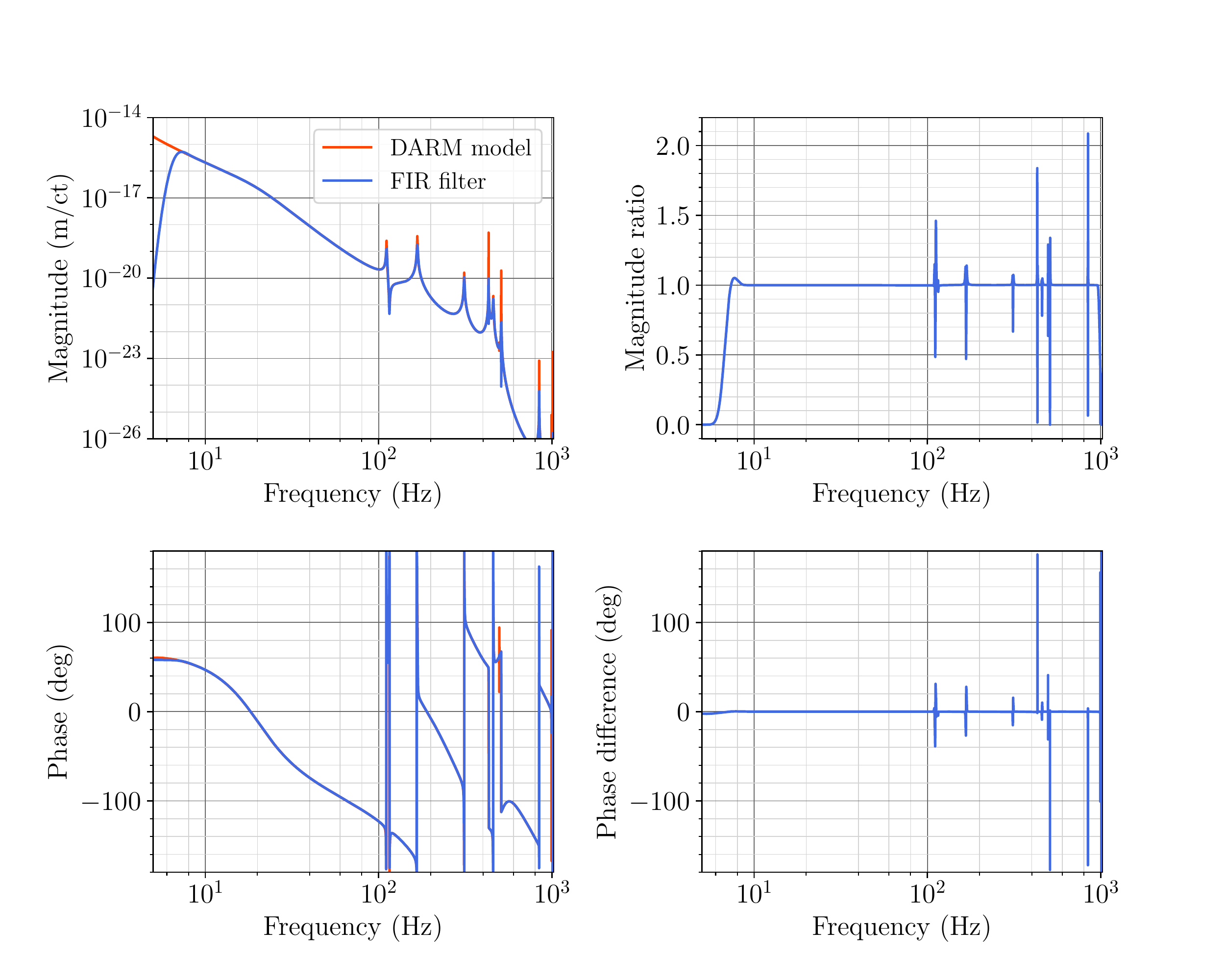}}
        \caption[]{Comparison between the frequency response of the UIM actuation filter (blue) and the frequency-domain UIM model (red) at Hanford. Panels on the left and right display the two UIM transfer functions and the fractional residual between them, respectively. The top and bottom panels are for magnitude and phase, respectively. The sharp features from $\sim$100--500~Hz are difficult to resolve with short FIR filters.}
        \label{fig:UIM_filter}
\end{figure}

\item{Below 10~Hz and above 6~kHz, the detector sensitivity degrades rapidly, and the data become dominated by $\Delta L_{\rm free}$ that is not of astrophysical origin.
For convenience in data handling, as well as the prevention of spectral leakage or aliasing, aggressive high-pass and low-pass FIR filters (with corner frequencies of 9~Hz and 6~kHz, respectively) are applied in post-processing in both the low-latency and high-latency calibration pipelines.
The well-understood systematic error resulting from these additional FIR filters only impacts these extreme frequency regions.}

\item{The physical construction of the detectors minimizes the cross-coupling between auxiliary control loops and the DARM loop.
The auxiliary loop control designs are adjusted to further reduce this cross-coupling as the detectors sensitivity is improved.
\Sref{sec:detuning} describes the first evidence of undesirable interactions between auxiliary control loops and the DARM loop during an observation period.
While more sophisticated models that account for these interactions are academically interesting, future improvements to the detector hardware and control system parameters will render complex models unnecessary.
Residual errors from cross-coupling effects are accounted for with techniques described in~\sref{sec:residual}.}

\item{The UIM-to-TST, force-to-displacement (i.e., force from UIM stage to displacement at TST stage) transfer function shows a number of resonant features above 50~Hz (as discussed in \sref{sec:residual}; see \fref{fig:gpr_uim}). 
In the mid-frequency band (50--250~Hz), the features result from the twisting and/or bending of the UIM stage vertical blade springs in the longitudinal direction as a result of the force producing $\Delta L_{\rm ctrl}$. 
Between O2 and O3, damping mechanisms were modified on the UIM vertical isolation blade springs (see~\cite{torrie2016nmbds} for details).
The increased weight of these improved dampers lowered the bending mode frequencies of the blades, changing the force-to-displacement transfer function for the UIM.
While these improved dampers change the bending modes in similar ways for both Hanford and Livingston detectors, the impact is only significant on the Hanford response function because of different choices in $F_U$ between two detectors [compare figures~\ref{fig:h1_contri} and \ref{fig:l1_contri}].
These changes were not included in the UIM actuator model, resulting in an underestimate of the contribution from $A_U$ to $R$ around the frequency of the bending mode resonances.
This results in three narrow, resonant features appearing in spectra of the Hanford calibrated data stream in the band 150--155~Hz.
These narrow features have a maximum excursion ($\sim $1~Hz width) of $\pm$3\% in magnitude and 3~deg in phase.
Careful inspection of \fref{fig:RRnom_with_pcal2darm} (discussed later in \sref{sec:results}) at $\sim$150~Hz hints at this error but does not resolve it in the overall systematic error estimate.
Recent investigations and efforts \change{(after O3A finished)} have resolved this error. \change{The correction will be applied in the UIM actuator model after O3A.} But it remains an uncompensated systematic error in the Hanford O3A data.}

\item{The estimate of the overall residual time delay from each actuation stage is determined by the MCMC fit. 
However, as shown in \fref{fig:gpr_uim} and discussed above, the data input to the MCMC may include discrepancy between the model and measurement unrelated to a time delay.
In that case, fitting for only a scalar $H_{i}$ and a delay $\tau_{i}$ is incorrect. 
We thus, after accounting for all understood time delays, restrict the frequencies of the MCMC to a band where an actuator transfer function appears to be frequency-independent in order to determine $H_{i}$.
Any remaining uncertainty in timing for each actuator stage ($\tau_{i}$) or in the sensing function ($\tau_{C}$) is determined via GPR as described in \sref{sec:residual}.}

\item{In \eref{eqn:static_sensing} we assume the SRC detuning effect is small enough that the coupled-cavity, single-pole response of the coupled arm and SRC cavities (the first parenthetical term) and the detuned SRC response (the second parenthetical term) can be separated and parameterized independently by $f_{cc}$ and $f_{s}$.
The physical model from which \eref{eqn:static_sensing} was derived~\cite{hall2017long,Hall2019}, however, suggests that the response at frequencies $\sim$300~Hz and above may no longer be described by $f_{cc}$ alone when SRC detuning is sufficiently large.
The amount of detuning is proportional to any modification between the GW signal phase and the phases determined by two physical quantities: (a) the homodyne phase $\zeta$, which could deviate from its nominal value due to unintended small imperfections in the instrument, and (b) the signal extraction phase $\phi_{\rm SRC}$ determined by the SRC cavity length.
Both $\zeta$ and $\phi_{\rm SRC}$ are nominally 90~deg. 
The residual between the physical model response and the approximate model response in \eref{eqn:static_sensing} shows the systematic error is frequency dependent, but does not exceed 1\% in magnitude or 1~deg in phase at frequencies below 1~kHz for the measured extremes of detuning, $|f_{s}^2| \lesssim 75$~Hz$^2$ (or equivalently, $|\phi_{\rm SRC} - 90^\circ| < 1^\circ$ and $|\zeta - 90^\circ| < 3^\circ$).}

\item{\change{The above mentioned} single-pole response is also an approximation to the complete response of the Fabry-P\'{e}rot arm cavities fluctuation in their lengths~\cite{RAKHMANOV2002}.
This approximation leads to errors in the sensing function at high frequencies above 1~kHz in both magnitude and phase (larger in phase).
The resulting phase error is compensated by an artificial time delay of $-11.7$~microseconds, included in $\tau_{C}$ [the last term in \eref{eqn:static_sensing}].
The magnitude error, increasing with frequency up to 4\% at 5~kHz, is left uncompensated~\cite{KiwamuSinglepole}.
The systematic errors in magnitude and phase resulting from these two approximations \change{[(f) and (g)]} of the detector full opto-mechanical response are accounted for in the uncertainty of unknown systematic error via GPR.}

\item{During Pcal excitation ($x_{\rm Pcal}$), the actuation forces deform the test masses in their natural bending modes, producing a deformation-induced, arm cavity length variation (not equivalent to the displacement of center of the mass) sensed by the interferometer.
This phenomenon impacts the accuracy of $\Delta L_{\rm Pcal}$ at high frequencies ($\gtrsim 1$~kHz) to a level depending on the positions on the test mass surface where the Pcal beams reflect~\cite{Karki2016,karki2019accurate}. 
We estimate that the reflecting positions are within $\pm$2~mm of their optimal locations (close to the nodal circle of the dominant mode). 
The magnitude error in the estimate of $H_{\rm Pcal}$ due to the deformation is $\lesssim0.1\%$ below 1~kHz, increases with frequency, and reaches at most $\sim5\%$ at 5~kHz~\cite{karki2019accurate}.
The phase error may also increase with frequency, but is expected to be less than 0.5~deg even at 5~kHz.
This may limit the accuracy of the long-duration measurements used to characterize the sensing function.
We see no evidence for this error exceeding all other known and unknown systematic errors above 1~kHz (e.g., see \fref{fig:gpr_sensing}).
As such, this effect has been excluded from $H_{\rm Pcal}$. }

\item{Finally, within a given detector, the analog and digital components of the DARM loop are synchronized via a sophisticated timing system~\cite{bartos2010advanced}.
The uncertainty in synchronization of these components is less than 1~microsecond throughout O3A~\cite{Asali2019}.
The frequency-dependent phase impact from timing uncertainty on an individual detector is believed to be within the bounds of the unknown systematic error, ${\eta}_{\rm C}$ or ${\eta}_{\rm A}$, estimated via GPR, and hence is not explicitly accounted for.
The GW detectors within the network are synchronized to each other via the GPS receivers of the timing systems.
The network timing uncertainty, estimated to be at the level of 10~microseconds~\cite{2008GPSSPS}, is negligible compared to the uncertainty in estimates of the time-of-arrival for any GW event (typically at the level of 1 millisecond).}

\end{enumerate}

\section{Combined error and uncertainty estimate}
\label{sec:results}
\change{In this section, we quantify the overall combined systematic error and uncertainty present in the detector response via a numerical approach.
As a reminder, systematic errors presented here are not corrected in the final estimated $h$ data stream (see \sref{sec:introduction}).
\Sref{sec:resultsatagiventime} describes the method of estimating the combined error and uncertainty at a given time.
With the collection of time-specific statistics obtained using the method described in \sref{sec:resultsatagiventime} at a 1-hour cadence throughout the observing run, we evaluate the variation of the combined error and uncertainty over time in \sref{sec:resultsovertime}.
In \sref{sec:resultsdicussion}, we discuss the features seen in the numerically estimated error and uncertainty, and briefly describe how the calibrated data stream and these estimated error and uncertainty are used in astrophysical analysis.}

\subsection{Estimate at a given time}
\label{sec:resultsatagiventime}

We numerically estimate the combined estimate of systematic error and uncertainty, ${\eta}_{\rm R}(f;t)$, in each detector's response function at a given time $t$ as follows. 

Ten thousand response functions, ${R}_{i}(f;t)$, are constructed with
\begin{equation}
{R_i} (f;t)  =  \eta_{{\rm Pcal}_i}\left[\frac{1}{{\eta}_{C_i}(f) {C}(\bm{\lambda}_{i}^C;f;t)}+  {\eta}_{A_i}(f) {A}(\bm{\lambda}_{i}^A;f;t) {D}(f)\right].
\label{eq:rrnom}
\end{equation}
Here, $i$ indexes each response function and all draws associated with it. 
The $i$th sensing and actuation functions ${C}(\bm{\lambda}_{i}^C;f;t)$ and ${A}(\bm{\lambda}_{i}^A;f;t)$ are constructed using \eref{eqn:tdcf_sensing} and \eref{eqn:tdcf_actuation}, with the $i$th draw from the MCMC posterior distributions of the reference model parameters, $\bm{\lambda}^C$ and $\bm{\lambda}^A$ (\sref{sec:mcmc}).
Within the time-dependent ${C}(\bm{\lambda}_{i}^C;f,t)$ and ${A}(\bm{\lambda}_{i}^A;f,t)$, TDCFs at time $t$ are applied (\sref{sec:tdcf}).
To account for the uncertainties of the TDCFs, we draw TDCF samples from normal distributions centered at the values recorded at time $t$ with $1\sigma$ standard deviation calculated using \eref{eq:unc_meas}.
The complex-valued, fractional, frequency-dependent residual functions, ${\eta}_{C_i}(f)$ and ${\eta}_{A_i}(f)$, are drawn from the sensing and actuation GPR posterior distributions, respectively (\sref{sec:residual}). \change{Note that here we do not explicitly split out the three stages in $A$, and use $i$ in ${\eta}_{A_i}$ to index the samples of the residual in total $A$.}
By drawing samples from the MCMC and GPR posterior distributions, the covariance between parameters in $\bm{\lambda}^C$ or $\bm{\lambda}^A$, and covariance between frequency points of ${\eta}_{C_i}(f)$ or ${\eta}_{A_i}(f)$, is preserved.
Finally, $\eta_{{\rm Pcal}_i}$ is an overall multiplicative real-valued scale factor drawn from a normal distribution centered at $\eta_{\rm Pcal}$ for each detector, with $1\sigma$ standard deviation equal to the Pcal system uncertainty (\sref{sec:precursory}). This factor accounts for the Pcal uncertainty and systematic error common to all interferometric measurements and TDCF computations for a given detector. Therefore it is convenient to apply $\eta_{{\rm Pcal}_i}$ to each ${R_i} (f;t)$ rather than equivalently to $H_{\rm Pcal}$, or to each interferometric transfer function and TDCF calculations that involve $\Delta L_{\rm Pcal}$.

The time-dependent MAP response function ${R}_{\rm MAP}(f;t)$ is constructed with the MAP parameters of the sensing and actuation functions ($\bm{\lambda}_{\rm MAP}^C$ and $\bm{\lambda}_{\rm MAP}^A$), similar to~\eref{eqn:response}, given by
\begin{equation}
\label{eq:Rmap}
{R}_{\rm MAP}(f;t)  =  \frac{1}{{C}(\bm{\lambda}_{\rm MAP}^C;f;t)} +  {A}(\bm{\lambda}_{\rm MAP}^A;f;t) {D}(f).
\end{equation}
\change{At time $t$, if the time-dependent systematic error is removed from the estimated $h$ data stream, it is equivalent to having a corrected ${R}_{\rm MAP}$ function with the TDCFs recorded at that time applied to ${C}(\bm{\lambda}_{\rm MAP}^C;f;t)$ and ${A}(\bm{\lambda}_{\rm MAP}^A;f;t)$. Otherwise the reference TDCF values are used when computing \eref{eq:Rmap}, i.e., systematic errors due to the uncorrected TDCFs are left in the estimated $h$ for that time.}

We then divide each ${R}_{i}(f;t)$ by ${R}_{\rm MAP}(f;t)$ to create the probability distribution of ${\eta}_{R}(f;t)$.
At any given time $t$ and frequency $f$, the median (50th percentile) value of the distribution ${\eta}_{R}(f;t)$ represents the total systematic error in ${R}^{\rm (model)}(f;t)$ at that time and frequency \change{(generally not expected to equal zero due to the residual systematic error)}. 
The 16th and 84th percentiles represent the lower and upper bounds, respectively, of the combined systematic error and $1\sigma$ statistical uncertainty in ${R}^{\rm (model)}(f;t)$. 
As such, these percentiles of ${\eta}_{R}(f;t)$ represent the complex-valued, frequency-dependent, overall uncertainty and systematic error bounds of $h$ at time $t$.

\begin{figure}[!tbh]
	\centering
	\includegraphics[width=0.75\textwidth]{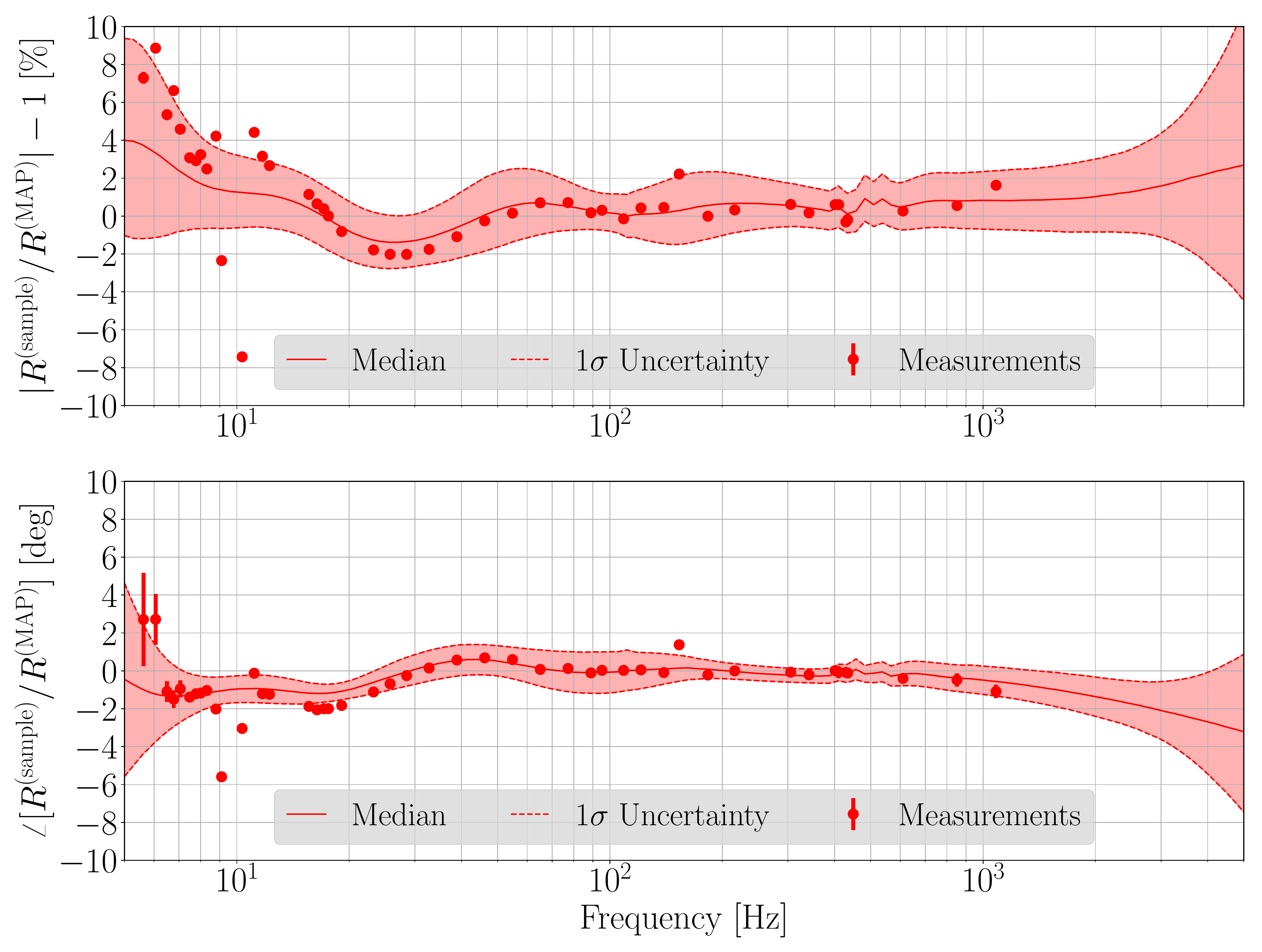}
	\caption[]{Combined error and uncertainty estimate at the reference time of Epoch~(c) for the Hanford detector. The top and bottom panels show the frequency-dependent excursions of the response from unity magnitude and zero phase compared to ${R}_{\rm MAP}$, respectively. The dashed curves indicate the 16th and 84th percentiles of the ${\eta}_R$ distribution. The solid curve is the median of the ${\eta}_R$ distribution, indicating the best estimated frequency-dependent systematic error in ${R}^{\rm (model)}$. The shaded region bounded by the dashed curves represents the $1\sigma$ uncertainty bounds on the systematic error. The red dots show a set of validating measurement taken on September 16, 2019, which are generally consistent with the overall uncertainty estimate. Vertical error bars crossing the markers indicate uncertainties of the measurements, most of which are too small to be seen by eye.}
	\label{fig:RRnom_with_pcal2darm}
\end{figure}

An example of the combined uncertainty and error estimate, ${\eta}_R(f;t)$, for the Hanford detector is shown in \fref{fig:RRnom_with_pcal2darm}.
The vertical axes indicate the excursions of ${\eta}_R(f;t)$ from zero systematic error, i.e., unity magnitude (top panel) and zero phase (bottom panel).
The solid curve shows the median value of ${\eta}_R(f;t)$, indicating the best estimate of the frequency-dependent systematic error in the response function at that time. 
The dashed curves bounding the shaded region represent the collection of $1\sigma$ uncertainties, including that of the systematic error.
The red dots show a swept-sine measurement of $hL/\Delta L_{\rm Pcal}$ taken on September 16, 2019, that aligns with the estimate of ${\eta}_R(f;t)$ at that time \change{within the frequency band 20--1000~Hz}.
Some measured data points at frequencies below 20~Hz deviate from the median curve and exceed the $1\sigma$ uncertainty bounds.
This is \change{a hint} of the systematic error induced by, e.g. detuning between the SRC and the arm cavities (see \sref{sec:detuning}), or resonant modes of the quadrupole suspension actuator stages that are not sufficiently accounted for when estimating ${\eta}_R(f;t)$.
\change{The single outlying data point around 150~Hz is caused by the imperfect dynamical model of the UIM stage at Hanford (see \sref{sec:uncompensated}). 
The distribution of ${\eta}_R$ at any given frequency and  time is generally Gaussian. In \ref{appendix:eta_dist}, we show the ${\eta}_R$ distribution at a cross section of 20.57~Hz in \fref{fig:RRnom_with_pcal2darm}, close to the lower end of the 20--2000~Hz band. The median and mean values of ${\eta}_R$ overlap each other, as shown in \fref{fig:eta_R_20Hz}. Other sample cross sections at 99.58~Hz and 509.15~Hz are examined and the distributions are similar. We show one example in \ref{appendix:eta_dist} for brevity.}

\subsection{Estimate over time}
\label{sec:resultsovertime}
\begin{figure}[!tbh]
	\centering
	\subfigure[]
	{
		\label{fig:h1_chunk1}
		\scalebox{0.20}{\includegraphics{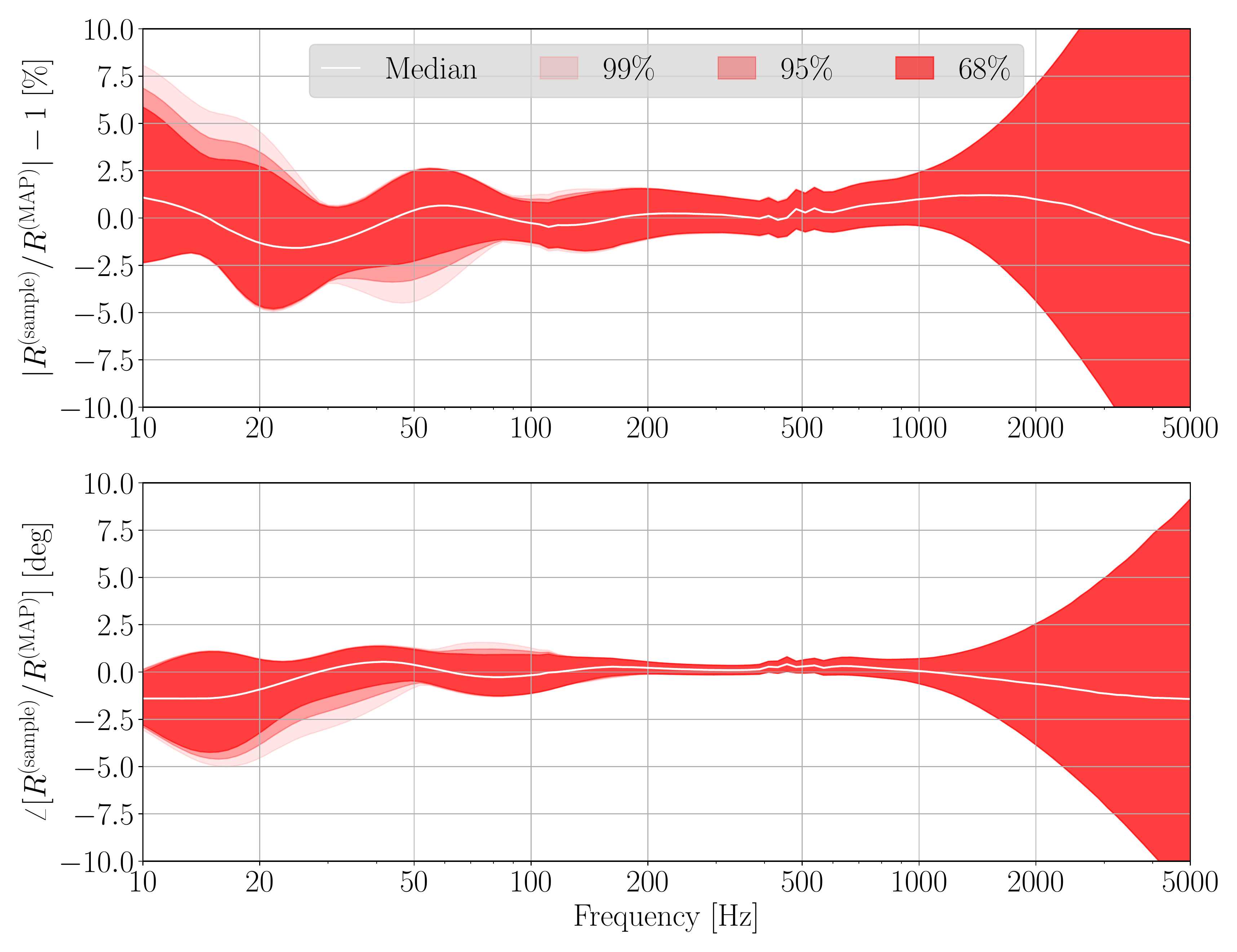}}
		\scalebox{0.20}{\includegraphics{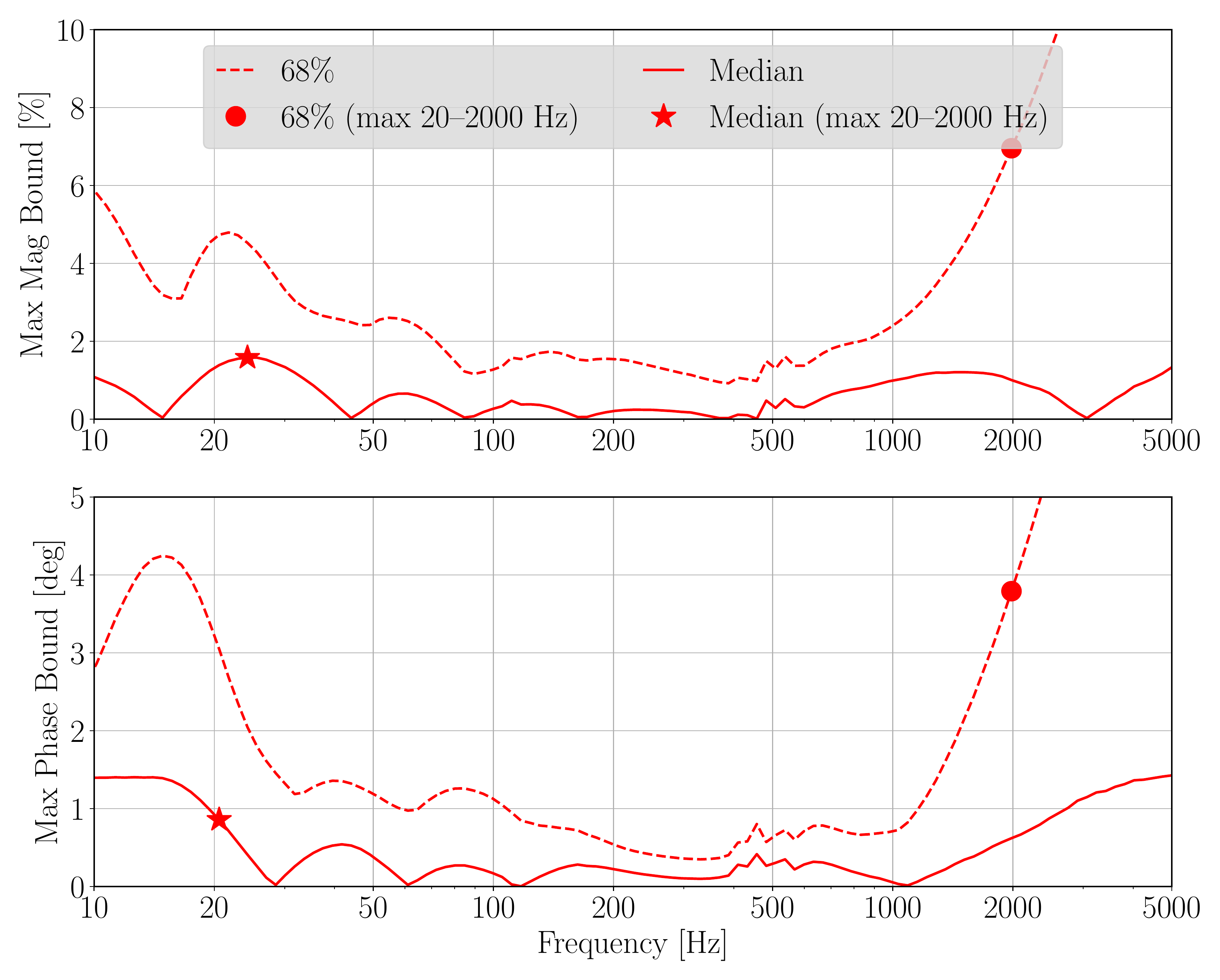}}
	}
	\subfigure[]
	{
		\label{fig:h1_chunk2}
		\scalebox{0.20}{\includegraphics{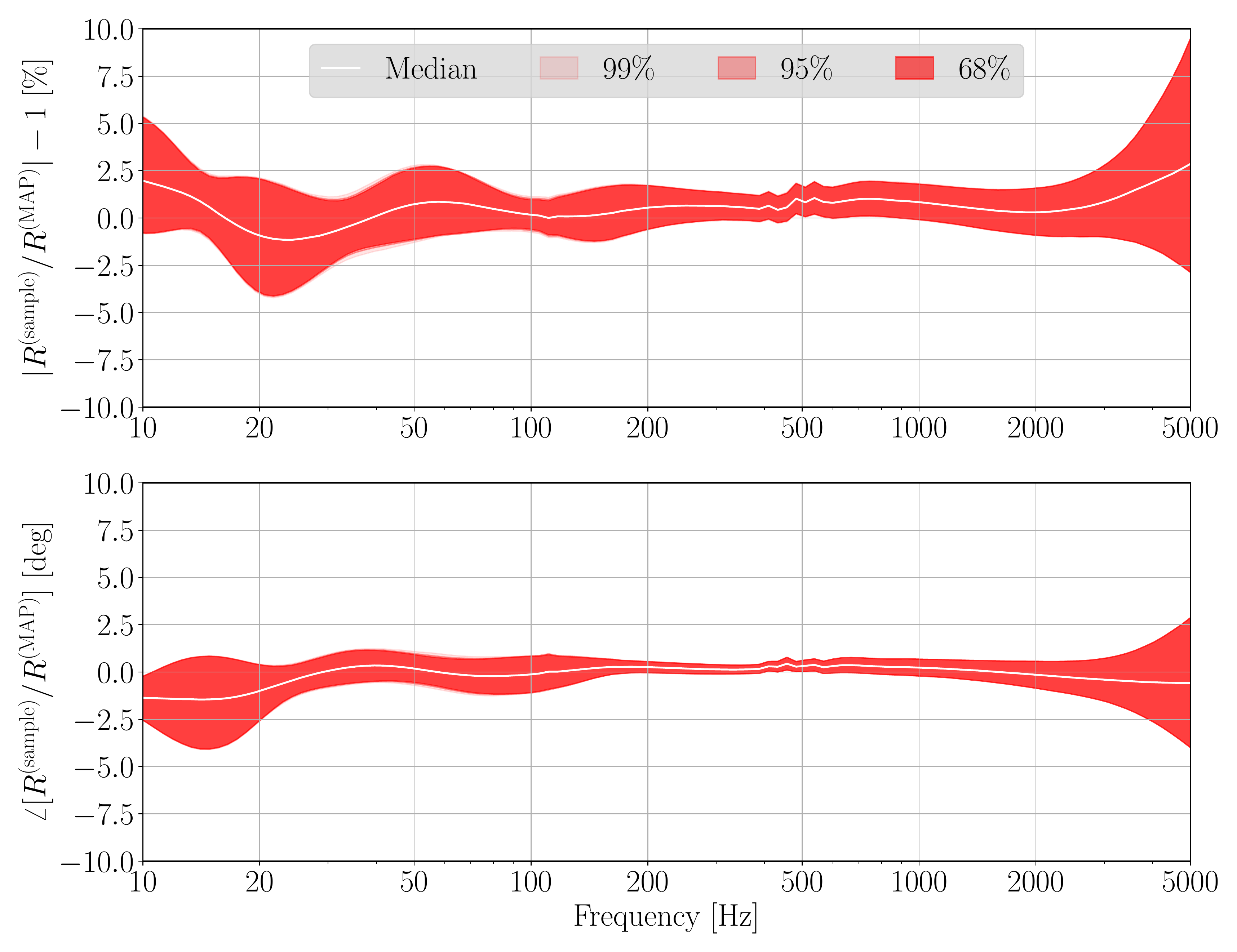}}
		\scalebox{0.20}{\includegraphics{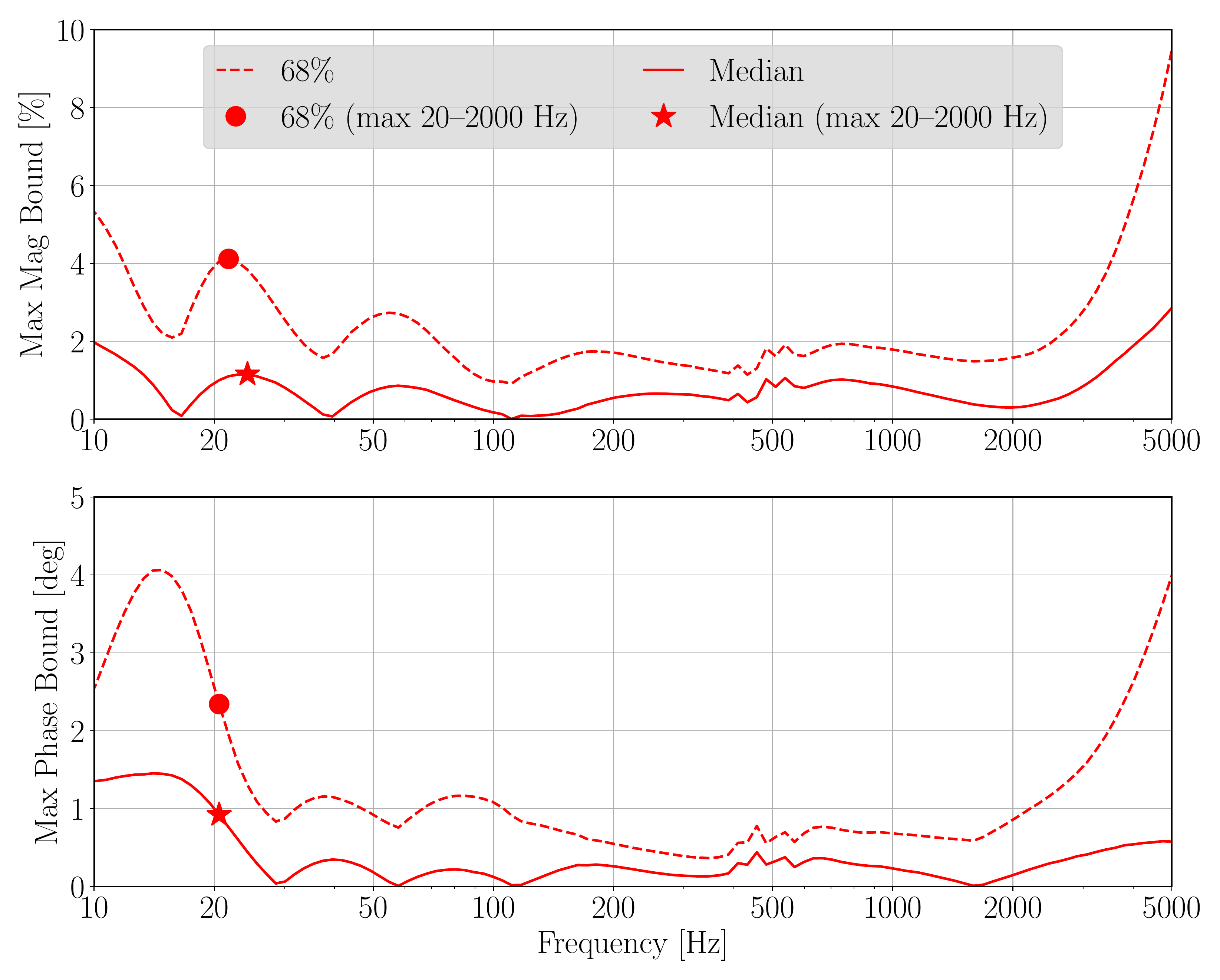}}
	}
	\subfigure[]
	{
		\label{fig:h1_chunk3}
		\scalebox{0.20}{\includegraphics{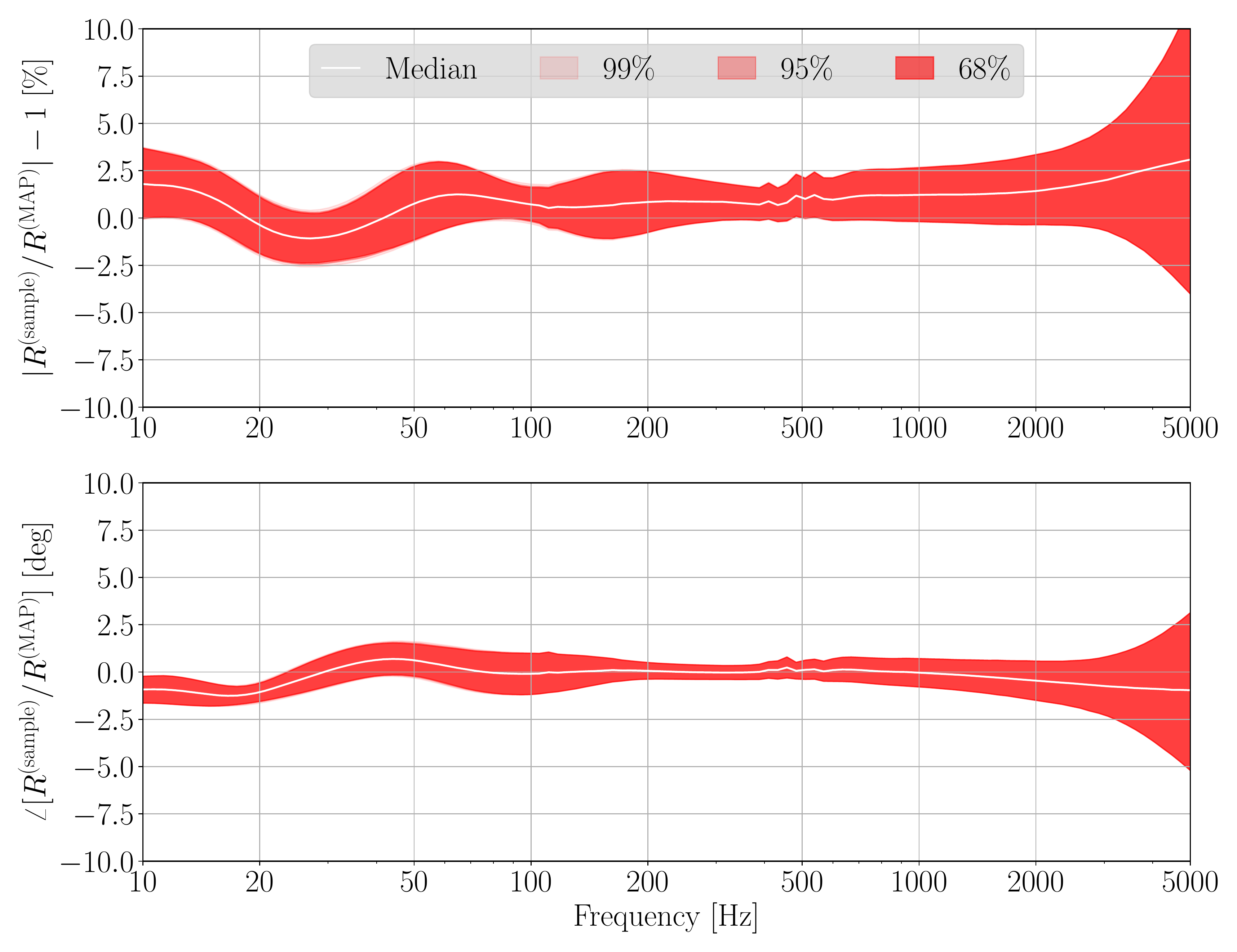}}
		\scalebox{0.20}{\includegraphics{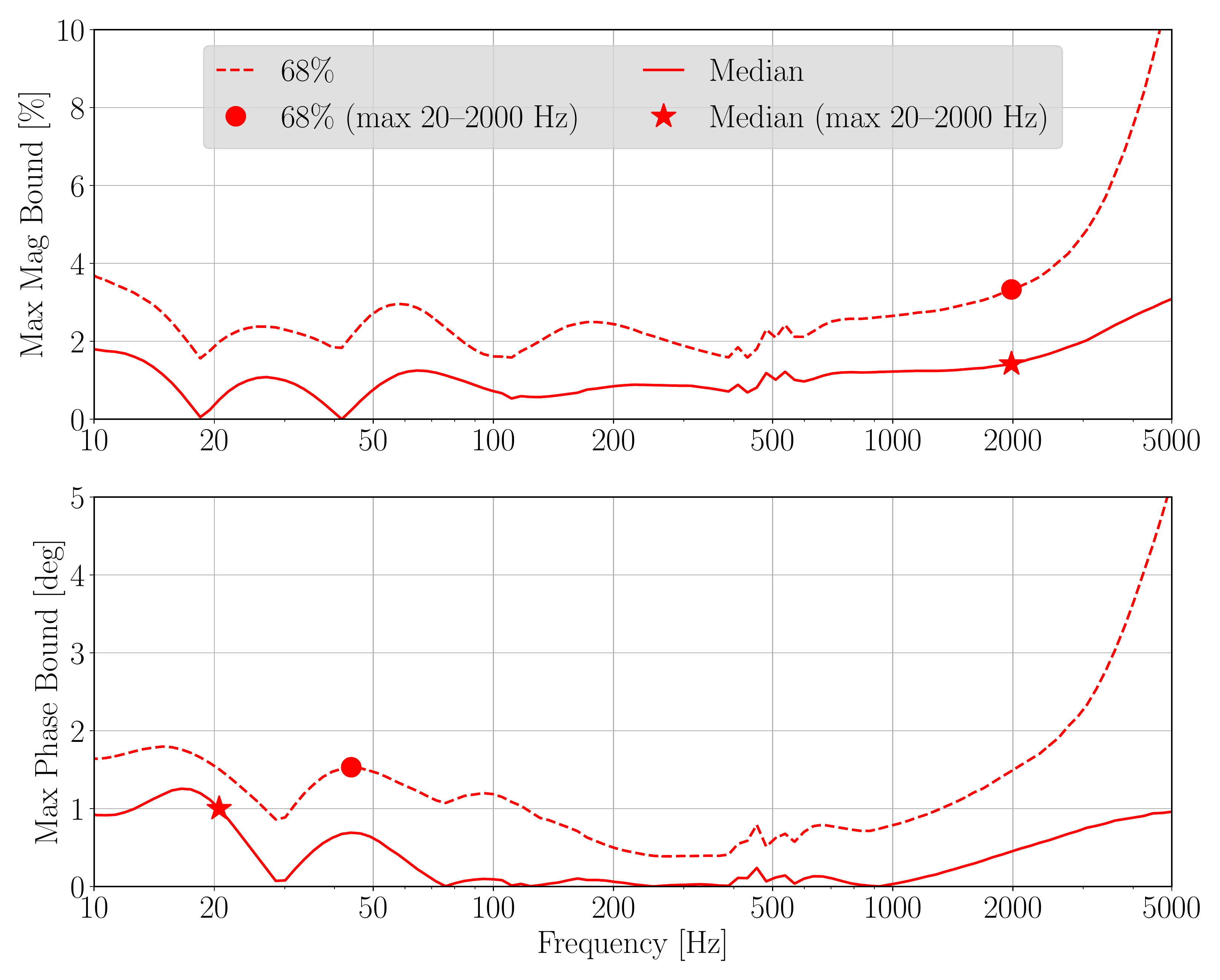}}
	}
	\caption[]{Variation of the combined systematic error and uncertainty (left) and the maximum bounds (right) for Hanford. The three subfigures correspond to Hanford epochs (a)--(c) in \tref{tab:C01_results}. 
		The top and bottom panels of each subfigure show the frequency dependent excursions of response from unity magnitude and zero phase compared to ${R}_{\rm MAP}$, respectively. The percentiles are obtained from all the hourly evaluated ${\eta}_R(f;t_k)$ over each epoch. 
		In the left panels, the colors represent $1\sigma$ uncertainty for 68\%, 95\%, and 99\% of the run time, as indicated in the legend. The white curve indicate the median excursion. 
		The absolute values of the boundaries (median and 68\%) in the left panels are plotted on the right. The star and dot markers indicate the median and $1\sigma$ maximum excursions in the frequency band 20--2000~Hz, respectively.}
	\label{fig:h1_results}
\end{figure}

\begin{figure}[!tbh]
	\centering
	\subfigure[]
	{
		\label{fig:l1_chunk1}
		\scalebox{0.20}{\includegraphics{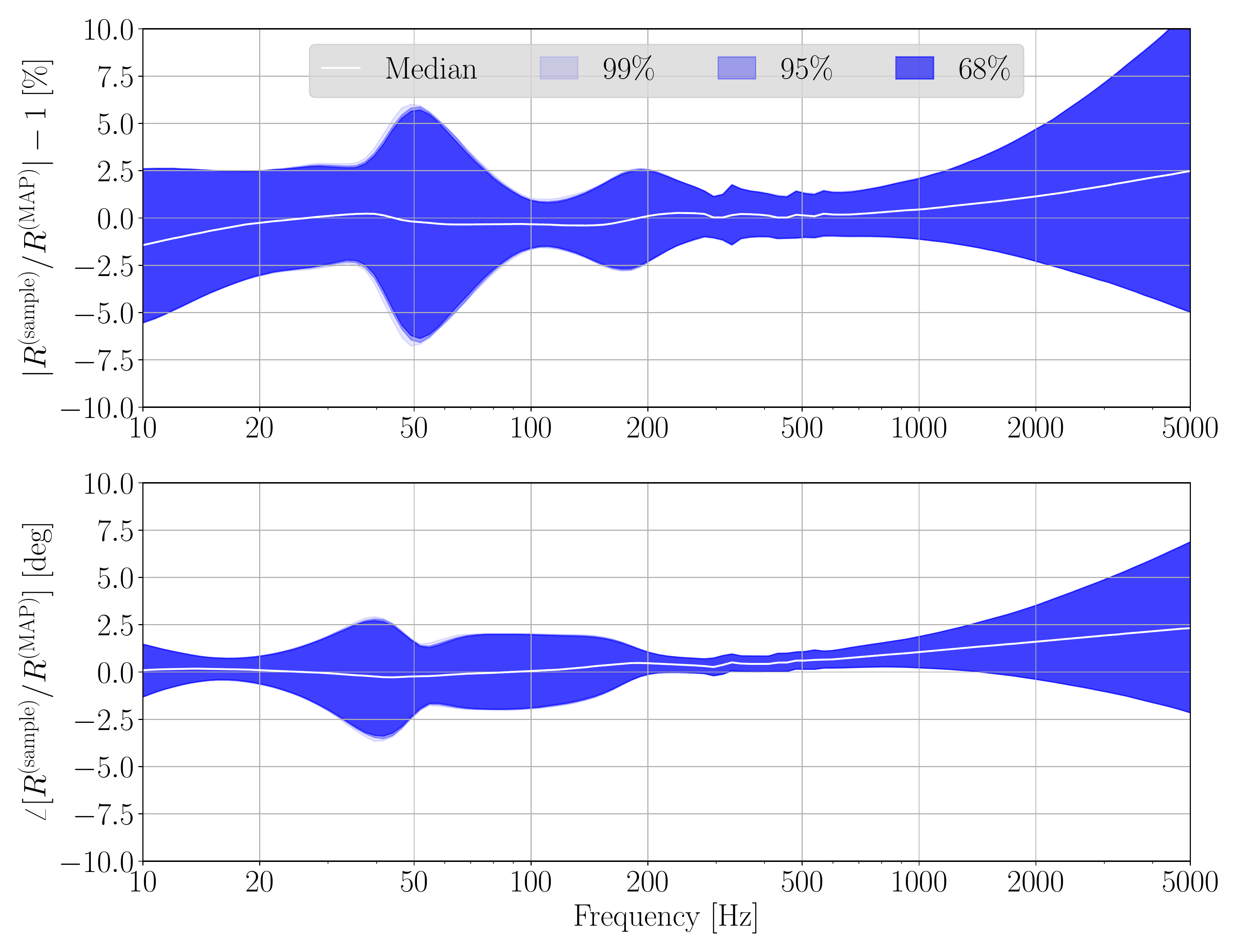}}
		\scalebox{0.20}{\includegraphics{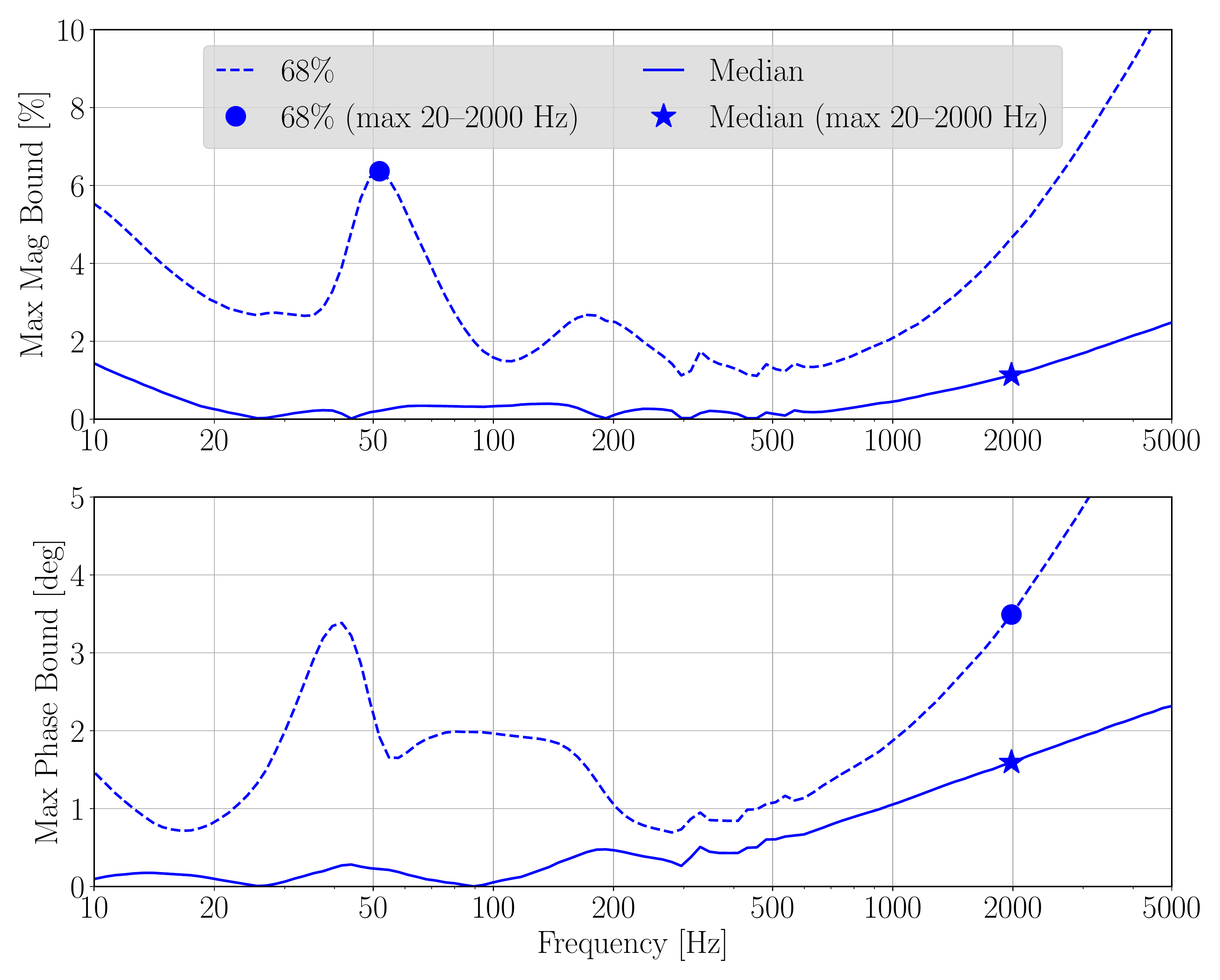}}
	}
	\subfigure[]
	{
		\label{fig:l1_chunk2}
		\scalebox{0.20}{\includegraphics{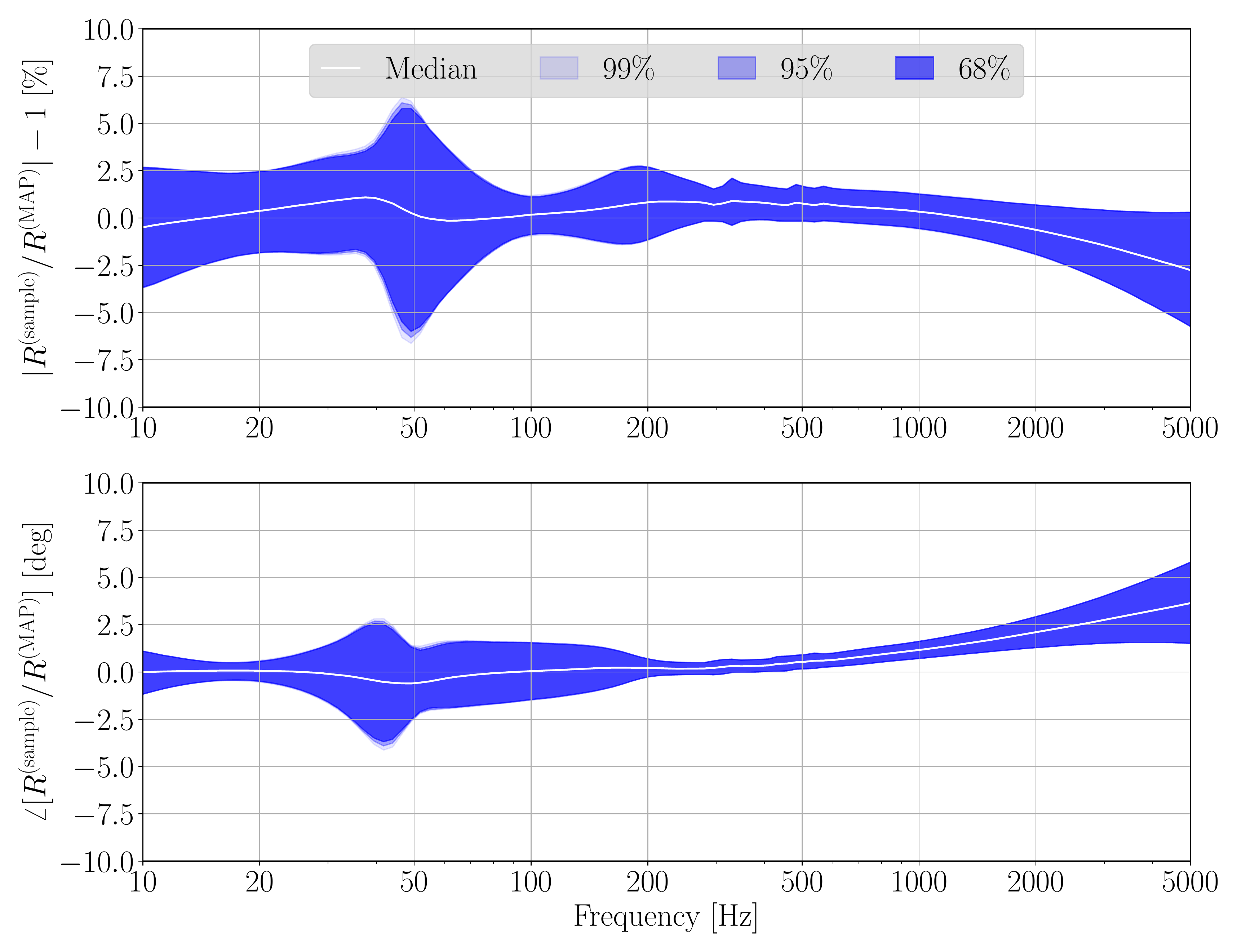}}
		\scalebox{0.20}{\includegraphics{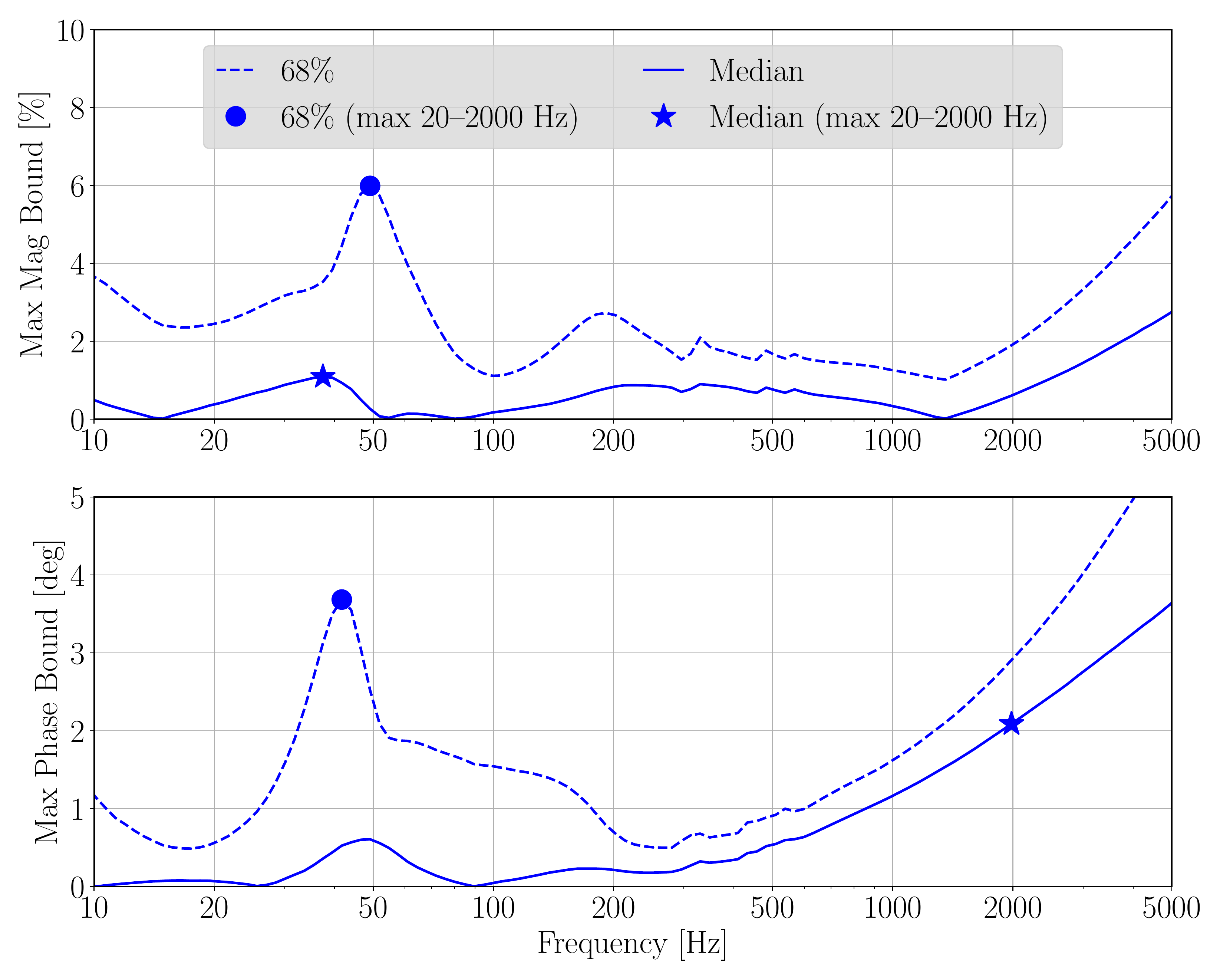}}
	}
	\caption[]{Variation of the combined systematic error and uncertainty (left) and the maximum bounds (right) for Livingston. The two subfigures correspond to Livingston epochs (a)--(b) in \tref{tab:C01_results}. 
		The top and bottom panels of each subfigure show the frequency dependent excursions of response from unity magnitude and zero phase compared to ${R}_{\rm MAP}$, respectively. The percentiles are obtained from all the hourly evaluated ${\eta}_R(f;t_k)$ over each epoch.
		In the left panels, the colors represent $1\sigma$ uncertainty for 68\%, 95\%, and 99\% of the run time, as indicated in the legend. The white curve indicate the median excursion. 
		The absolute values of the boundaries (median and 68\%) in the left panels are plotted on the right. The star and dot markers indicate the median and $1\sigma$ maximum excursions in the frequency band 20--2000~Hz, respectively.}
	\label{fig:l1_results}
\end{figure}

Estimates of the combined systematic error and uncertainty over longer periods are generated using the collections of time-specific estimates described in \sref{sec:resultsatagiventime}.
To quantify the final calibration accuracy and precision in O3A, the entire duration is split into three epochs for Hanford, and two for Livingston (see \tref{tab:epoch}).
Each epoch is defined by a physical configuration change in the detector. Within each epoch only TDCFs vary.
Previous shorter-duration observing runs did not require intra-run epochs, hence estimates of systematic error and associated uncertainty were constructed for those entire observing runs~\cite{Cahillane2017}.
In O3A, the combined uncertainty and systematic error for each epoch is quantified using the collection of percentile curves of ${\eta}_{R}(f;t)$ described above, and shown in figures~\ref{fig:h1_results} and \ref{fig:l1_results}.

The procedure of calculating these results is as follows.
First, the distribution of \change{${\eta}_{R}(f;t_k)$} is computed with a 1-hour cadence during observing periods, \change{i.e., $t_k$ takes discrete values with 1-hour cadence}.
Second, we compute the 16th, 50th, and 84th percentile curves from each of these distributions representing the systematic error and $1\sigma$ uncertainty bounds at that time (i.e., the lower, median, and upper curves shown in \fref{fig:RRnom_with_pcal2darm}). \change{We denote the 16th, 50th, and 84th percentile curves at time $t_k$ by $-\sigma_{\eta_{R}}(f;t_k)$, $\tilde{\eta}_R (f;t_k)$, and $+\sigma_{\eta_{R}}(f;t_k)$.}
Third, all hourly percentile curves within an epoch form a complex-valued, frequency-dependent ``epoch distribution" of systematic error and uncertainty estimates (the medians and the upper and lower uncertainty bounds). 
\change{For computational reasons, we make use of the condensed statistics, i.e., $-\sigma_{\eta_{R}}(f;t_k)$, $\tilde{\eta}_R  (f;t_k)$, and $+\sigma_{\eta_{R}}(f;t_k)$, rather than saving all $10^4$ samples collected at each $t_k$. 
Finally, within each epoch, the median of $\tilde{\eta}_R (f;t_k)$ and the distribution of $\pm \sigma_{\eta_{R}}(f;t_k)$ for all $t_k$} are used to determine 
the variability of \change{those hourly condensed statistics} and the rate that the upper and lower bounds exceed a given value (frequency-dependent; in magnitude and phase). \change{For the rest of the paper, we call these distributions constructed from hourly condensed statistics, ``epoch distributions".}

In the left panels of figures~\ref{fig:h1_results} and \ref{fig:l1_results}, 
the white curves, \change{i.e., the median of $\tilde{\eta}_R (f;t_k)$,} indicate the estimated frequency-dependent systematic error for each epoch. 
The 68\%, 95\%, and 99\% \change{confidence intervals} of the $1\sigma$ uncertainty boundaries in the epoch distributions are shown as dark, moderate, and light shaded regions, respectively. 
\change{The upper and lower bounds of the 68\% shaded region are, respectively, the 84th percentile of $+\sigma_{\eta_{R}}(f;t_k)$ and the 16th percentile of $-\sigma_{\eta_{R}}(f;t_k)$. Similarly, the 95\% and 99\% shaded regions can be constructed from the distributions of $\pm \sigma_{\eta_{R}}(f;t_k)$.}
These \change{epoch distributions} quantify the time-dependent variation of the combined uncertainty and systematic error bounds over the entire epoch.
Figures~\ref{fig:h1_chunk2}, \ref{fig:h1_chunk3}, \ref{fig:l1_chunk1}, and \ref{fig:l1_chunk2} show that the variation of the overall uncertainty bounds is generally negligible (i.e., the 68\%, 95\%, and 99\% interval boundaries almost overlap in each epoch).
\Fref{fig:h1_chunk1}, however, shows that the variation of the uncertainty bounds during the first epoch of O3A for the Hanford detector is not negligible.
The 95\% and 99\% intervals deviate from the 68\% interval due to uncorrected $\kappa_U(t)$ and $\kappa_P(t)$ variations during the first 16 days of the first epoch at Hanford (see \sref{sec:tdcf}).

In the right panels, we introduce a simplified presentation of the results, \change{for the convenience of discussions and comparisons in astrophysical communities}.
For brevity, the systematic error and uncertainty estimate for a given epoch across a given frequency band is \change{quoted by two numbers (one for magnitude and the other for phase), which indicate the maximum excursions from zero systematic error in that band.}
The maximum excursion values are determined as follows.
\change{First, in each epoch and at all frequencies, the absolute values of both the upper and lower bounds of the 68\% epoch-distribution interval (dark shaded region in the left panels) are computed.} 
Then, a frequency-dependent curve (dashed) is formed by taking the larger of the two absolute values at any given frequency. 
The solid curve in each of the right panels represents the absolute values of the white median curve on the left.
Finally, the maximum value of these curves \change{in the right panels} is determined within a frequency band, over which a given GW analysis is conducted.
\change{We give an example in this figure for the most sensitive frequency band of 20--2000~Hz.}
The star and dot markers indicate the maximum excursions in the frequency band 20--2000~Hz, \change{corresponding to the 68\% bounds and median value of the epoch distribution at the indicated frequencies}, respectively. 
The values of these markers in each epoch for each detector are listed in \tref{tab:C01_results}. The maximum median values represent the best estimate of the systematic error bounds in the band 20--2000~Hz. 

\begin{table}[!tbh]
	\caption{\label{tab:C01_results} O3A calibration epochs and the maximum $1\sigma$ and median excursions of response from unity magnitude and zero phase compared to ${R}_{\rm MAP}$, in the frequency band 20--2000~Hz. The maximum median values represent the best estimate of the systematic error bounds.}
	\begin{indented}
		\item[]\begin{tabular}{@{}lllll}
			\br
			Hanford epoch & Max $1\sigma$  & Max $1\sigma$  & Max median & Max median\\
			& magnitude [\%] & phase [deg] & magnitude [\%] & phase [deg]\\
			\mr
			(a) Mar 28--Jun 11 & 6.96  & 3.79 & 1.58& 0.86 \\
			(b) Jun 11--Aug 28 & 4.11  & 2.34  &1.15&0.92\\
			(c) Aug 28--Oct 1 & 3.33  & 1.53  &1.42&1.00\\
			\br
			Livingston epoch & Max $1\sigma$  & Max $1\sigma$  & Max median & Max median\\
			& magnitude [\%] & phase [deg] & magnitude [\%] & phase [deg]\\
			\mr
			(a) Mar 28--Jun 11 & 6.37  & 3.49  &1.13&1.59\\
			(b) Jun 11--Oct 1 & 5.99  & 3.68  &1.09&2.09\\
			\br
		\end{tabular}
	\end{indented}
\end{table}

As detectors have become more sensitive and more transient GW events are observed, it is desirable to frequently deliver offline-calibrated data and estimates of the systematic error and uncertainties to GW analyses.
Therefore, in O3A (and for future observing periods), that data and the overall uncertainty for collections of epochs as described above are delivered in $\sim$3-month intervals, the boundaries of which are coincidentally aligned with those of the Hanford epochs.
Balancing the requirements of (a) delivering high-quality data and uncertainty estimates quickly and (b) maintaining systematic error at a level that does not impact astrophysical parameter estimation requires that we do not intend to revise data or estimates of previously vetted intervals, unless circumstances are extraordinary.

\subsection{Interpretation and discussion}
\label{sec:resultsdicussion}
Astrophysical parameter estimation for any GW event in O3A has used the most accurate, offline-calibrated data~\cite{GW190425,GW190412,GW190814,O3A-Catalog}.
The calibration systematic error and uncertainty folded into the parameter estimation is informed by the single hourly ${\eta}_{R}(f;t)$ distribution for the time closest to the event (e.g., \fref{fig:RRnom_with_pcal2darm}).
A five-point interpolation of the frequency-dependent 68\% confidence boundaries is used as an approximation to the full ${\eta}_{R}(f;t)$ distribution \change{at the sample time closest to the event}~\cite{CBCnote-T1400682,Vitale2012}.
In searches for persistent astrophysical signals, the offline calibrated data and the 68\% confidence bounds from each epoch distribution is used as representative of the uncertainty and systematic error estimate for the entire duration of the search (e.g., figures~\ref{fig:h1_results} and \ref{fig:l1_results}).

Throughout all epochs of O3A, the systematic error is less than 2\% in magnitude and 2~deg in phase in the band 20--2000~Hz at both detectors (as indicated by the solid curves in the right panels of figures~\ref{fig:h1_results} and \ref{fig:l1_results}; see \tref{tab:C01_results} also).
As discussed in \sref{sec:resultsovertime}, we expect GW events within a given epoch to have the same estimated systematic error defined by the physical configuration of the detector.
In the first epoch at Hanford [\fref{fig:h1_chunk1}], the uncertainty on the systematic error is larger than usual in the 1--4~kHz frequency band because no measurement had yet been made in that band.
The uncertainty estimate is also larger in the first epoch at Livingston because measurements in the 1--4~kHz band were sparse [\fref{fig:l1_chunk1}]. 
Also at Livingston, the increase in the contribution of the TST actuator to the response function at $\sim$50~Hz that occurred between O2 and O3A [as shown in \fref{fig:l1_contri}] leads to the relatively higher uncertainty around 50~Hz, as shown in \fref{fig:l1_results}.
We anticipate reduction of this uncertainty at $\sim$50~Hz in future observing runs.

The LIGO Scientific Collaboration and Virgo Collaboration use near real-time analyses to quickly process data in search of transient GW sources, enabling multi-messenger astrophysics~\cite{multimessenger-Abbott_2019}.
These analyses use the low-latency estimate of $h$ for detection of GW events and preliminary parameter estimation.
Low-latency data, however, occasionally contains increased systematic errors due to a variety of factors.
The increased systematic errors are often reduced after a short period ($\sim$weeks).
The maximum systematic error in the low-latency data does not exceed 6\% in magnitude and 5~deg in phase at Hanford, and does not exceed 10\% in magnitude and 6~deg in phase at Livingston across the frequency band 20--2000 Hz at any time during O3A. 
This is verified by comparing the low-latency product $hL$ to $\Delta L_{\rm Pcal}$ during various forms of Pcal measurements in the 5--1200~Hz band made periodically throughout O3A (e.g., as shown in \fref{fig:RRnom_with_pcal2darm}).
Although the systematic error and associated uncertainty is larger in the low-latency calibrated data than in the high-latency data, analyses using the low-latency data can nevertheless confidently detect GW events and make rapid astrophysical parameter estimates.
Near real-time analyses have been shown to be robust against calibration errors of this scale~\cite{abbott2016gw150914b}.

\section{Conclusion}
\label{sec:conclusion}
In this paper, we (1) review the procedure for creating a model of the DARM loop used to produce the calibrated data streams, $h$, for the Advanced LIGO detectors; (2) present the systematic errors incurred at each stage of that procedure; and (3) quantify the resulting overall accuracy and precision for the most accurate, offline version of $h$ used for GW astrophysical parameter estimation in O3A. 
The discussion of systematic error includes all known sources, and, where possible, how they have been accounted for in $h$ or in the overall systematic error and uncertainty estimate. 
In O3A, the overall, combined systematic error and associated uncertainty of the most accurate, offline-calibrated data is within 7\% in magnitude and 4~deg in phase in the frequency band 20--2000~Hz.
In this same band, the systematic error alone is estimated to be below 2\% in magnitude and 2~deg in phase.
This is similar to the accuracy and precision as achieved by LIGO in O2~\cite{Cahillane2017}.
Current detection of GW events and estimation of their astrophysical parameters are not yet limited by such levels of uncertainty and systematic error~\cite{O3A-Catalog,CBCnote-T1400682}.

As the global GW detector network sensitivity increases, however, detector calibration systematic error and uncertainty plays an increasingly important role.
Limitations caused by calibration systematics on estimated GW source parameters, precision astrophysics, population studies, cosmology, and tests of general relativity are possible. 
For example, correlated systematic errors in the estimated luminosity distance of high-SNR GW events due to calibration systematic errors could bias estimates of the cosmological Hubble constant, $H_0$.
Efforts to \change{better} integrate the work presented in this paper into future GW event astrophysical parameter estimation are ongoing, including the use of the full, numerically evaluated, distribution of systematic error and uncertainty.
These efforts will enable quantifying the impact of calibration systematics on individual GW events as well as studies that rely on a population of GW events.
Additionally, these efforts guide research and development of new techniques currently underway to further reduce combined calibration systematic error and uncertainty below the 1\% level, a key milestone towards minimizing impacts of calibration systematics on astrophysical and cosmological results.

\section{Acknowledgments}

The authors gratefully acknowledge the operators, commissioners, and LSC fellows at Hanford and Livingston for their help in setting up the detector configurations and taking measurements needed for this work.
The authors also gratefully acknowledge Jonathan Gair, Keita Kawabe and Lo\"{i}c Rolland for the review and comments.
LIGO was constructed by the California Institute of Technology and Massachusetts Institute of Technology with funding from the United States National Science Foundation (NSF), and operates under cooperative agreement PHY--1764464. Advanced LIGO was built under award PHY--0823459. 
The authors gratefully acknowledge the support of the United States NSF for the construction and operation of the
LIGO Laboratory and Advanced LIGO as well as the Science and Technology Facilities Council (STFC) of the
United Kingdom, the Max-Planck-Society (MPS), and the State of
Niedersachsen/Germany for support of the construction of Advanced LIGO 
and construction and operation of the GEO600 detector. 
Additional support for Advanced LIGO was provided by the Australian Research Council (ARC).
LS, DB, VB, PBC, LEHD, RG, TM, EP, AB and CC acknowledge the LSC Fellows program for supporting their research at LIGO sites.
EG acknowledges the support of the Natural Sciences and Engineering Research Council (NSERC) of Canada.
DB and SK are supported by NSF award PHY--1921006. 
AV is supported by NSF award PHY--1841480.
MW is supported by NSF awards PHY--1607178 and PHY--1847350.
VB and EP acknowledge the supported of the ARC Centre of Excellence for Gravitational Wave Discovery (OzGrav), grant number CE170100004.
PBC acknowledges the support of the Spanish Agencia Estatal de Investigaci{\'o}n and Ministerio de Ciencia, Innovaci{\'o}n y Universidades grants FPA2016-76821-P, the Vicepresidencia i Conselleria d'Innovaci{\'o}, Recerca i Turisme del Govern de les Illes Balears (grant FPI-CAIB FPI/2134/2018), the Fons Social Europeu 2014-2020 de les Illes Balears, the European Union FEDER funds, and the EU COST actions CA16104, CA16214, CA17137 and CA18108.
The authors would like to thank all of the essential workers who put their health at risk during the COVID-19 pandemic, 
without whom we would not have been able to complete this work.
This paper carries LIGO Document Number \dcc.

\clearpage

\appendix
\section{Impact of uncorrected TDCFs}
\label{appendix:carpet}

The figures in this appendix show the impact of uncorrected TDCFs, $f_{cc}$, $\kappa_{T}$, $\kappa_{P}$, and $\kappa_{U}$, on Hanford detector response (see details in section~\ref{sec:tdcf}).

\begin{figure}[!tbh]
	\centering
	\scalebox{0.3}{\includegraphics{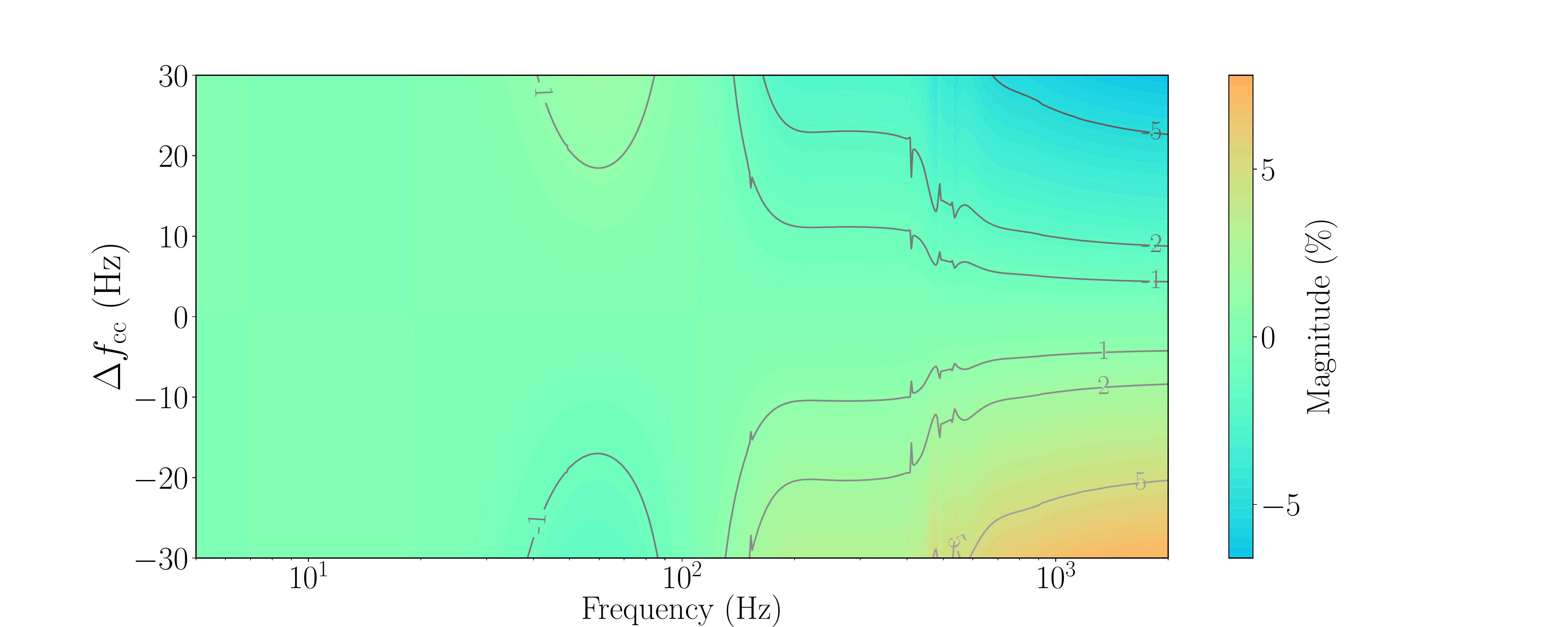}}
	\scalebox{0.3}{\includegraphics{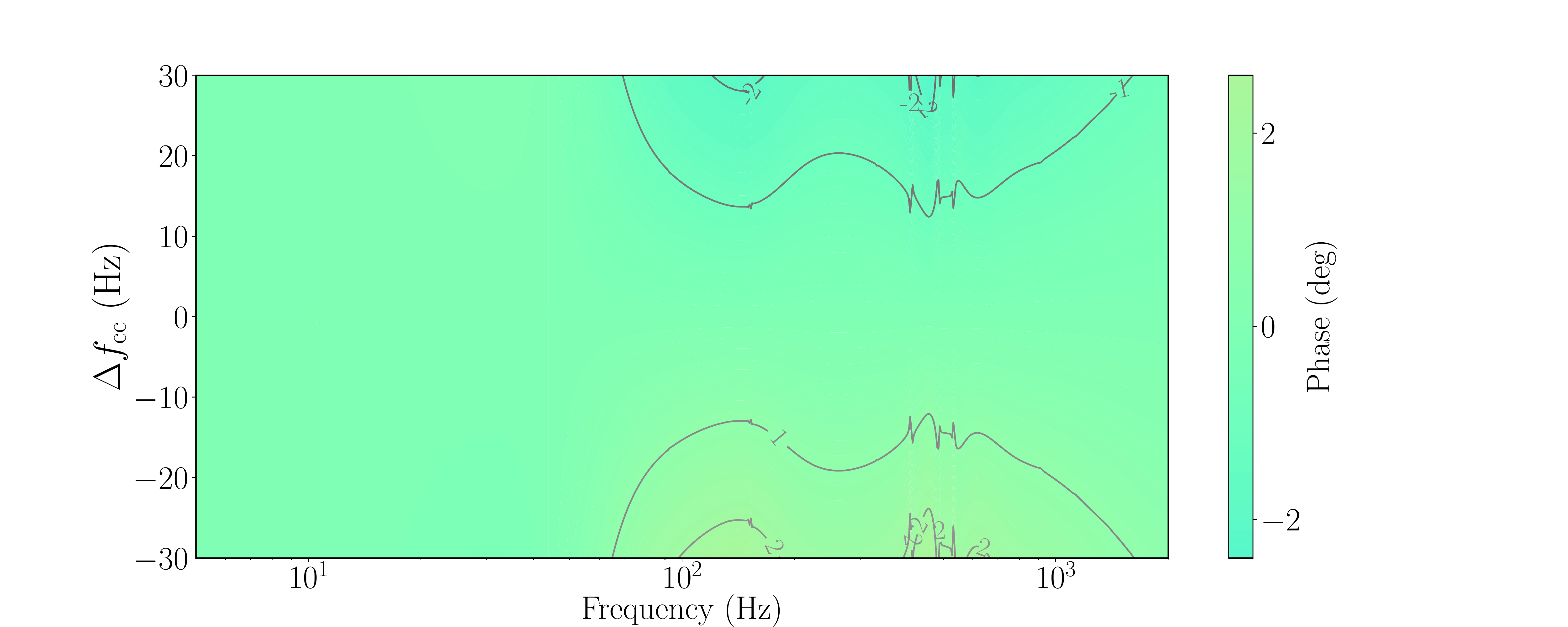}}
	\caption[]{Magnitude (top) and phase (bottom) of the fractional error ${\eta}_{R;C}-1$ in the Hanford detector response [O3A Epoch (c)] as a function of frequency due to uncorrected time-dependent coupled cavity pole frequency, $f_{cc}$. The reference value of $f_{cc}$ is 410.6~Hz.}
	\label{fig:carpet_fcc}
\end{figure}

\pagebreak

\begin{figure}[!tbh]
	\centering
	\scalebox{0.3}{\includegraphics{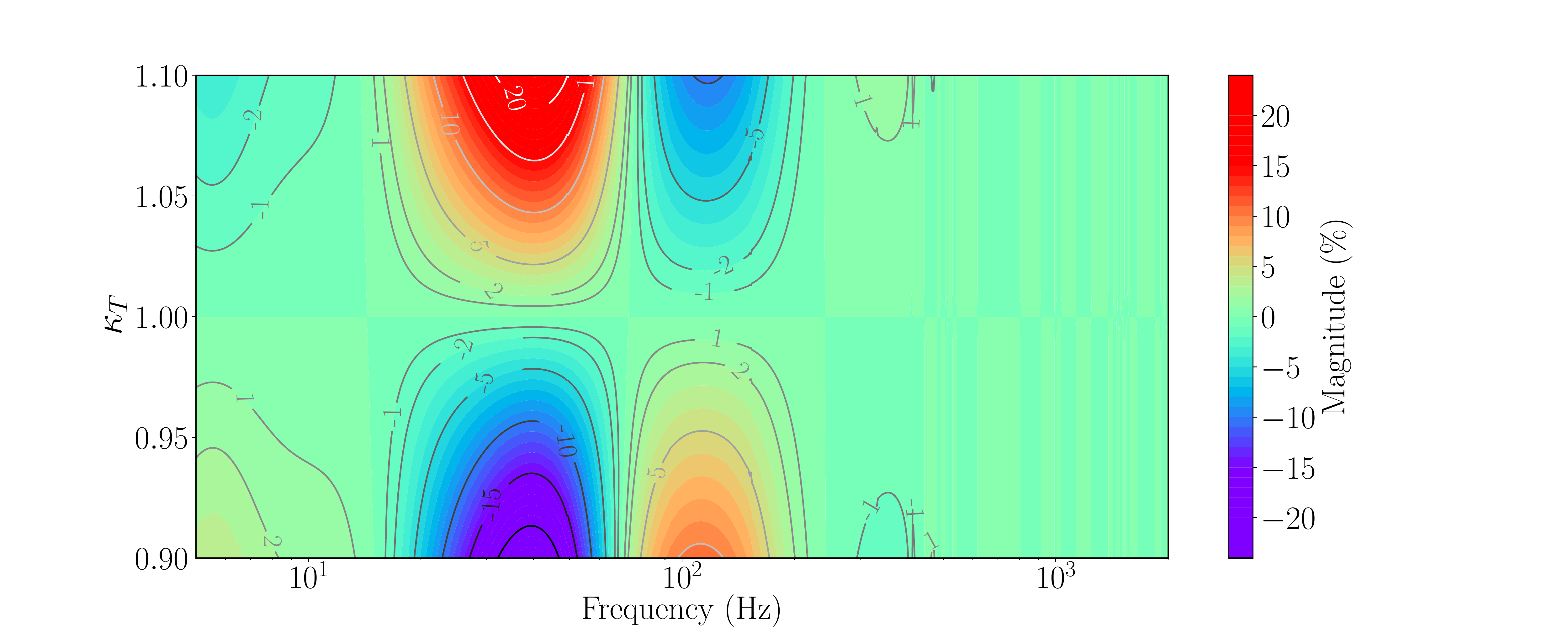}}
	\scalebox{0.3}{\includegraphics{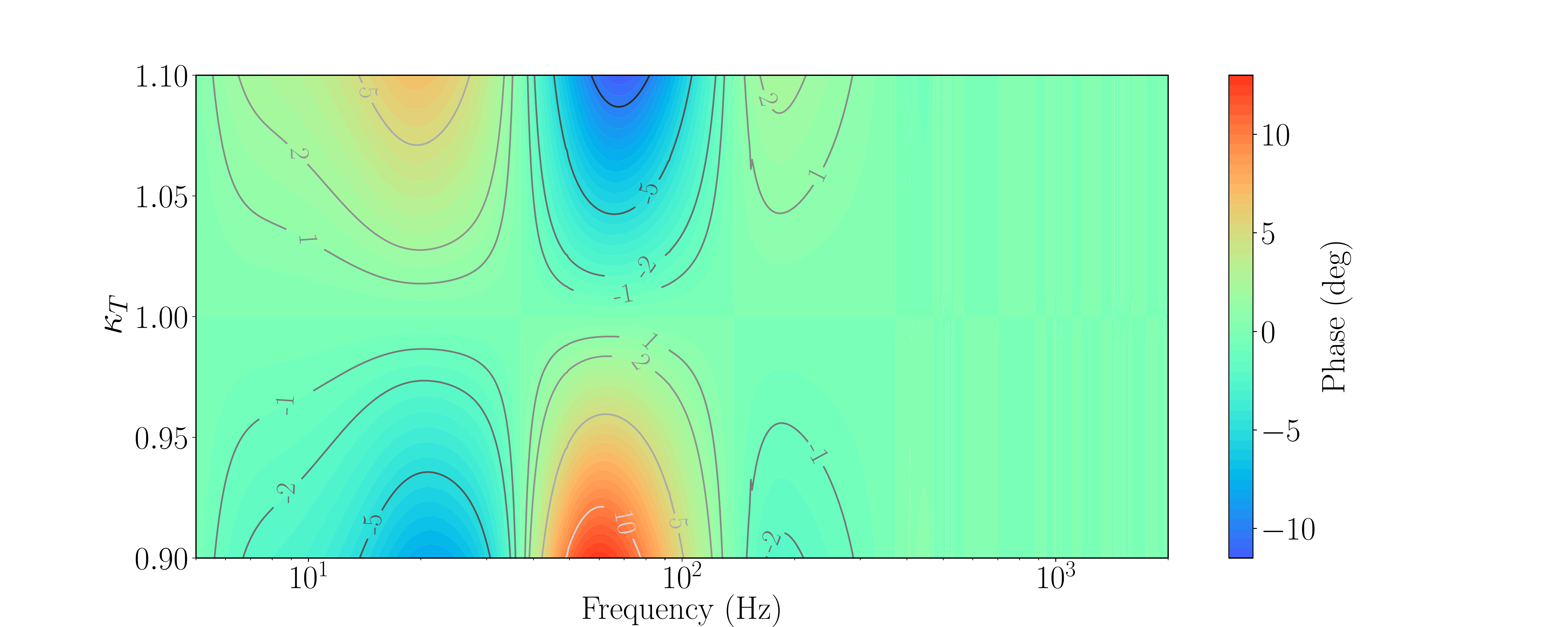}}
	\caption[]{Magnitude (top) and phase (bottom) of the fractional error ${\eta}_{R;A_T}-1$ in the Hanford detector response [O3A Epoch (c)] as a function of frequency due to uncorrected gain variations in the TST actuation stage, tracked by the scalar time-dependent factor, $\kappa_{T}$.}
	\label{fig:carpet_kt}
\end{figure}

\pagebreak

\begin{figure}[!tbh]
	\centering
	\scalebox{0.3}{\includegraphics{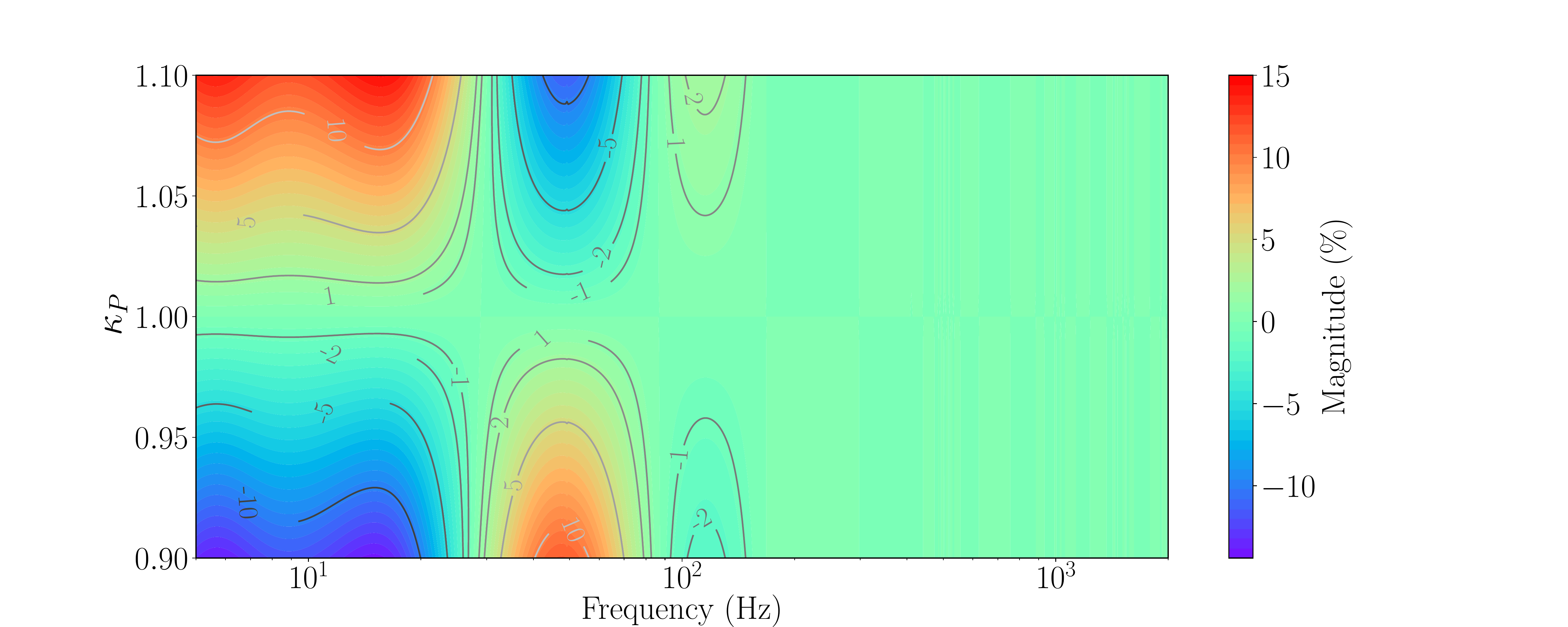}}
	\scalebox{0.3}{\includegraphics{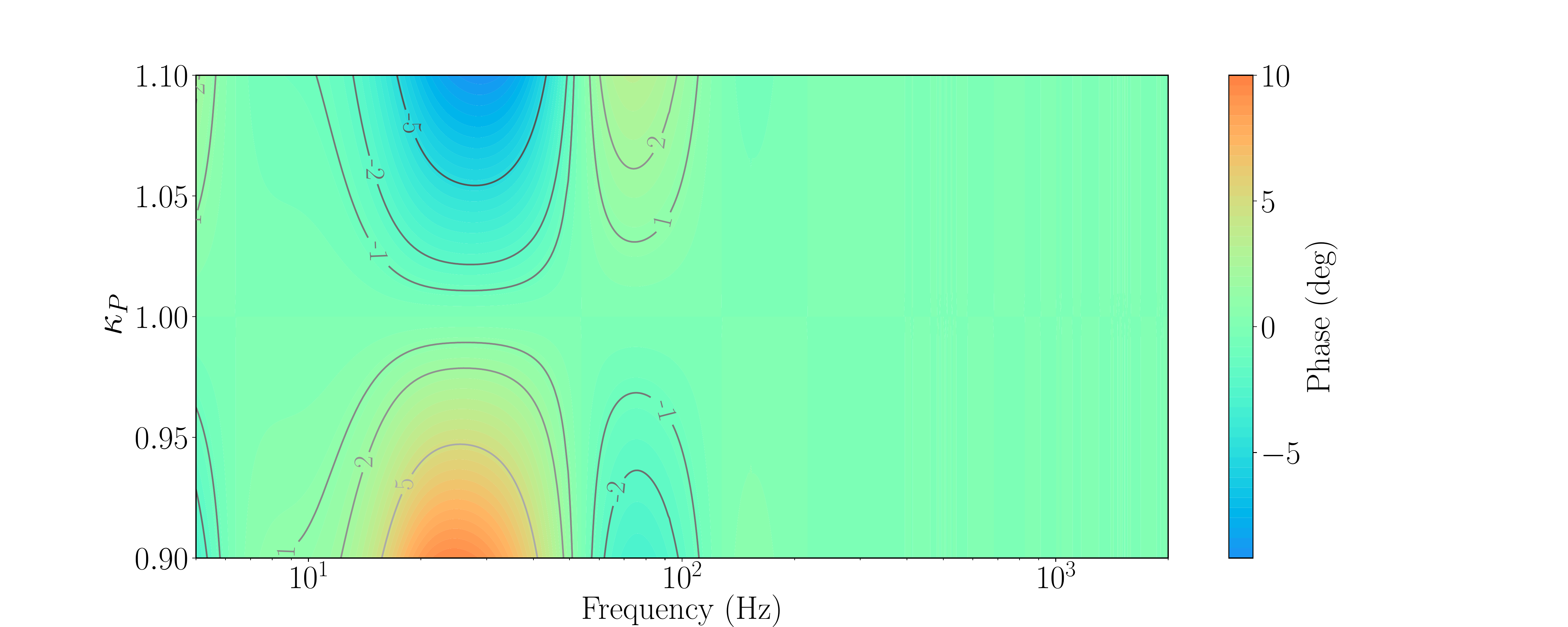}}
	\caption[]{Magnitude (top) and phase (bottom) of the fractional error ${\eta}_{R;A_P}-1$ in the Hanford detector response [O3A Epoch (c)] as a function of frequency due to uncorrected gain variations in the PUM actuation stage, tracked by the scalar time-dependent factor, $\kappa_{P}$.}
	\label{fig:carpet_kp}
\end{figure}

\pagebreak

\begin{figure}[!tbh]
	\centering
	\scalebox{0.3}{\includegraphics{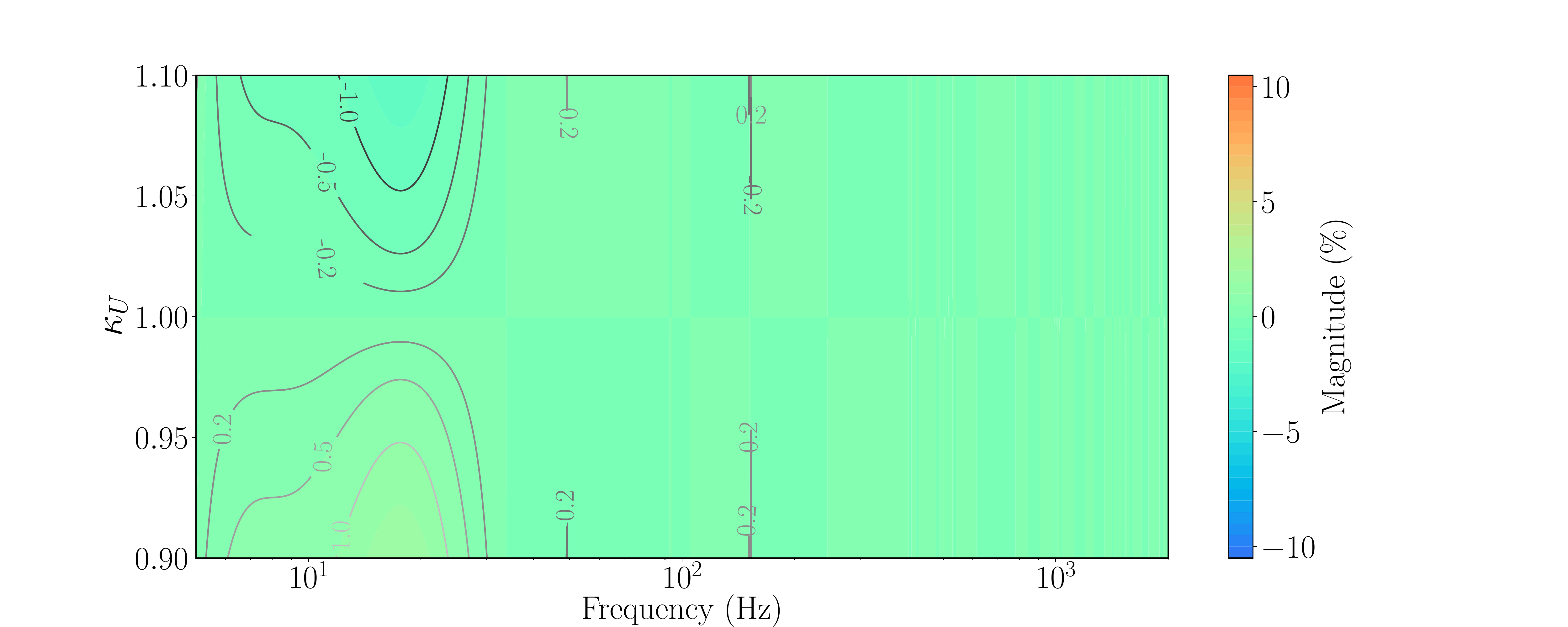}}
	\scalebox{0.3}{\includegraphics{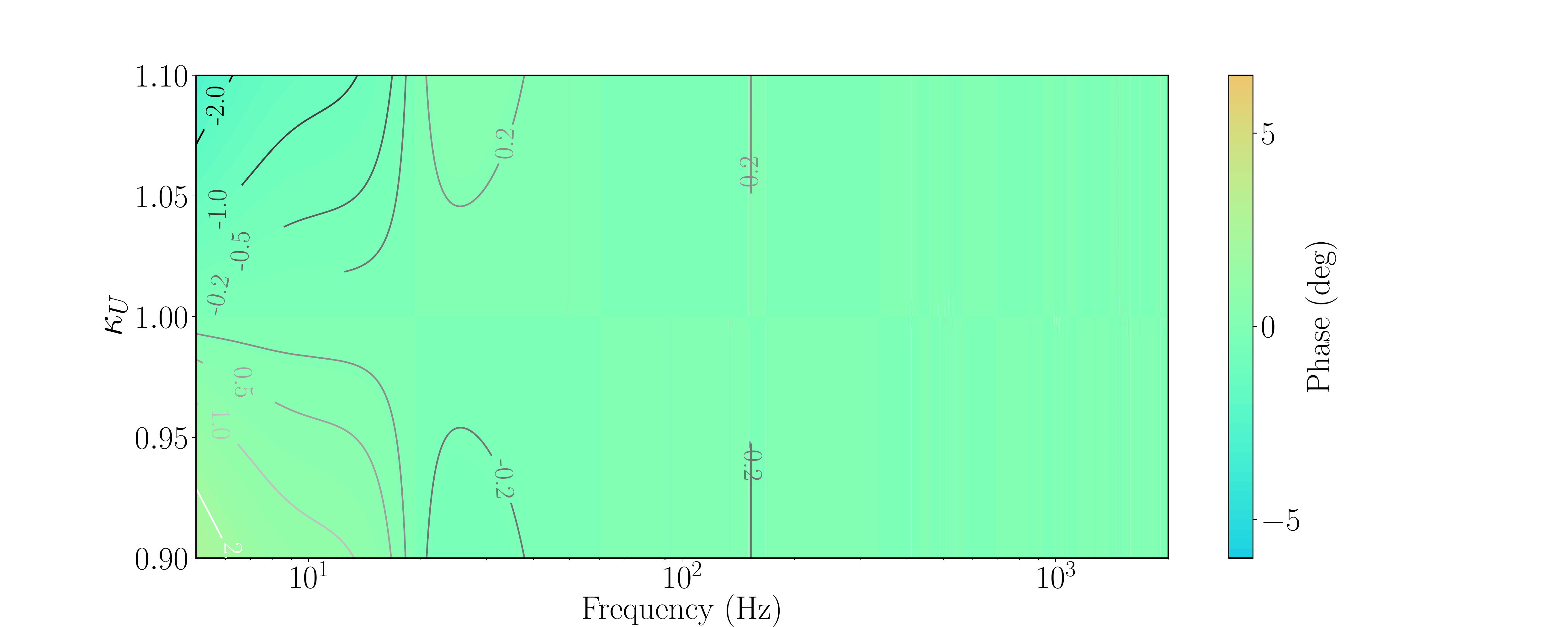}}
	\caption[]{Magnitude (top) and phase (bottom) of the fractional error ${\eta}_{R;A_U}-1$ in the Hanford detector response [O3A Epoch (c)] as a function of frequency due to uncorrected gain variations in the UIM actuation stage, tracked by the scalar time-dependent factor, $\kappa_{U}$.}
	\label{fig:carpet_ku}
\end{figure}

\section{Impact of complex-valued actuator TDCFs in early O3A}
\label{appendix:imag_kappa}

The figure in this appendix shows the estimates of systematic error and associated uncertainty using the collection of percentile curves of ${\eta}_{R}(f;t)$ in the first epoch. 
The comparison of two cases are shown side-by-side: only the real-valued actuator TDCFs are applied (left) and the full complex-valued actuator TDCFs are applied from April 16 to June 11, 2019 at Hanford and from April 1 to June 11, 2019 at Livingston (right).
See discussions in section~\ref{sec:tdcf}.

\begin{figure}[!tbh]
	\centering
	\subfigure[]
	{
		\label{fig:h1_epoch1_real}
		\scalebox{0.20}{\includegraphics{percentiles_O3_C01_190328-190611_LHO.pdf}}
		\scalebox{0.20}{\includegraphics{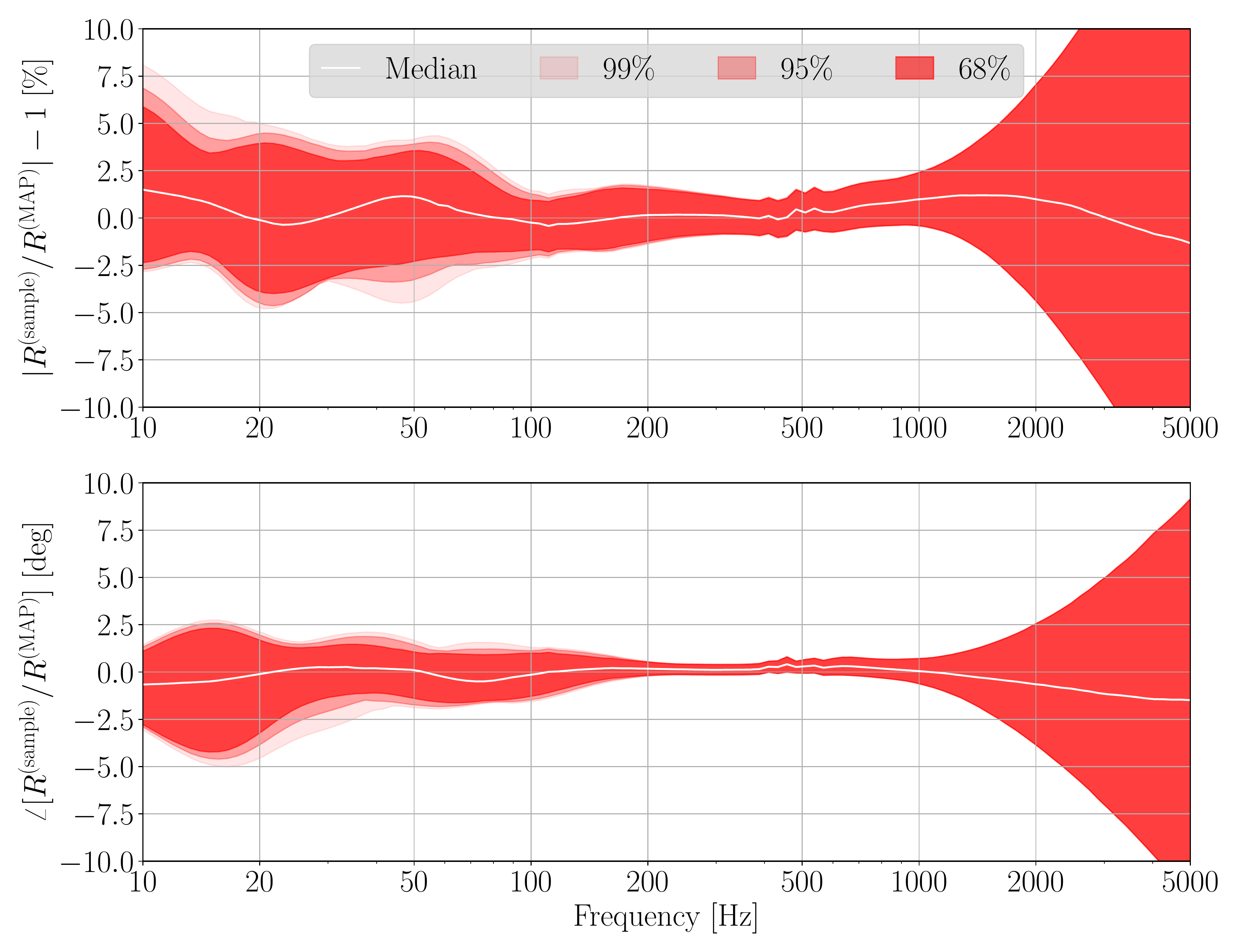}}
	}
	\subfigure[]
	{
		\label{fig:h1_epoch1_imag}
		\scalebox{0.20}{\includegraphics{percentiles_O3_C01_190327-190611_LLO.pdf}}
		\scalebox{0.20}{\includegraphics{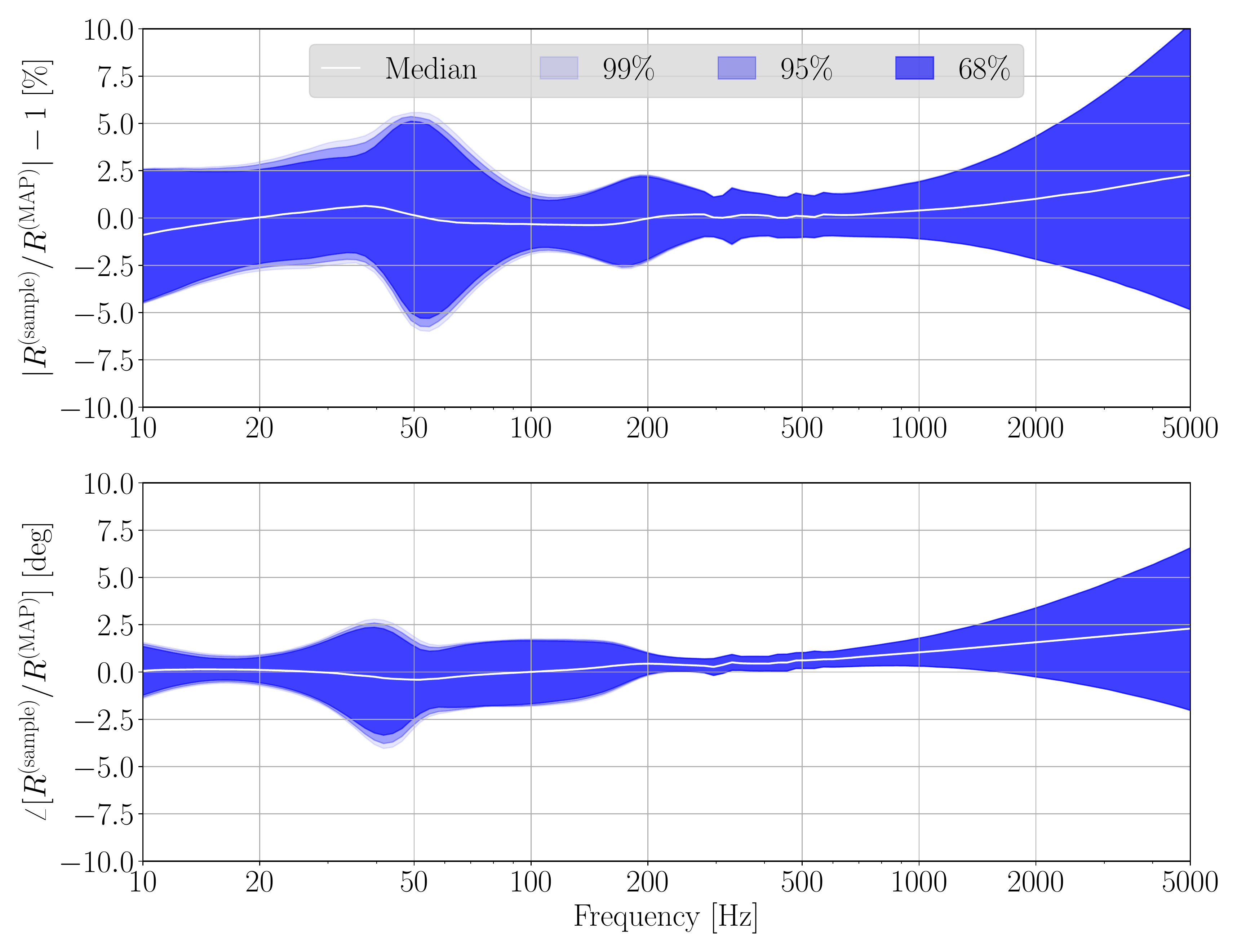}}
	}
	\caption[]{Variation of the combined systematic error and uncertainty for (a) Hanford and (b) Livingston in the first epoch in O3A. The top and bottom panels of each subfigure show the frequency dependent excursions of response from unity magnitude and zero phase compared to ${R}_{\rm MAP}$, respectively. The percentiles are obtained from all the hourly evaluated ${\eta}_R(f;t_k)$ over the first epoch. 
	In the left panels, the results are obtained when only the real-valued actuator TDCFs are applied [equivalent to figures \ref{fig:h1_chunk1} and \ref{fig:l1_chunk1} in \sref{sec:resultsovertime}]. 
	In the right panels, the results are obtained when complex-valued actuator TDCFs are applied from April 16 to June 11, 2019 at Hanford and from April 1 to June 11, 2019 at Livingston. 
	The colors represent $1\sigma$ uncertainty for 68\%, 95\%, and 99\% of the run time, as indicated in the legend. The white curve indicate the median excursion. }
	\label{fig:imag_kappas_impact}
\end{figure}

\pagebreak

\section{\change{Distribution of $\eta_{R}$ at a given time and a given frequency}}
\label{appendix:eta_dist}

\begin{figure}[!tbh]
	\centering
		\scalebox{0.31}{\includegraphics{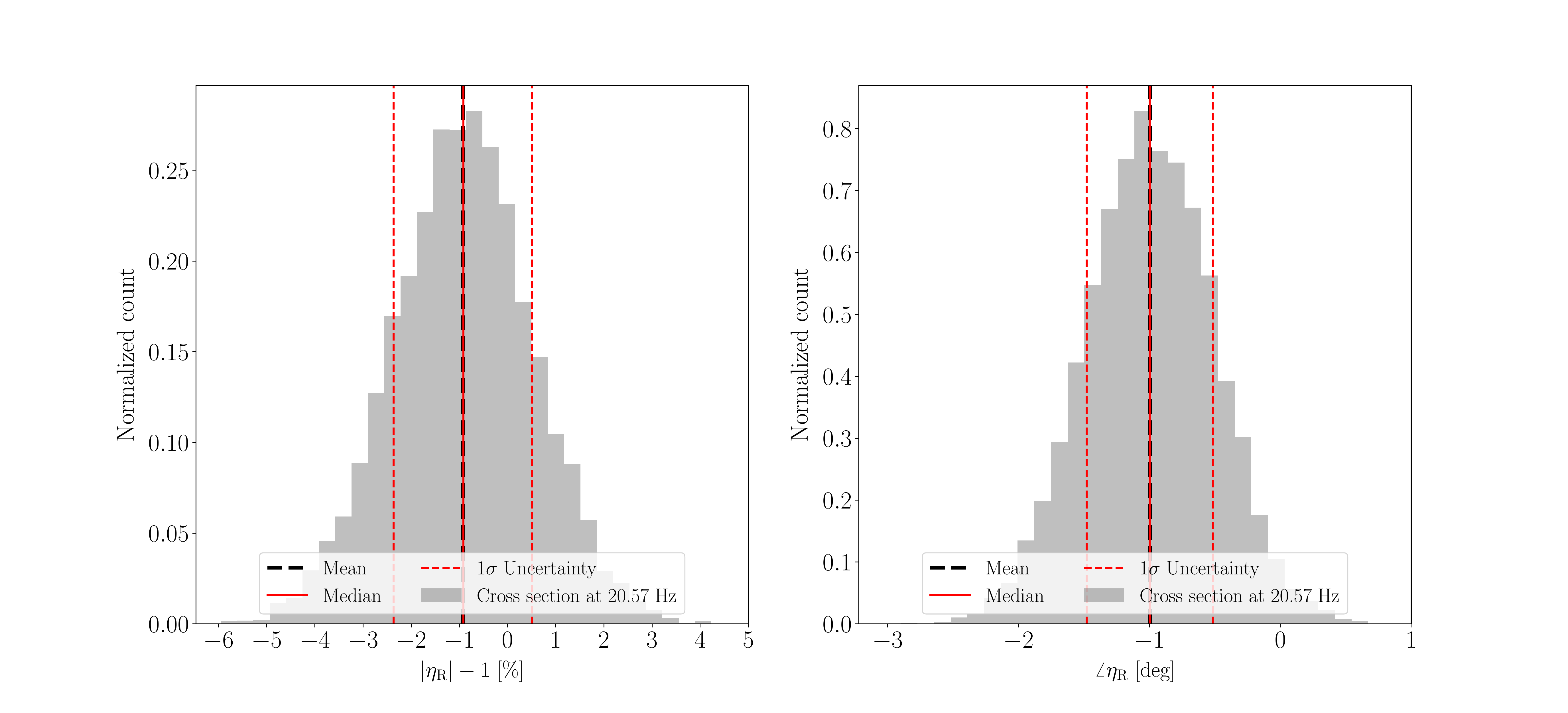}}
	\caption[]{\change{Distribution of $\eta_{R}$ at a cross section of 20.57~Hz at the reference time of Epoch (c) for the Hanford detector. The left and right panels show the excursions of the response from unity magnitude and zero phase compared to ${R}_{\rm MAP}$ at 20.57~Hz, respectively. The red solid and dashed lines indicate the median and $1\sigma$ values of the distribution. The black dashed line indicates the mean value of the distribution. The median and mean values generally overlap each other. This figure is equivalent to the $\eta_{R}$ distribution at a cross section of 20.57~Hz in \fref{fig:RRnom_with_pcal2darm}.}}
	\label{fig:eta_R_20Hz}
\end{figure}

\clearpage
\newpage

\providecommand{\newblock}{}

\end{document}